\documentclass[numbook,natbib,final,runningheads]{svjour3}
\usepackage{graphicx,epsfig,amssymb,mathptmx} 
\usepackage{makeidx} 
\makeindex

\newcommand{\degr}{\ifmmode^\circ\else$^\circ$\fi}

\newcommand{\lapprox} {\, \lower3pt\hbox{$\sim$}\llap{\raise2pt\hbox{$<$}}\,}
\newcommand{\gapprox} {\, \lower3pt\hbox{$\sim$}\llap{\raise2pt\hbox{$>$}}\,}

\begin{document}

\newcommand{\diff}[2]{\frac{\partial #1}{\partial #2}}
\newcommand{\difff}[2]{\frac{\partial^2 #1}{\partial #2^2}}

\title{Implications of X-ray Observations for Electron Acceleration 
and Propagation in Solar Flares}

\author{G.D. Holman$^{1}$,
        M.~J.~Aschwanden$^{2}$,
        H.~Aurass$^{3}$,
        M.~Battaglia$^{4}$,
        P.~C.~Grigis$^{5}$,
        E.~P.~Kontar$^{6}$,
        W.~Liu$^{1}$,
        P.~Saint-Hilaire$^{7}$, and
        V.~V.~Zharkova$^{8}$}

\institute{$^{1}$ NASA Goddard Space Flight Center, Code 671,
                  Greenbelt, MD 20771, U.S.A.
                  \email{Gordon.D.Holman@nasa.gov}\\
           $^{2}$ Lockheed Martin Advanced Technology Center,
                  Solar and Astrophysics Laboratory, Organization ADBS,
                  Building 252, 3251 Hanover Street, Palo Alto, CA 94304, U.S.A.\\
           $^{3}$ Astrophysikalisches Institut Potsdam \\
           $^{4}$ Department of Physics \& Astronomy, University of Glasgow, Glasgow, G12 8QQ, Scotland, UK\\
	   $^{5}$ Harvard-Smithsonian Center for Astrophysics P-148, 
		  60 Garden St., Cambridge, MA 02138, U.S.A.\\
           $^{6}$ Department of Physics and Astronomy, University
                  of Glasgow, Kelvin Building, Glasgow, G12 8QQ, U.K. \\
           $^{7}$ Space Sciences Lab, UC Berkeley, CA, U.S.A. \\
           $^{8}$ Department of Computing and Mathematics, University of Bradford, Bradford, BD7 1DP, UK \\
             }

\date{\today}

\authorrunning{Holman et al.}
\titlerunning{Electron Acceleration and Propagation}

\maketitle

\begin{abstract}

High-energy X-rays and $\gamma$-rays
from solar flares were discovered just over fifty years ago.  Since
that time, the standard for the interpretation of spatially integrated
flare X-ray spectra at energies above several tens of keV has been
the collisional thick-target model.  After the launch of the {\it
Reuven Ramaty High Energy Solar Spectroscopic Imager} ({\it RHESSI}) in
early 2002, X-ray spectra and images have been of sufficient quality
to allow a greater focus on the energetic electrons responsible for
the X-ray emission, including their origin and their interactions
with the flare plasma and magnetic field.  The result has been new
insights into the flaring process, as well as more quantitative
models for both electron acceleration and propagation, and for the
flare environment with which the electrons interact.  In this article
we review our current understanding of electron acceleration, energy
loss, and propagation in flares.  Implications of these new results
for the collisional thick-target model, for general flare models,
and for future flare studies are discussed.  

\end{abstract}


\tableofcontents

\section{Introduction} 
\label{sec:holman_introduction}

A primary characteristic of solar flares is the acceleration of
electrons to high, suprathermal energies.  These electrons are
observed directly in interplanetary space, and indirectly at the
Sun through the X-ray, $\gamma$-ray, and radio emissions they emit
\citep{1995ARA&A..33..239H}.  Understanding how these electrons are
produced and how they evolve is fundamental to obtaining an
understanding of energy release in flares.  Therefore, one of the
principal goals of solar flare research is to determine when, where,
and how these electrons are accelerated to suprathermal energies,
and what happens to them after they are accelerated to these high
energies.

A major challenge to obtaining an understanding of electron
acceleration in flares is that the location where they are accelerated
is not necessarily where they are most easily observed.  The
flare-accelerated electrons that escape the Sun are not directly
observed until they reach the instruments in space capable of
detecting them, usually located at the distance of the Earth from
the Sun.  The properties of these electrons are easily modified
during their long journey from the flaring region to the detecting
instruments \citep[e.g., ][]{2009A&A...507..981A}.  Distinguishing
flare-accelerated electrons from electrons accelerated in interplanetary
shock waves\index{shocks!interplanetary!electron acceleration} 
is also difficult \citep{2007SSRv..129..359K}.

The electrons that are observed at the Sun through their X-ray or
$\gamma$-ray emissions radiate most intensely where the density of
the ambient plasma is highest (see Section~\ref{sec:holman_thick}).
Therefore, the radiation from electrons in and near the acceleration
region may not be intense enough to be observable.  Although these
radiating electrons are much closer to the acceleration region than
those detected in interplanetary space, their properties can still
be significantly modified as they propagate to the denser regions
where they are observed.\index{acceleration region!distinguished from energy-loss region}
The radio emission
\index{radio emission!gyrosynchrotron}
from the accelerated electrons also depends on the plasma environment,
especially the magnetic field strength for the gyrosynchrotron
radiation observed from flares \citep{1998ARA&A..36..131B}.  Therefore,
determining when, where, and how the electrons were accelerated
requires a substantial amount of deductive reasoning.

Here we focus primarily on the X-ray emission from the accelerated
electrons\index{hard X-rays}.  Interplanetary electrons and low-energy emissions are
addressed by \citet{Chapter2}, while the $\gamma$-ray emission is
addressed by \citet{Chapter4}, and the radio by \citet{Chapter5}.
The X-rays are predominantly {\em electron-ion
bremsstrahlung}\index{bremsstrahlung} (free-free radiation), emitted
\index{free-free emission}
when the accelerated electrons scatter off ions in the ambient
thermal plasma.  Issues related to the emission mechanism and
deducing the properties of the emitting electrons from the X-ray
observations are primarily addressed by \citet{Chapter7}.  Here we
address the interpretation of the X-ray observations in terms of
flare models, and consider the implications of the observations for
the acceleration process, energy release in flares, and electron
propagation.  Specific models for particle acceleration and energy
release in flares are addressed by \citet{Chapter8}.

The accelerated electrons interact with both ambient electrons and
ions, but lose most of their energy through {\em electron-electron
Coulomb collisions}.  Consequently, the brightest X-ray sources are
associated with high collisional energy losses.
\index{collisions!particle energy losses}
These losses in turn change the energy distribution
of the radiating electrons.  When the accelerated electrons lose
their suprathermal energy to the ambient plasma as they radiate,
the source region is called a {\em thick target}.\index{hard X-rays!thick-target}\index{thick-target model!description}\index{suprathermal populations!and thick-target model}
Electrons streaming downward into the higher densities in lower
regions of the solar atmosphere, or trapped long enough in lower
density regions, will emit thick-target X-rays.  Hence, thick-target
models are important to understanding the origin and evolution of
accelerated electrons in flares.  Thick-target X-ray emission is
addressed in Section~\ref{sec:holman_thick}.

The total energy carried by accelerated electrons is important to
assessing acceleration models, especially considering that these
electrons carry a significant fraction of the total energy released
in flares\index{electrons!accelerated!energy content}.  
Also, the energy carried by electrons that escape the
acceleration region is deposited elsewhere, primarily to heat the
plasma in the thick-target source regions.\index{acceleration region!distinguished from energy-loss region}
The X-ray flux from
flares falls off rapidly with increasing photon energy, indicating
that the number of radiating electrons decreases rapidly with
increasing electron energy.  Therefore, the energy carried by the
accelerated electrons is sensitive to the value of the low-energy
cutoff
\index{electrons!distribution function!low-energy cutoff}
\index{low-energy cutoff}
to the electron distribution.  The determination of this low-energy
cutoff and the total energy in the accelerated electrons is addressed
in Section~\ref{sec:holman_cutoffs}.

In the standard thick-target model, the target plasma is assumed
to be fully ionized.\index{ionization state}\index{thick-target model!standard} 
If the target ionization
is not uniform\index{non-uniform ionization}, so that the accelerated
electrons stream down to cooler plasma that is partially ionized
or un-ionized, the X-ray spectrum is modified.  This is addressed
in Section~\ref{sec:kontar_nonuniform}.

Observations of the radiation from hot flare plasma have shown this
plasma to primarily be confined to magnetic loops or arcades of
magnetic loops \citep[cf.\ ][]{2004psci.book.....A}.  The observations
also indicate that the heating of this plasma and particle acceleration
initially occur in the corona above these hot loops \citep[see
Section~\ref{sec:holman_discussion_models} and][]{Chapter2}.  When
the density structure in these loops is typical of active region
loops, or at least not highly enhanced above those densities, the
highest intensity, thick-target X-ray emission will be from the
footpoints\index{hard X-rays!footpoint sources}
\index{footpoints!hard X-rays}
of the loops, as is most often observed to be the case.  If accelerated
electrons alone, unaccompanied by neutralizing ions, stream down
the legs of the loop from the acceleration region to the footpoints,
they will drive a co-spatial return current\index{return current}\index{acceleration region!return current}
in the ambient plasma to neutralize the high current associated
with the downward-streaming electrons.
We refer to this primarily
downward-streaming distribution of energetic (suprathermal) electrons
as an electron beam\index{electrons!distribution function!beam}.
The electric field\index{electric fields!self-field} associated with the return current will
decelerate electrons in the beam, which can in turn modify the X-ray
spectrum from the accelerated electrons.  This is addressed in
Section~\ref{sec:zharkova_return_current}.

Both the primary beam of accelerated electrons and the return current
can become unstable and drive the growth of waves in the ambient
plasma.  These waves can, in turn, interact with the electron beam
and return current, altering the energy and angular distributions
of the energetic electrons.  These plasma instabilities are discussed
in Section~\ref{sec:holman_instabilities}.

The collisional energy loss rate\index{collisions!particle energy losses}
is greater for lower energy electrons.  Therefore, for suprathermal
electrons streaming downward to the footpoints of a loop, the
footpoint X-ray sources observed at lower energies should be at a
higher altitude than footpoint sources observed at higher X-ray
energies.  
The height dispersion\index{hard X-rays!height dispersion} of these sources provides information
about the height distribution of the plasma density in the footpoints.
The spatial resolution\index{RHESSI@\textit{RHESSI}!spatial resolution} of the {\it
Reuven Ramaty High Energy Solar Spectroscopic Imager} ({\it RHESSI} -- see
Lin et al. 2002)\index{RHESSI@\textit{RHESSI}} has made such a study
possible.  
\index{satellites!RHESSI@\textit{RHESSI}}
This is described in Section~\ref{sec:aschwanden_height}.
\nocite{2002SoPh..210....3L}
{\it RHESSI} has observed X-ray sources that move downward from the
loop top and then upward from the footpoints during some flares.
This source evolution in time is also discussed in
Section~\ref{sec:aschwanden_height}.

If electrons of all energies are simultaneously injected, the
footpoint X-ray emission from the slower, lower energy electrons
should appear after that from faster, higher energy electrons.  
The length of this time delay\index{hard X-rays!time delays} provides an
important test for the height of the acceleration region\index{acceleration region!height}. 
Longer time delays can result from magnetic trapping\index{trapping!magnetic} of the electrons.  The evolution of the thermal plasma in flares can also exhibit time delays
associated with the balance between heating and cooling processes.
These various time delays and the information they provide are
addressed in Section~\ref{sec:aschwanden_timing}.

An important diagnostic of electron acceleration and propagation
in flares is the time evolution of the X-ray spectrum during flares.
In most flares, the X-ray spectrum\index{hard X-rays!soft-hard-soft} 
becomes harder (flatter, smaller spectral
index\index{hard X-rays!spectral index}) and then softer (steeper, larger
spectral index) as the X-ray flux evolves from low to high intensity
and then back to low intensity.  There are notable exceptions to
this pattern, however.  Spectral evolution is addressed in
Section~\ref{sec:grigis_spectral_evolution}.

One of the most important results from the {\it Yohkoh}\index{Yohkoh@\textit{Yohkoh}}
mission is the discovery in some flares of a hard (high energy)
X-ray source above the top of of the thermal (low energy) X-ray
loops\index{flare (individual)!SOL1992-01-13T17:25 (M2.0)!above-the-looptop source}.  
This, together with the {\it Yohkoh} observations of cusps\index{magnetic structures!cusps}
at the top of flare X-ray loops, provided strong evidence that
energy release occurs in the corona above the hot X-ray loops (for
some flares, at least).  
Although several models have been proposed,
the origin of these ``above-the-looptop'' hard X-ray sources\index{hard X-rays!above-the-looptop source} is not well understood.  
We need to determine how their properties and
evolution compare to the more intense footpoint hard X-ray sources.
These issues are addressed in
Section~\ref{sec:battaglia_footpoint_coronal}.

Radio observations\index{radio emission} provide another view of
accelerated electrons and related flare phenomena.  Although radio
emission and its relationship to flare X-rays are primarily addressed
in \citet{Chapter5}, some intriguing radio observations that bear
upon electron acceleration in flares are presented in
Section~\ref{sec:aurass_implications}.

{\it RHESSI} observations of flare X-ray emission have led to
substantial progress, but many questions remain unanswered.  Part
of the progress is that the questions are different from those that
were asked less than a decade ago.  The primary context for
interpreting the X-ray emission from suprathermal electrons is still
the thick-target model, but the ultimate goal is to understand how
the electrons are accelerated.  In Section~\ref{sec:holman_discussion}
we summarize and discuss the implications of the X-ray observations
for the thick-target model and electron acceleration mechanisms,
and highlight some of the questions that remain to be answered.
Implications of these questions for future flare studies are
discussed.  

\section{Thin- and thick-target X-ray emission} 
\label{sec:holman_thick}
\index{thin target}\index{thick-target model!summary}

As was summarized in
Section~\ref{sec:holman_introduction}, the electron-ion bremsstrahlung
X-rays from a beam of accelerated electrons will be most intense
where the density of target ions is highest, as well as where the
flux of accelerated electrons is high.  The local emission (emissivity)
at position ${\bf r}$ of photons of energy $\epsilon$ by electrons
of energy $E$ is given by the plasma ion density, $n({\bf r})$ ions
cm$^{-3}$, times the electron-beam flux density distribution,
$F({E,\bf r})$ electrons
cm$^{-2}$ s$^{-1}$ keV$^{-1}$, times the differential electron-ion
bremsstrahlung cross-section, $Q(\epsilon,E)$ cm$^2$ keV$^{-1}$\index{electrons!distribution function!flux density}\index{bremsstrahlung!cross-section}.
For simplicity, we do not
consider here the angular distribution of the beam electrons or of
the emitted photons, topics addressed in \citet{Chapter7}.  

The emissivity of the radiation at energy $\epsilon$ from all the
electrons in the beam is obtained by integrating over all contributing
electron energies, which is all electron energies above the photon
energy.  The photon flux emitted per unit energy is obtained by
integrating over the emitting source volume ($V$) or, for an imaged
source, along the line of sight through the source region.  Finally,
assuming isotropic emission, the observed spatially integrated flux
density of photons of energy $\epsilon$ at the X-ray detector,
$I(\epsilon)$ photons cm$^{-2}$ s$^{-1}$ keV$^{-1}$, is simply the
flux divided by the geometrical dilution factor $4 \pi R^2$, where
$R$ is the distance to the X-ray detector.  Thus,

\begin{equation} I(\epsilon) = {1 \over 4 \pi R^2} \int_V \,
\int_\epsilon^\infty \, n({\bf r}) \, F(E, {\bf r}) \, Q(\epsilon,
E) \, dE \, dV.  \label{eqn:holman_obsflux} \end{equation}

We refer to $I(\epsilon)$ as the X-ray flux spectrum, or simply the
X-ray spectrum.  The spectrum obtained directly from an X-ray
detector is generally a spectrum of counts versus energy loss in
the detector, which must be converted to an X-ray flux spectrum by
correcting for the detector response as a function of photon energy
\citep[see, for example,][]{2002SoPh..210...33S}.

Besides increasing the X-ray emission, a high plasma density also
means increased Coulomb energy losses for the beam electrons.  
In a plasma, the brems\-strah\-lung losses are small compared
to the collisional losses to the plasma electrons.  
\index{collisions!particle energy losses} 
For a fully ionized plasma and beam electron
speeds much greater than the mean speed of the thermal electrons,
the (nonrelativistic) energy loss rate is 
\begin{equation} dE/dt =
- (K/E) \, n_e({\bf r}) \, {\rm v}(E), \label{eqn:holman_colloss}
\end{equation} 
where $K = 2 \pi e^4 \, \Lambda_{ee}$, $\Lambda_{ee}$
is the Coulomb logarithm\index{Coulomb logarithm!electron-electron collisions} 
for electron-electron collisions, $e$~is the electron charge, $n_e({\bf
r})$ is the plasma electron number density, and v$(E)$ is the speed
of the electron \citep[see][]{1971SoPh...18..489B,1978ApJ...224..241E}.
The coefficient $K$ is usually taken to be constant, although
$\Lambda_{ee}$ depends weakly on the electron energy and plasma
density or magnetic field strength, typically falling in the range
20 -- 30 for X-ray-emitting electrons.  Taking a value of 23 for
$\Lambda_{ee}$, the energy loss rate in keV~s$^{-1}$ or erg~s$^{-1}$
with $E$ in keV is numerically determined by 
\begin{equation} K = 3.00 \times 10^{-18} \left(\frac{\Lambda_{ee}}{23}\right)\,\; {\rm
keV} {\rm cm}^2 = 4.80 \times 10^{-27}
\left(\frac{\Lambda_{ee}}{23}\right)\,\; {\rm erg\,} {\rm
cm}^2.  \label{eqn:holman_colloss_K} \end{equation} 
Here and in
equations to follow, notation such as $(\Lambda_{ee}/23)$ is used
to show the scaling of a computed constant (here, $K$) with an
independent variable and the numerical value taken for the independent
variable ($\Lambda_{ee} = 23$ in Equation~\ref{eqn:holman_colloss_K}).
If the plasma is not fully ionized, $K$ also depends on the ionization
state (see Section~\ref{Ko2_non-uniform_ionized_plasma}).
\index{ionization state!particle energy loss rate} 

Noting that v$dt = dz$, Equation~\ref{eqn:holman_colloss} can be
simplified to $dE/dN_e = -K/E$, where $dN_e(z) = n_e(z)dz$ and
$N_e(z)$ (cm$^{-2}$) is the plasma electron {\em column
density}.
\index{column density}  
(Here we treat this as a one-dimensional
system and do not distinguish between the total electron velocity
and the velocity component parallel to the magnetic field.)  Hence,
the evolution of an electron's energy with column density is simply
\begin{equation} E^2 = E_0^2 - 2KN_e, \label{eqn:holman_colevol}
\end{equation} where $E_0$ is the initial energy of the electron
where it is injected into the target region.  For example, a 1~keV
electron loses all of its energy over a column density of $1/(2K)
= 1.7 \times 10^{17}$~cm$^{-2}$ (for $\Lambda_{ee} = 23$).  A 25~keV
electron loses 20\% of its energy over a column density of $3.8
\times 10^{19}$~cm$^{-2}$.

If energy losses are not significant within a spatially unresolved
X-ray source region, the emission is called {\em
thin-target}.
\index{hard X-rays!thin-target}  
If, on the other hand, the
non-thermal electrons lose all their suprathermal energy within the
spatially unresolved source during the observational integration
time, the emission is called {\em thick-target}.
\index{hard X-rays!thick-target}
We call a model that assumes these energy losses are from Coulomb
collisions (equations \ref{eqn:holman_colloss} \&
\ref{eqn:holman_colloss_K}) a {\em collisional thick-target model}.\index{thick-target model!collisional}
Collisional thick-target models have been applied to flare
x-ray/$\gamma$-ray emission since the discovery of this emission
in 1958 \citep{1958PhRvL...1..205P,1959JGR....64..697P}.

The maximum information that can be obtained about the accelerated
electrons from an X-ray spectrum alone is contained in the {\em
mean electron flux distribution},
the plasma-density-weighted, target-averaged
electron flux density distribution
\citep[][]{2003ApJ...595L.115B,Chapter7}.
\index{electrons!distribution function!mean electron flux}\index{mean electron flux}\index{electrons!mean electron flux} 
The mean electron flux
distribution is defined as \begin{equation} \bar{F}(E) =
\frac{1}{\bar{n}V}\int_V\, n({\bf r}) \, F(E, {\bf r}) \, dV\;\;\;\;{\rm
electrons}\; {\rm cm}^{-2}\, {\rm s}^{-1}\, {\rm keV}^{-1},
\label{eqn:holman_mef} \end{equation} where $\bar{n}$ and $V$ are
the mean plasma density and volume of the emitting region.  As can
be seen from Equation~\ref{eqn:holman_obsflux}, the product
$\bar{n}V\bar{F}$ can, in principle, be deduced with only a knowledge
of the bremsstrahlung cross-section, $Q(\epsilon,E)$.  Additional
information is required to determine if the X-ray emission is
thin-target, thick-target, or something in between.  The flux
distribution of the emitting electrons and the mean electron flux
distribution
are equivalent for a homogeneous, thin-target source region\index{electrons!distribution function!mean electron flux}.

Equation~\ref{eqn:holman_obsflux} gives the observed X-ray flux in
terms of the accelerated electron flux density distribution throughout
the source.  However, we are interested in the electron distribution
injected into the source region, $F_0(E_0,{\bf r_0})$, since that
is the distribution produced by the unknown acceleration mechanism,
including any modifications during propagation to the source region.
To obtain this, we need to know how to relate $F(E,{\bf r})$ at all
locations within the source region to $F_0(E_0,{\bf r_0})$.  Since
we are interested in the X-rays from a spatially integrated,
thick-target source region, the most direct approach is to first
compute the bremsstrahlung photon yield from a single electron of
energy $E_0$, $\nu(\epsilon,E_0)$ photons keV$^{-1}$ per electron
\citep{1971SoPh...18..489B}.  As long as the observational integration
time is longer than the time required for the electrons to radiate
all photons of energy $\epsilon$ (i.e., longer than the time required
for energy losses to reduce all electron energies to less than
$\epsilon$), the thick-target X-ray spectrum is then given by
\begin{equation} I_{thick}(\epsilon) = {1 \over 4 \pi R^2} \,
\int_{\epsilon}^\infty \, {\cal F}_0(E_0) \, \nu(\epsilon, E_0) \,
dE_0, \label{eq:holman_ithick1} \end{equation} where ${\cal F}_0(E_0)$
is the electron beam flux distribution
(electrons s$^{-1}$ keV$^{-1}$)\index{electrons!distribution function!flux}.
${\cal F}_0(E_0)$
is the integral of $F_0({\bf r_0},E_0)$ over the area at the injection
site through which the electrons stream into the thick-target region.

The rate at which an electron of energy $E$ radiates bremsstrahlung
photons of energy $\epsilon$ is $n({\bf r}){\rm v}(E)Q(\epsilon,
E)$.  The photon yield is obtained by integrating this over time.
Since the electrons are losing energy at the rate $dE/dt$, the time
integration can be replaced by an integration over energy from the
initial electron energy $E_0$ to the lowest energy capable of
radiating a photon of energy $\epsilon$: 
\begin{equation}
\nu(\epsilon,E_0) = \int_{E_0}^\epsilon {n({\bf r}) \, {\rm v}(E)
\, Q(\epsilon, E) \, dE \over dE/dt }.  \label{eqn:holman_nu}
\end{equation}

Using Equation~\ref{eqn:holman_colloss} for $dE/dt$,
Equation~\ref{eq:holman_ithick1} becomes
\begin{equation} I_{thick}(\epsilon) = \frac{1}{4 \pi R^2} \,
\frac{1}{\overline{Z}K} \int_{E_0 = \epsilon}^\infty \, {\cal
F}_0(E_0) \, \int_{E = \epsilon}^{E_0} E \, Q(\epsilon, E) \, dE
\, dE_0.  
\label{eqn_holman_ithick2} 
\end{equation}\index{hard X-rays!thick-target}We have used
the relationship $n_e = \overline{Z}n$, where $\overline{Z} \simeq
1.1$ is the ion-species-number-density-weighted (or, equivalently,
relative-ion-abundance-weighted) mean atomic number of the target
plasma.\index{abundances!bremsstrahlung efficiency}
Thus, the thick-target X-ray flux spectrum does not depend
on the plasma density.  However, the plasma must be dense enough
for the emission to be thick-target, i.e., dense enough for all the
electrons to be thermalized in the observation time interval.
Integration of Equation~\ref{eqn:holman_colloss} shows that this
typically implies a plasma density $\gapprox 10^{11} - 10^{12}$
cm$^{-3}$ for an observational integration time of 1~s (see Sections
\ref{sec:battaglia_theory} and \ref{sec:holman_discussion_tt} for
more about this).  This condition is well satisfied at loop
footpoints\index{hard X-rays!footpoint sources}.

Observed non-thermal X-ray spectra from solar flares can usually be
well fitted with a model photon spectrum that is either a single
or a double power-law.  For a single power-law electron flux
distribution of the form ${\cal F}(E) = AE^{-\delta}$, the photon
spectrum is also well approximated by the power-law form $I(\epsilon)
= I_0\epsilon^{-\gamma}$.  The relationship between the electron
and photon spectral indices\index{hard X-rays!spectral index} $\delta$
and $\gamma$ can most easily be obtained from equations
\ref{eqn:holman_obsflux} and \ref{eqn_holman_ithick2} using the
Kramers\index{bremsstrahlung!Kramers approximation} approximation
to the nonrelativistic Bethe-Heitler (NRBH) bremsstrahlung cross
section (see Koch \& Motz 1959 for bremsstrahlung cross-sections).
\nocite{1959RvMP...31..920K}
\index{cross-sections!Bethe-Heitler}
\index{cross-sections!NRBH}
\index{bremsstrahlung!Bethe-Heitler cross-section}\index{bremsstrahlung!NRBH} 
The NRBH cross-section is given by: \begin{equation} Q_{NRBH}(\epsilon,
E) = \frac{\overline{Z^2} Q_0}{\epsilon E}\: \ln
\left(\frac{1+\sqrt{1-\epsilon/E}}{1-\sqrt{1-\epsilon/E}}\right)\;\;\;\;{\rm
cm}^2\, {\rm keV}^{-1}, \label{eq:holman_NRBHCS} \end{equation}
where $Q_0 = 7.90 \times 10^{-25}$~cm$^2$~keV and $\overline{Z^2}
\simeq 1.4$ is the ion-species-number-density-weighted mean square
atomic number of the target plasma.  
\index{cross-sections!Kramers approximation}
The Kramers approximation to
this cross-section is Equation~\ref{eq:holman_NRBHCS} without the
logarithmic term.  The bremsstrahlung cross-section is zero for
$\epsilon > E$, since an electron cannot radiate a photon that is
more energetic than the electron.  Analytic expressions for the
photon flux from both a uniform thin-target source and a thick-target
source can be obtained with the Kramers and the NRBH cross-sections
when the electron flux distribution has the single-power-law form
\citep[][]{1971SoPh...18..489B,1988psf..book.....T}.  The thin-target
result also generalizes to the photon flux from a single-power-law
mean electron flux distribution.
\index{electrons!distribution function!mean electron flux}

For a uniform thin-target source\index{hard X-rays!thin-target} and
${\cal F}(E) = AE^{-\delta}$, \begin{equation} I_{thin}(\epsilon)
= 3.93 \times 10^{-52} \left(\frac{1\;{\rm AU}}{R}\right)^2
\left(\frac{\overline{Z^2}}{1.4}\right) N A \beta_{tn}(\delta)
\epsilon^{-(\delta+1)}, \label{eqn:holman_ithinpl} \end{equation}
giving $\gamma_{thin} = \delta + 1$.  The photon energy $\epsilon$
is in keV, the distance from the source to the X-ray detector is
taken to be one Astronomical Unit (1 AU), a typical value of 1.4
is taken for $\overline{Z^2}$, and $N$ is the ion column density.
\index{cross-sections!NRBH}
The power-law-index-dependent coefficient for the NRBH cross-section
is \begin{equation} \beta_{tn}(\delta) = \frac{B(\delta,1/2)}{\delta},
\label{eq:holman_beta_tn} \end{equation} where $B(x,y)$ is the
standard Beta function.  In the Kramers approximation, $\beta_{tn}(\delta)
= 1/\delta$.  Typical values for the ion column density and $A$,
the differential electron flux at 1~keV, are $N = 10^{18}$--$10^{20}$
cm$^{-2}$ and $A = 10^{34}$--$10^{38}$ electrons s$^{-1}$ keV$^{-1}$.\index{cross-sections!Kramers approximation}

For a thick-target source\index{hard X-rays!thick-target} region,
\begin{equation} I_{thick}(\epsilon) = 1.17 \times 10^{-34}
\left(\frac{1\;{\rm AU}}{R}\right)^2
\left(\frac{\overline{Z^2}/\overline{Z}}{1.25}\right)
\left(\frac{23}{\Lambda_{ee}}\right) A \beta_{tk}(\delta)
\epsilon^{-(\delta-1)}, \label{eqn:holman_ithickpl} \end{equation}
giving $\gamma_{thick} = \delta - 1$.  The power-law-index-dependent
coefficient for the NRBH cross-section is \begin{equation}
\beta_{tk}(\delta) = \frac{B(\delta-2,1/2)}{(\delta-1)(\delta-2)}.
\label{eq:holman_beta_tk} \end{equation} In the Kramers approximation,
$\beta_{tk}(\delta) = 1/[(\delta-1)(\delta-2)]$.

\begin{figure}[h] 
\vspace{-0.2in} \hspace{-3.4in}
\includegraphics[width=20cm]{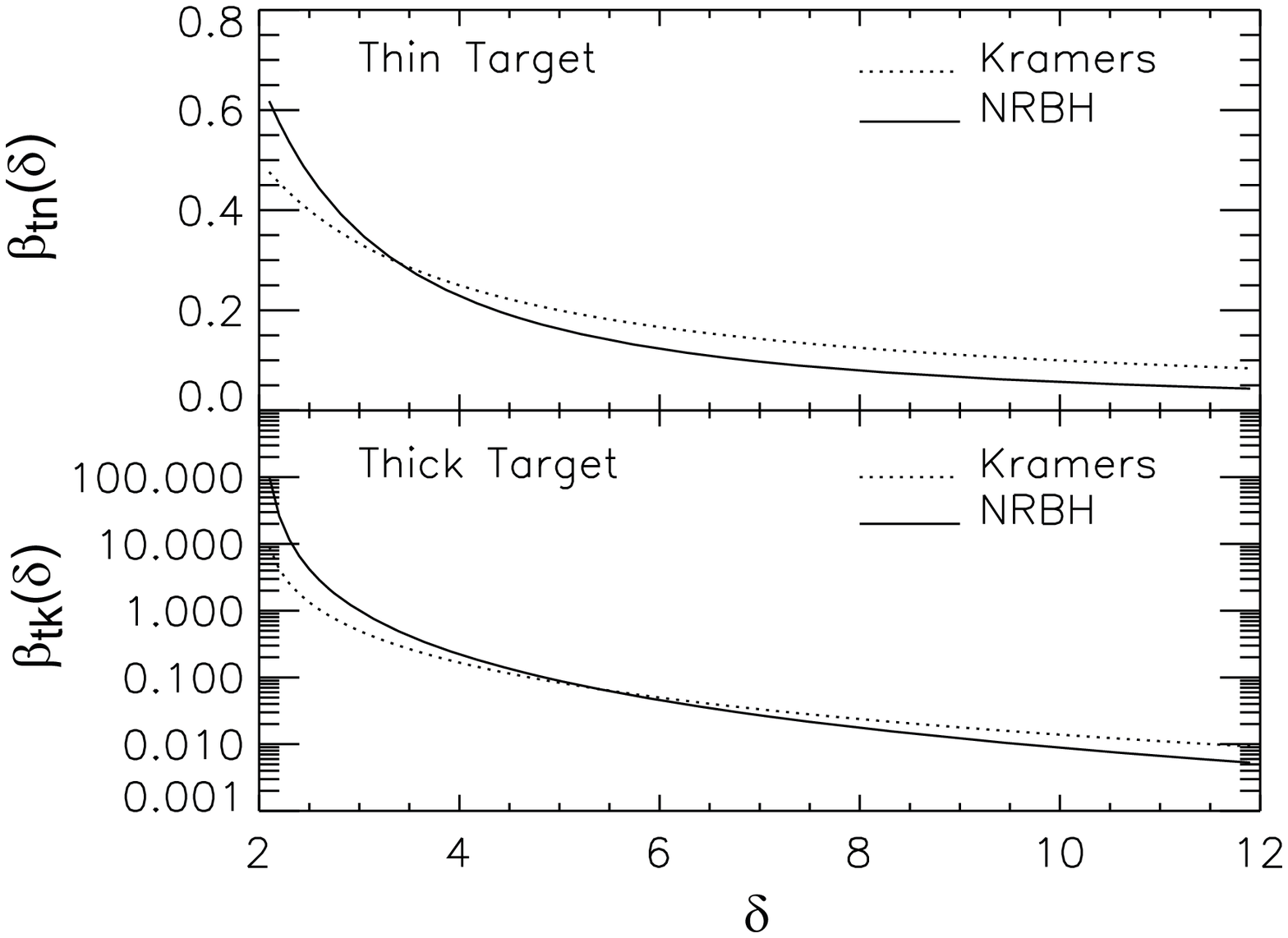} \vspace{-1.25in}
\caption{ The power-law-index-dependent terms $\beta_{tn}(\delta)$
(equation \ref{eq:holman_beta_tn}) and $\beta_{tk}(\delta)$ (equation
\ref{eq:holman_beta_tk}) in the analytic expressions for the
thin-target\index{hard X-rays!thin-target} and the
thick-target\index{hard X-rays!thick-target} photon flux from a power-law
electron flux distribution (equations \ref{eqn:holman_ithinpl} and
\ref{eqn:holman_ithickpl}).  
The {\it solid} curves are for the
nonrelativistic Bethe-Heitler\index{bremsstrahlung!Bethe-Heitler cross-section} bremsstrahlung cross-section and the {\it
dotted} curves are for the Kramers\index{bremsstrahlung!Kramers approximation} 
cross-section.  
Note that the $\beta$ axis is
linear for the thin-target coefficient, and logarithmic for the
thick-target coefficient.
\index{cross-sections!NRBH!illustration}
\index{cross-sections!Kramers approximation!illustration}
} 
\label{fig:holman_beta_tn_tk} 
\end{figure}

The coefficients $\beta_{tn}(\delta)$ and $\beta_{tk}(\delta)$ for
both the NRBH and the Kramers cross-sections are plotted as a
function of $\delta$ in Figure~\ref{fig:holman_beta_tn_tk}.
\index{cross-sections!NRBH!comparison with Kramers} 
The Kramers and NRBH results are equal for
thin-target\index{hard X-rays!thin-target} emission when $\delta \simeq
3.4$, and for thick-target\index{hard X-rays!thick-target} emission when
$\delta \simeq 5.4$.  For the plotted range of $\delta$, the Kramers
approximation can differ from the NRBH result by over 90\%.  For
$\delta$ in the range 3--10, the Kramers result can differ from the
NRBH result by as much as 76\% and 57\% for thin- and thick-target
emission, respectively.

It is important to recognize that the above power-law relationships
are only valid if the electron {\em flux density distribution},
$F({\bf r},E)$ electrons cm$^{-2}$ s$^{-1}$ keV$^{-1}$, or the
electron {\em flux distribution}, ${\cal F}(E)$ electrons s$^{-1}$
keV$^{-1}$, is assumed to have a power-law energy dependence.\index{spectrum!power-law}
It is sometimes convenient to work with the electron {\em density
distribution},
$f({\bf
r},E)$ (electrons cm$^{-3}$ keV$^{-1}$), rather than the flux density
distribution, especially when considering thin-target emission alone
or comparing X-ray spectra with radio spectra.
\index{radio emission!microwaves}
\index{electrons!distribution function!density} 
The flux density and density distributions are related through
$F({\bf r},E) = f({\bf r},E){\rm v}(E)$.  If the electron density
distribution rather than the flux or flux density distribution is
assumed to have a power-law index $\delta'$, so that $f({\bf r},E)
\propto E^{-\delta'}$, the relationships between this power-law
index and the photon spectral index\index{hard X-rays!spectral index}
become $\gamma_{thin} = \delta' + 0.5$ and $\gamma_{thick} = \delta'
- 1.5$.

The simple power-law relationships are {\em not valid} if there is
a break or a cutoff in the electron distribution at an energy less
than $\sim$2 orders of magnitude above the photon energies of
interest\index{electrons!high-energy cutoff}.  
Since all electrons with energies above a given photon
energy $\epsilon$ contribute to the bremsstrahlung at that photon
energy, for the power-law relationships to be valid the break energy
must be high enough that the deficit (or excess) of electrons above
the break energy does not significantly affect the photon flux at
energy $\epsilon$.  The power-law relationship is typically not
accurate until photon energies one to two orders of magnitude below
the break energy, depending on the steepness of the power-law
electron distribution \citep[see Figures 9 \& 10 of
][]{2003ApJ...586..606H}.  Thus, for example, these relationships
are not correct for the lower power-law index of a double power-law
fit to a photon spectrum at photon energies within about an order
of magnitude below the break energy in the double power-law electron
distribution.  Equation~\ref{eqn:holman_obsflux} or
\ref{eqn_holman_ithick2} can be used to numerically compute the
X-ray spectrum from an arbitrary flux distribution in electron
energy.

When electrons with kinetic energies approaching or exceeding 511~keV
significantly contribute to the radiation, the relativistic
Bethe-Heitler brems\-strah\-lung cross-section
\index{bremsstrahlung!Bethe-Heitler cross-section} 
\citep[Equation 3BN of][]{1959RvMP...31..920K}
or a close approximation \citep{1997A&A...326..417H} must be used.
\citet{1997A&A...326..417H} has shown that the maximum error in the
NRBH cross-section relative to the relativistic Bethe-Heitler cross
section becomes greater than 10\% at electron energies of 30~keV
and above. 
\index{cross-sections!NRBH!comparison with relativistic}
Numerical computations using the relativistic Bethe-Heitler
cross-section have been incorporated into the {\it RHESSI} spectral
analysis software (OSPEX) for both thin- and thick-target emission
from, in the most general case, a broken-power-law electron flux
\index{low-energy cutoff}
distribution with both low-\index{electrons!distribution function!low-energy
cutoff} and high-energy cutoffs\index{electrons!distribution
function!high-energy cutoff} (the functions labeled ``thin'' and
``thick'' using the IDL programs \verb|brm_bremspec.pro| and
\verb|brm_bremthick.pro| -- see \citeauthor{2003ApJ...586..606H}
\citeyear{2003ApJ...586..606H}).  Faster versions of these programs
are now available in OSPEX and are currently labeled ``thin2'' and
``thick2'' and use the IDL programs \verb|brm2_thintarget.pro| and
\verb|brm2_thicktarget.pro|.

The analytic results based on the NRBH cross-section have been
generalized to a broken-power-law electron flux distribution with
cutoffs by \citet{2008A&A...486.1023B}.  They find a maximum error
of 35\% relative to results obtained with the relativistic Bethe-Heitler
cross-section for the range of parameters they consider.  These
results provide the fastest method for obtaining thin- and thick-target
fits to X-ray spectra in the {\it RHESSI} spectral analysis software,
where they are labeled ``photon\_thin'' and ``photon\_thick'' and
use the IDL programs \verb|f_photon_thin.pro| and
\verb|f_photon_thick.pro|\index{bremsstrahlung!software}.


\section{Low-energy cutoffs and the energy in non-thermal electrons} 
\label{sec:holman_cutoffs}


One of the most important aspects of the distribution of accelerated
electrons is the low-energy cutoff\index{electrons!distribution
function!low-energy cutoff}.  The acceleration
\index{acceleration!from thermal plasma}
of charged particles out of the thermal plasma typically involves
a competition between the collisions that keeps the particles
thermalized and the acceleration mechanism.  The particles are
accelerated out of the tail of the thermal distribution, down to
the lowest particle energy for which the acceleration mechanism can
overcome the collisional force.  Thus, the value of the low-energy
cutoff can provide information about the force of the acceleration
mechanism.  More generally, as discussed below, the electron
distribution must have a low-energy cutoff (1) so that the number
and energy flux of electrons is finite and reasonable, and (2)
because electrons with energies that are not well above the thermal
energy of the plasma through which they propagate will be rapidly
thermalized.  Knowledge of the low-energy cutoff and its evolution
during a flare is critical to determining the energy flux and energy
in non-thermal electrons and, ultimately, the efficiency of the
acceleration process.

\subsection{Why do we need to determine the low-energy cutoff of
non-thermal electron distributions?}

An important feature of the basic thick-target model is that the
photon spectrum $I(\epsilon)$ is directly determined by the injected
electron flux distribution ${\cal F}_0(E_0)$.  As can be seen from
Equation~\ref{eqn_holman_ithick2}, no additional parameters such
as source density or volume need to be determined.  Consequently,
by integrating over all electron energies, we can also determine
the total flux of non-thermal electrons, $N_{nth}$ electrons s$^{-1}$,
the power in non-thermal electrons, $P_{nth}$ erg s$^{-1}$, and,
integrating over time, the total number of, and energy in, non-thermal
electrons.

The total non-thermal electron number flux and power are computed
as follows:\index{electrons!distribution function!total flux}
\index{electrons!distribution function!total power}
\begin{equation} N_{nth} = \int_{E_c}^{+\infty} {\cal F}_0(E_0)
\,dE_0 = \frac{A}{\delta-1} {E_c}^{-\delta+1} \;\;\;\; {\rm electrons\;
s}^{-1} \label{eqn:sainthilaire_nnth} \end{equation}
\begin{equation} P_{nth} = \kappa_{E} \, \int_{E_c}^{+\infty} E_0
\cdot {\cal F}_0(E_0) \,dE_0 = \frac{\kappa_{E}A}{\delta-2}
{E_c}^{-\delta+2} \;\;\;\; {\rm erg\; s}^{-1} \label{eqn:sainthilaire_pnth}
\end{equation} 
The last expression in each equation is the result
for a power-law electron flux distribution of the form ${\cal
F}_0(E_0) = A \cdot E_0^{-\delta}$.  The constant $\kappa_{E} =
1.60 \times 10^{-9}$ is the conversion from keV to erg.  $E_c$ is
a low-energy cutoff\index{electrons!distribution function!low-energy
cutoff} to the electron flux distribution\index{low-energy cutoff}.
These expressions are
valid and finite for $\delta > 2$ and $E_c > 0$. We call this form
of low-energy cutoff a {\em sharp low-energy cutoff}.  An electron
distribution that continues below a transition energy $E_c$ that
has a positive slope, is flat, or in general has a spectral index
$\delta_{low} < 1$ also provides finite electron and energy fluxes,
but these fluxes are somewhat higher than those associated with the
sharp low-energy cutoff\index{low-energy cutoff!sharp}.

For this single-power-law electron flux distribution with a sharp
low-energy cutoff, the non-thermal {\it power} (erg s$^{-1}$), and
ultimately the non-thermal {\it energy} (erg), from the power-law
electron flux distribution depends on only three parameters: $\delta$,
$A$, and $E_c$.
    Observations indicate that $\delta$ is greater than 2
    \citep{1985SoPh..100..465D,1987ApJ...312..462L,1991ApJ...375..366W,2003ApJ...595L..97H}.
    Hence, were $E_c = 0$, the integral would yield an infinite
    value, a decidedly unphysical result!  Therefore, the power-law
    electron distribution cannot extend all the way to zero energy
    with the same or steeper slope, and some form of {\it low-energy
    cutoff} in the accelerated electron spectrum must be present.
    As we will see, the determination of the energy at which this
    cutoff occurs is not a straightforward process, but it is the
    single most important parameter to determine (as the other two
    are generally more straightforward to determine -- see
    Section~\ref{sec:holman_thick} and Kontar et al. 2011).
    \nocite{Chapter7}  
    For example, with $\delta$~=~4 (typical during the peak time of
    strong flares), a factor of 2 error in $E_c$ yields a factor
    of 4 error in $P_{nth}$.  For larger $\delta$ (as found in small
    flares, or rise/decay phases of large flares), such an error
    quickly leads to an order of magnitude (or even greater)
    difference in the injected power $P_{nth}$ and in the total
    energy in the non-thermal electrons accelerated during the flare!

\subsection{Why is the low-energy cutoff difficult to determine?}

	    \begin{figure}[h] \centering
	    \includegraphics[width=11cm]{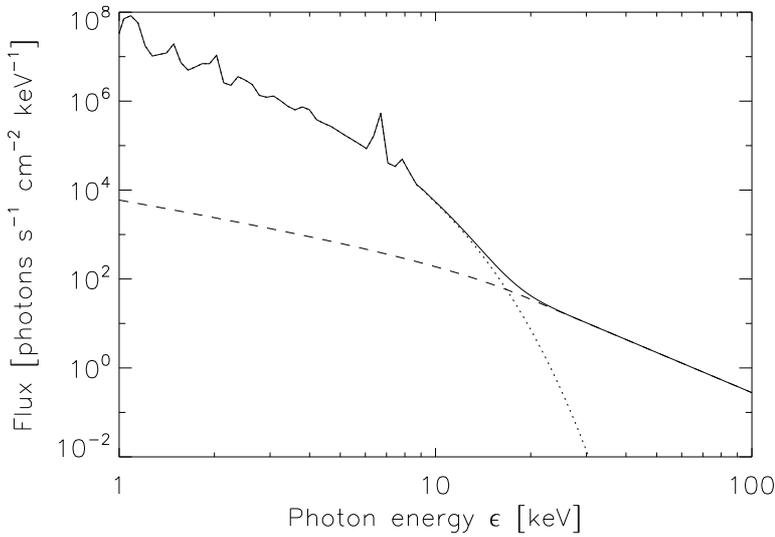} \caption{
		Typical full-Sun flare spectrum.  {\it Dashed:}
		Nonthermal thick-target spectrum from an accelerated
		electron distribution with $\delta$=4, and a
		low-energy cutoff of 20~keV.  {\it Dotted:} Thermal
		spectrum, from a plasma with temperature $T=20$~MK
		and emission measure $EM=10^{49}$ cm$^{-3}$.  {\it
		Solid:}  Total radiated spectrum.  The multiple
		peaks in the thermal spectrum are from spectral
		lines, as observed by an instrument with
		$\sim 1$~keV spectral resolution.
	    } \label{fig:sainthilaire:1} \end{figure}

    The essence of the problem in many flare spectra is summarized
    in Figure~\ref{fig:sainthilaire:1}: the non-thermal power-law is
    well-observed above $\sim$20 keV, but any revealing features
    that it might possess at lower energies, such as a low-energy
    cutoff, are masked by the thermal emission.

Even if a spectrum does show a flattening at low energies that could
be the result of a low-energy cutoff, other mechanisms that could
produce the flattening must be ruled out (see
Section~\ref{sec:sainthillaire_caveats}).  The low-energy cutoff
has the characteristic feature, determined by the photon energy
dependence of the bremsstrahlung cross-section (see
Equation~\ref{eq:holman_NRBHCS}), that the X-ray spectrum eventually
approaches a spectral index of $\gamma \approx 1$ at low energies
\citep[cf.][]{2003ApJ...586..606H}.  It is currently impossible,
however, to observe a flare spectrum to low enough photon energies
to see that it does indeed become this flat.  Generally we can only
hope to rule out the other mechanisms based on additional data and
detailed spectral fits.

\subsection{What is the shape of the low-energy cutoff, and how
does it impact the photon spectrum and $P_{nth}$?}
\label{sec:SH_cutoff_shape}

    Bremsstrahlung photon spectra are obtained from convolution
    integrals over the electron flux distribution (equations
    \ref{eqn:holman_obsflux} and \ref{eqn_holman_ithick2}).  Hence,
    features in an electron distribution are smoothed out in the
    resulting photon spectrum \citep[see, e.g.,][]{2006ApJ...643..523B}.

	    \begin{figure}[h] \centering
	    \includegraphics[width=11cm]{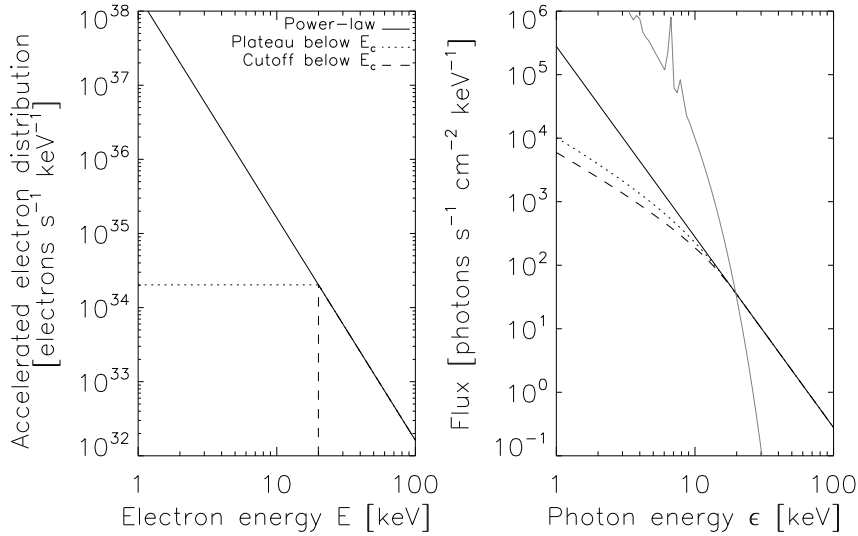} \caption{
		Different shapes of low-energy cutoff in the injected
		electron distribution {\it (left)} lead to slightly
		different photon spectra {\it (right)}.  The
		cutoff/turnover electron energy is $E_c$=20 keV.
		The thin curve in the right panel demonstrates how
		the cutoff can be masked by emission from thermal
		plasma.  See also \citet{2003ApJ...586..606H} for
		a thorough discussion of bremsstrahlung spectra
		generated from electron power-laws with cutoff.}
	    \label{fig:sainthilaire:shape} \end{figure}

    As can be seen in Figure~\ref{fig:sainthilaire:shape}, both a
    sharp cutoff at $E_c$ and a ``turnover''\index{low-energy cutoff!turnover} (defined here to be a
    constant $F_0(E)$ below $E_c$, a ``plateau'')\index{low-energy cutoff!plateau} in the injected
    electron distribution lead to similar thick-target photon
    spectra.  This subtle difference is difficult to discriminate
    observationally, and the problem is compounded by the dominance
    of the thermal component at low energies.

\begin{figure}[h] \centering
	    \includegraphics[width=11cm]{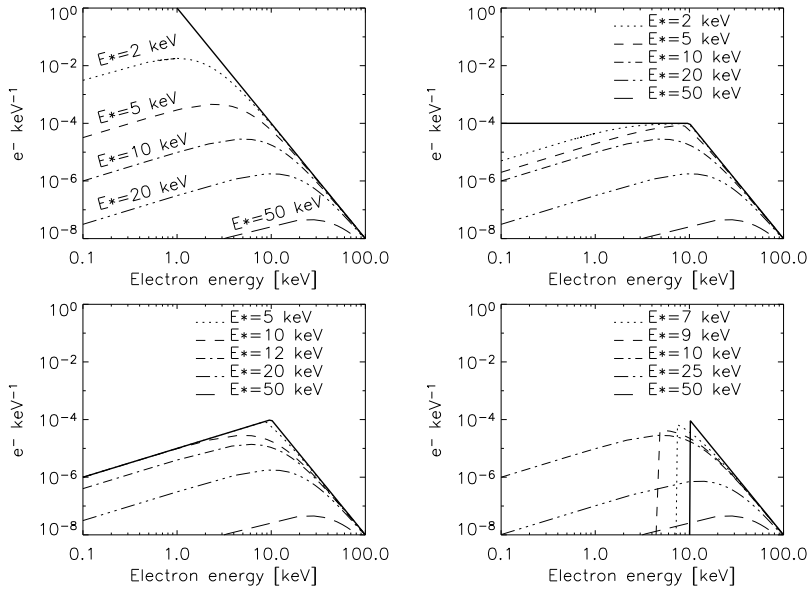}
	    \caption{The four plots show the Coulomb-collisional
	    evolution with column density of an injected electron
	    distribution ({\it thick, solid line}).  For the simple
	    power-law case ({\it upper left}), the low-energy end
	    of the distribution becomes linear, and the peak of the
	    distribution is found at $E_{peak}= E_* / \sqrt{\delta}$,
	    where $\delta$ is the injected distribution power-law
	    spectral index ($\delta$=4 in the plots), and $E_*=\sqrt{2
	    K \cdot N_*}$ is the initial energy that electrons must
	    possess in order not to be fully stopped by a column
	    density\index{column density} $N_*$
	    (Equation~\ref{eqn:holman_colevol}).  When a low-energy
	    cutoff is present, the peak of the distribution is seen
	    to first decrease in energy until $E_*$ exceeds the
	    cutoff energy \citep[from][]{2005PhDT........10S}.  }
	    \label{fig:sainthilaire:collevol} \end{figure}

    A sharp cutoff would lead to plasma instabilities that should
    theoretically flatten the distribution around and below the
    cutoff within microseconds\index{plasma instabilities!bump-on-tail}
    (see Section~\ref{sec:holman_instabilities}).  On the other
    hand, the electron flux distribution below the cutoff must be
    flatter than $E^{-1}$, as demonstrated by
    Equation~\ref{eqn:sainthilaire_nnth}, or the total electron
    number flux would be infinite.  Having a constant value for the
    distribution below $E_c$ (turnover case) seems like a reasonable
    middle ground and approximates a quasilinearly relaxed electron
    distribution \citep[Section~\ref{sec:holman_instabilities};][Chapter
    10]{1973ppp..book.....K}.  Coulomb collisional losses\index{electrons!collision losses}, on the
    other hand, yield an electron distribution that increases
    linearly at low energies (see Figure~\ref{fig:sainthilaire:collevol}),
    leading to a photon spectrum between the sharp cutoff case and
    the turnover case.

    Notice that the photon spectra actually flatten gradually to
    the spectral index of 1 at low energies from the spectral index
    of $\gamma=\delta+1$ at $E_c$ and higher energies.  Below $E_c$,
    it is {\em not} a power-law.  Fitting a double power-law model
    photon spectrum, and using the break (i.e., kink) energy as the
    low-energy cutoff typically leads to a large error in $E_c$
    \citep[e.g.,][]{2001ApJ...552..858G,2005A&A...435..743S}, and
    hence to an even larger error in $P_{nth}$.

    In terms of the energetics, \citet{2005A&A...435..743S} have
    shown that the choice of an exact shape for the low-energy
    cutoff as a model is not dramatically important.  For a fixed
    cutoff energy, from Equation~\ref{eqn:sainthilaire_pnth} it can
    be shown that the ratio of the power in the turnover model to
    the power in the sharp cutoff model without the flat component
    below the cutoff energy is $\delta/2$.  In obtaining spectral
    fits, however, the turnover model gives higher cutoff energies
    than the sharp cutoff model.  Using simulations,
    \citet{2005A&A...435..743S} found that assuming either a sharp
    cutoff model or a turnover model led to differences in $P_{nth}$
    typically less than $\sim$20\%.  Hence, the sharp cutoff, being
    the simplest, is the model of choice for computing flare
    energetics.  Nevertheless, knowing the shape of the low-energy
    cutoff would not only yield more accurate non-thermal energy
    estimates, but would be a source of information on the acceleration
    mechanism and/or propagation effects.

    Spectral inversion\index{inverse problem!for X-ray spectra} methods have recently been developed for
    deducing the {\it mean electron flux distribution}
    (Equation~\ref{eqn:holman_mef}) from X-ray spectra
\index{electrons!distribution function!mean electron flux}
\index{mean electron flux}
    \citep{1992SoPh..137..121J,2003ApJ...595L.115B,2006ApJ...643..523B}.
    A spectral ``dip'' has been found just above the presumed thermal
    component in some deduced mean electron flux distributions that
    may be associated with a low-energy cutoff
    \citep[e.g.,][]{2003ApJ...595L.127P}.  In the collisional
    thick-target model, the slope of the high-energy ``wall'' of
    this dip should be linear or flatter, with a linear slope
    indicating the absence of emitting electrons in the injected
    electron distribution at the energies displaying this slope.
    \citet{2006AdSpR..38..945K} have found evidence for slopes that
    are steeper than linear, but their spectra were not corrected
    for photospheric albedo\index{hard X-rays!albedo} 
\index{albedo!dip in spectrum}\index{thick-target model!and spectral ``dip''}
(see
    Section~\ref{sec:sainthilaire_Ec}).  Finding and understanding
    these dips is a crucial element for gaining an understanding
    of the low-energy properties of flare electron distributions
    \citep[see][]{Chapter7}.

    \citet{2003ApJ...595L.119E} has pointed out that the non-thermal
    electron distribution could seamlessly merge into the thermal
    distribution, removing the need for a low-energy cutoff.  As
    was shown by \citet{2003ApJ...595L..97H} for SOL2002-07-23T00:35 (X4.8)\index{flare (individual)!SOL2002-07-23T00:35 (X4.8)!energy in
    non-thermal electrons}, however, merger of the electron distribution
    into the typically derived $\sim$10--30 MK thermal flare plasma
    generally implies an exceptionally high energy in non-thermal
    electrons.  Thus, for a more likely energy content, a higher
    low-energy cutoff or a hotter plasma would need to be present
    in the target region.  Any emission from this additional ``hot
    core,'' because of its much lower emission measure, is likely
    to be masked by the usual $\sim$10--30~MK thermal emission.
    This merger of the non-thermal electron distribution into the
    thermal tail in the target region does not remove the need for
    a low-energy cutoff in the electrons that escape the acceleration
    region, however.

This section has dealt with the shape of the low-energy cutoff under
the assumptions that the X-ray photon spectra are not altered by
other mechanisms and that the bremsstrahlung emission is isotropic.
The next section lists the important caveats to these assumptions,
and their possible influence in the determination of the low-energy
cutoff to the electron flux distribution.

\subsection{Important caveats} \label{sec:sainthillaire_caveats}
\index{caveats!low-energy cutoff}

    As previously discussed, apparently minor features in the
    bremsstrahlung photon spectrum can have substantial implications
    for the mean electron flux and, consequently, the injected
    electron distribution.  This means that unknown or poorly-understood
    processes that alter the injected electron distribution
    (propagation effects, for example) or the photon spectrum
    (including instrumental effects) can lead to significant errors
    in the determination of the low-energy cutoff.  Known processes
    that affect the determination of the low-energy cutoff are
    enumerated below.

    \begin{enumerate}
	\item Detector pulse pileup\index{pulse pileup} effects
	\citep{2002SoPh..210...33S}, if not properly corrected for,
\index{RHESSI@\textit{RHESSI}!pulse pileup}
	can introduce a flattening of the spectrum toward lower
	energies that simulates the flattening resulting from a
	low-energy cutoff.  
	\item The contribution of Compton
	back-scattered photons (photospheric al\-be\-do)
\index{hard X-rays!albedo}
\index{albedo!and low-energy cutoff}
	to the measured X-ray spectrum can simulate the spectral
	flattening produced by a low-energy cutoff.
	\citet{2005SoPh..232...63K} have shown that the dip in a
	spectrum from SOL2002-08-20T08:25 (M3.4)
\index{flare (individual)!SOL2002-08-20T08:25 (M3.4)!low-energy cutoff \& albedo}
	becomes statistically insignificant when the spectrum is
	corrected for photospheric albedo \citep[also see][]{Chapter7}.
	\citet{2007A&A...466..705K} show that spectra in the
	15--20~keV energy band tend to be flatter near disk center
	when albedo from isotropically emitted photons is not taken
	into account, further demonstrating the importance of
	correcting for photospheric albedo. 
	\item The assumed
	differential cross-section and electron energy loss rate
	can influence the results \citep[for a discussion of this,
	see][]{2005A&A...435..743S}.  In some circumstances, a
	contribution from recombination radiation may significantly
	change the results (Brown et al. 2010; also see Kontar et al., 2011).
	\nocite{Chapter7}\nocite{2010A&A...515C...1B}
	\item Anisotropies in the electron beam
	directivity and the bremsstrahlung differential cross-section
	can significantly alter the X-ray spectrum
	\citep{2004ApJ...613.1233M}.  \item Non-uniform target
	ionization (the fact that the chromosphere's ionization
	state varies with depth, see Section~\ref{sec:kontar_nonuniform})
	can introduce a spectral break that may be confused with
	the break associated with a low-energy cutoff.  
	\item Energy
	losses associated with a return current produce a low-energy
	flattening of the X-ray spectrum
	(Section~\ref{sec:zharkova_return_current}).\index{return current}  
	This is a
	low-energy ``cutoff'' in the electron distribution injected
	into the thick target, but it is produced between the
	acceleration region and the emitting source region.\index{acceleration region!spectral cutoff}
       \item
	A non-power-law distribution of injected electrons or
	significant evolution of the injected electron distribution
	during the observational integration time could affect the
	deduced value of the low-energy cutoff.
    \end{enumerate}

    For all the above reasons, the value of the low-energy cutoff
    in the injected electron flux distribution has not been determined
    with any degree of certainty except perhaps in a few special
    cases.  Even less is known about the shape of the low-energy
    cutoff.  The consensus in the solar physics community for now
    is to assume the simplest case, a sharp low-energy cutoff.
    Existing studies, presented in the next section, tend to support
    the adequacy of this assumption for the purposes of estimating
    the total power and energy in the accelerated electrons.

\subsection{Determinations of $E_c$ and electron energy content
from flare data} \label{sec:sainthilaire_Ec}

    Before {\it RHESSI}, instruments did not cover well (if at all)
    the $\sim$10--40~keV photon energies where the transition from
    thermal emission to non-thermal emission usually occurs.
    Researchers typically assumed an arbitrary low-energy cutoff
    at a value at or below the instrument's observing range (one
    would talk of the ``injected power
\index{electrons!distribution function!total power} 
in electrons above $E_c$ keV'' instead
    of the total non-thermal power $P_{nth}$).  An exception is
    \citet{1990ApJ...353..313N}.  They argued that spectral flattening
    observed in two flares with the {\it Solar Maximum Mission}
\index{Solar Maximum Mission@\textit{Solar Maximum Mission}} 
\index{satellites!SMM@\textit{SMM}} 
and {\it Hinotori}
\index{Hinotori@\textit{Hinotori}} 
\index{satellites!Hinotori@\textit{Hinotori}} 
indicated
    a cutoff energy of $\gtrsim$50~keV.  Also, \citet{2001ApJ...552..858G}
    interpreted spectral breaks at $\sim$80~keV in {\it Compton
    Gamma-Ray Observatory (CGRO)\index{Compton Gamma-Ray Observatory.}}\index{Compton@\textit{Compton Gamma Ray Observatory (CGRO)}} flare spectra
\index{satellites!CGRO@\textit{CGRO}}
    as the low-energy cutoff\index{low-energy cutoff} in estimating flare energetics, resulting
    in rather small values for the non-thermal energy in the analyzed
    flares.  The relatively low resolution of the spectra from these
    instruments prevented the quantitative evaluation of any spectral
    flattening toward lower energies, however.

The only high-resolution flare spectral data before the launch of
{\it RHESSI} was the balloon data of \citet{1981ApJ...251L.109L}
for SOL1980-06-27T16:17 (M6.7)\index{flare (individual)!SOL1980-06-27T16:17 (M6.7)!low-energy cutoff} 
along with $\sim$25 microflares observed
during the same balloon flight.  \citet{1994ApJ...435..469B} applied
a direct electric field\index{electric fields} electron acceleration
model\index{acceleration!DC electric field} to the SOL1980-06-27 flare
data.  They derived, along with other model-related parameters, the
time evolution of the critical energy above which runaway acceleration
occurs -- the model equivalent to the low-energy cutoff.  They found
this critical energy to range from $\sim$20--40~keV.

It is now possible in most cases to obtain a meaningful upper limit
on $E_c$, thanks to {\it RHESSI}'s high-spectral-resolution coverage
of the 10--40~keV energy range and beyond.  \citet{2003ApJ...595L..97H},
\citet{2004JGRA..10910104E}, and \citet{2005A&A...435..743S}, in
determining the low-energy cutoff, obtained the ``highest value for
$E_c$ that still fits the data.''  In many solar flare spectra,
because of the dominance of radiation from thermal plasma at low
energies, a range of values for $E_c$ fit the data equally well,
up to a certain critical energy, above which the $\chi^2$ goodness-of-fit
parameter becomes unacceptably large. The low-energy cutoff is taken
to be equal to this critical value.  This upper limit on the cutoff
energy results in a lower limit for the non-thermal power and energy.
The results obtained for the maximum value of $E_c$ were typically
in the 15--45~keV range, although late in the development of 
SOL2002-07-23T00:35 (X4.8)\index{flare (individual)!SOL2002-07-23!00:35 (X4.8)!low-energy cutoff}
some values as high as $\sim$80~keV were obtained for $E_c$.  
The minimum non-thermal energies thus determined were comparable to or
somewhat larger than the calculated thermal energies.

	    \begin{figure}[h] \centering
	    \includegraphics[width=12cm]{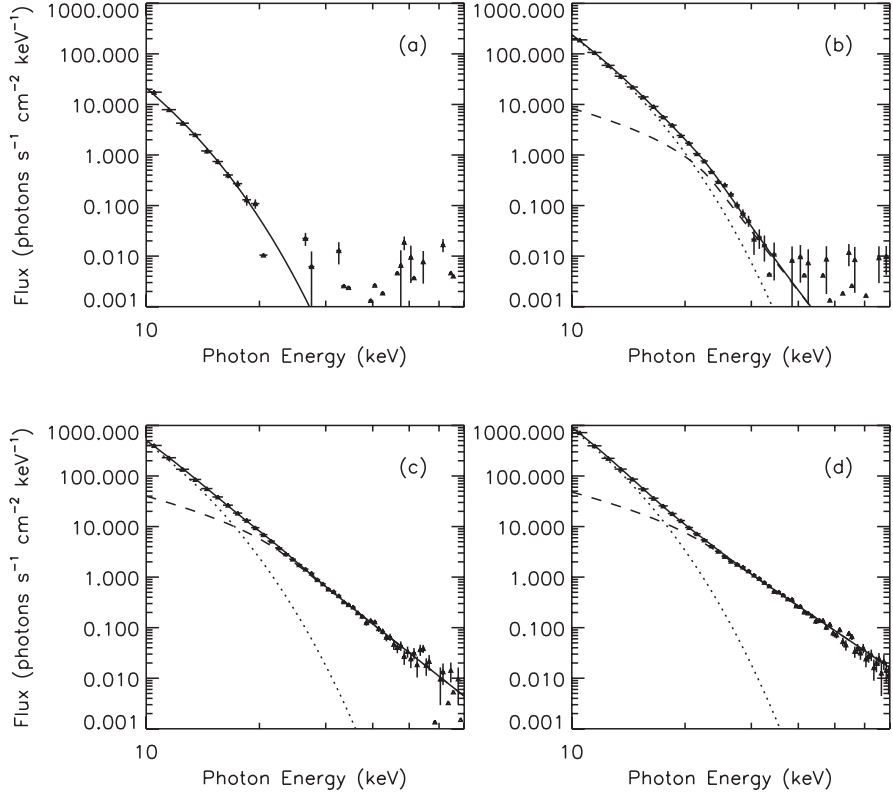}
	    \caption{
		{\it RHESSI} spatially integrated spectra in four
		time intervals during SOL2002-04-15T03:55 (M1.2). (a)
		Spectrum at 23:06:20--23:06:40 UT (early rise phase).
		(b) Spectrum at 23:09:00-23:09:20 UT (just before
		impulsive phase).  (c) Spectrum at 23:10:00--23:10:20
		UT (soon after the impulsive rise).  (d ) Spectrum
		at 23:11:00--23:11:20 UT (at the hard X-ray peak).
		The plus signs with error bars represent the spectral
		data. The lines represent model spectral fits: the
		dashed lines are non-thermal thick-target bremsstrahlung,
		the dotted lines are thermal bremsstrahlung, and
		the solid lines are the summation of the two
		\citep[from][]{2005ApJ...626.1102S}.
\index{flare (individual)!SOL2002-04-15T03:55 (M1.2)!spectral fits}
	    } \label{fig:holman_Apr15spectra} \end{figure}

    One of the best determinations of the low-energy cutoff so far
    was obtained by \citet{2005ApJ...626.1102S}.   They complemented
    the spatially-integrated spectral data for the 
    SOL2002-04-15T03:55 (M1.2)\index{flare (individual)!SOL2002-04-15T03:55 (M1.2)!low-energy cutoff}
    limb flare with imaging and lightcurve information.  Four spectra
    from this flare are shown in Figure~\ref{fig:holman_Apr15spectra}.
    The earliest spectrum, before the impulsive rise of the higher
    energy X-rays, was well fitted with an isothermal model.  The
    last spectrum, from the time of the hard X-ray peak, clearly
    shows a thermal component below $\sim$20~keV.  Of particular
    interest is the second spectrum, showing both thermal and
    non-thermal fit components.  As a consequence of the flattening
    of the isothermal component at low energies, the low-energy
    cutoff to the non-thermal component cannot extend to arbitrarily
    low energies without exceeding the observed emission.  This
    places a tight constraint on the value of the low-energy cutoff.
    The additional requirement that the time evolution of the derived
    temperature and emission measure of the thermal component be
    smooth and continuous throughout the flare constrains the value
    at other times.  Applying the collisional thick-target model
    with a power-law distribution of injected electrons, they found
    the best cutoff value to be $E_c$ = 24$\pm$2~keV (roughly
    constant throughout the flare).  The energy associated with
    these non-thermal electrons was found to be comparable to the
    peak energy in the X-ray-emitting thermal plasma, but an order
    of magnitude greater than the kinetic energy of the associated
    coronal mass ejection (CME)
\index{coronal mass ejections (CMEs)}
    \citep{2005ApJ...633.1175S}.  This contrasts with results
    obtained for large flares, where the minimum energy in non-thermal
    electrons is typically found to be less than or on the order
    of the energy in the CME \citep[e.g., ][]{2004JGRA..10910104E}.

The importance of correcting for the distortion of spectra by
albedo\index{albedo!low-energy cutoff} was revealed by a search for low-energy cutoffs
\index{hard X-rays!albedo}
\index{low-energy cutoff}
in a sample of 177 flares with relatively flat spectra ($\gamma
\leq 4$) between 15 and 20~keV \citep{2008SoPh..tmp..149K}.  Spectra
can be significantly flattened by the presence of albedo photons
in this energy range.  The X-ray spectra, integrated over the
duration of the impulsive phase of the flares, were inverted to
obtain the corresponding mean electron flux distributions.
\index{electrons!distribution function!mean electron flux}\index{mean electron flux}
Eighteen of the flares
showed significant dips in the mean electron flux distribution in
the 13-19~keV electron energy range that might be associated with
a low-energy cutoff (see Section \ref{sec:SH_cutoff_shape}).  However,
when the X-ray spectra were corrected for albedo\index{albedo} from
isotropically emitted X-rays, all of the dips disappeared.  Therefore,
the authors concluded that none of these flare electron distributions
had a low-energy cutoff above 12~keV, the lowest electron energy
in their analysis.

Low-energy cutoffs\index{low-energy cutoff} were identified in the spectra of a sample of
{\it early impulsive flares}\index{flare types!early impulsive} observed
by {\it RHESSI} in 2002 \citep{2007ApJ...670..862S}.
Early impulsive flares are flares in which
the $>$25~keV hard X-ray flux increase is delayed by less than 30~s after the flux increase at
lower energies. 
The pre-impulsive-phase heating of plasma to X-ray-emitting temperatures
is minimal in these flares, allowing the nonthermal part of the spectrum to be observed
to lower energies.
In the sample of 33 flares, 9~showed spectral flattening at low energies
in spectra obtained throughout the duration of each flare with a
4~s integration time.  After correcting for the albedo\index{albedo}
from isotropically emitted X-rays, the flattening in 3 of the 9
flares, all near Sun center, disappeared.  The flattening that
persisted in the remaining 6~flares was consistent with that produced
by a low-energy cutoff.  The values derived for the low-energy
cutoff ranged from 15~to 50~keV.  The authors found the evolution
of the spectral break and the corresponding low-energy cutoff in
these flares to be correlated with the non-thermal hard X-ray flux.
Further studies are needed to assess the significance of this
correlation.

Low-energy cutoffs\index{low-energy cutoff!above 100~keV} with values exceeding 100~keV were identified
in the spectra of the large flare SOL2005-01-19T08:22 (X1.3)
\index{flare (individual)!SOL2005-01-19T08:22 (X1.3)!high low-energy cutoffs}
\citep{2009ApJ...699..917W}.  The hard X-ray light curve of this
flare consisted of multiple peaks that have been interpreted as
quasi-periodic oscillations
\index{flare (individual)!SOL2005-01-19T08:22 (X1.3)!quasi-periodic oscillations}
\index{quasi-periodic pulsations!magnetoacoustic waves}
driven by either magnetoacoustic oscillations in a nearby loop 
\citep{2006A&A...452..343N} 
or by super-Alfv{\'e}nic beams in the vicinity of the reconnection\index{reconnection} region
\citep{2006ApJ...644L.149O}.\index{electrons!super-Alfv{\' e}nic beams}\index{reconnection!and super-Alfv{\' e}nic beams}\index{beams!super-Alfv{\' e}nic} 
The high low-energy cutoffs were found
in the last major peak of the series of hard X-ray peaks.  Unlike
the earlier peaks, this peak was also unusual in that it was not
accompanied by the Neupert effect\index{Neupert effect} (see
Section~\ref{sec:aschwanden_timing_thdelays}), consistent with the
high values of the low-energy cutoff\index{low-energy cutoff!and time-of-flight analysis}, and it exhibited soft-hard-harder\index{hard X-rays!spectral evolution!soft-hard-harder}\index{time-of-flight analysis}
\index{flare (individual)!SOL2005-01-19T08:22 (X1.3)!soft-hard-harder} 
rather than soft-hard-soft spectral
evolution (see Section~\ref{Gr_shs_observations}).  A change in the
character of the observed radio emission\index{radio emission} and
movement of one of the two hard X-ray footpoints\index{hard X-rays!footpoint
sources} into a region of stronger photospheric magnetic field were
also observed at the time of this peak.  These changes suggest a
strong connection between large-scale flare evolution and electron
acceleration\index{acceleration!and large-scale processes}.



\section{Nonuniform ionization in the thick-target region} 
\label{sec:kontar_nonuniform}
\index{non-uniform ionization}


In the interpretation of hard X-ray (HXR)
\index{hard X-rays!spectral interpretation} spectra in
terms of the thick-target model\index{hard X-rays!thick-target}, one
effect which has been largely ignored until recently is that of
varying ionization\index{ionization state!nonuniform distribution} along the path of
the thick-target beam.\index{thick-target model!variable ionization}
As first discussed by \citet{1973SoPh...28..151B},
the decrease of ionization with depth in the solar atmosphere reduces
long-range collisional energy losses.  
This enhances the HXR
bremsstrahlung efficiency\index{bremsstrahlung!efficiency!!and non-uniform ionization} there, elevating the high energy end of
the HXR spectrum by a factor of up to 2.8 above that for a fully
ionized target. The net result is that a power-law electron spectrum
of index $\delta$ produces a photon spectrum of index $\gamma=\delta
-1$ at low and high energies (see Equation~\ref{eqn:holman_ithickpl}),
but with $\gamma < \delta-1$ in between. The upward break, where
the spectrum begins to flatten toward higher energies, occurs at
fairly low energies, probably masked in measured spectra by the
tail of the thermal component.  The downward knee, where the spectrum
steepens again to $\gamma = \delta-1$, occurs in the few deka-keV
range, depending on the column depth of the transition zone.  Thus,
the measured X-ray spectrum may show a flattening similar to that
expected for a low-energy cutoff in the electron distribution.

\subsection{Electron energy losses and X-ray emission in a nonuniformly
ionized plasma} \label{Ko2_non-uniform_ionized_plasma}

The collisional energy-loss cross-section $Q_c(E)$ is dependent on
the ionization\index{ionization state!particle energy loss rate} of the background medium.
Flare-accelerated electron beams can propagate in the fully ionized
corona as well as in the partially ionized transition region and
chromosphere.\index{transition region}.\index{beams!propagation}  
Following \citet{1956PThPh..16..139H} and
\citet{1973SoPh...28..151B}, the cross-section $Q_c(E)$ can be
written for a hydrogen plasma ionization fraction $x$
\begin{equation}\label{Ko2_eq_col_ionized}
    Q_c(E)=\frac{2\pi e^4}{E^2}\left(x \Lambda _{ee}+ (1-x)\Lambda
    _{eH}\right)=\frac{2\pi e^4}{E^2}\Lambda (x+\lambda),
\end{equation} where $e$ is the electronic charge, $\Lambda _{ee}$
the electron-electron logarithm for fully ionized media and $\Lambda
_{eH}$ is an effective Coulomb logarithm\index{Coulomb logarithm!electron-hydrogen collisions} 
for electron-hydrogen atom
collisions. Numerically $\Lambda_{ee} \simeq 20$ and $\Lambda_{eH}
\simeq 7.1$, so $\Lambda = \Lambda_{ee}-\Lambda_{eH} \simeq 12.9$
and $\lambda = \Lambda _{eH}/\Lambda \simeq 0.55$.

Then, in a hydrogen target of ionization level $x(N)$ at column
density $N(z)$, the energy loss equation for electron energy $E$
\index{collisions!energy losses} is (cf.
Equation~\ref{eqn:holman_colloss}) \begin{equation}\label{Ko2_en_loss}
    \frac{dE}{dN}=-\frac{2\pi e^4 \Lambda}{E}(\lambda+x(N))=-\frac
    {K'}{E}(\lambda+x(N)),
\end{equation} where $K' = 2\pi e^4 \Lambda = (\Lambda/\Lambda_{ee})K
\simeq 0.65K$.

The energy loss of a given particle with initial energy $E_0$ depends
on the column density $N(z)=\int _0^z n(z')dz'$\index{column density},
so the electron energy at a given distance $z$ from the injection
site can be written $E^2=E_0^2-2K'M(N(z))$ (cf.
Equation~\ref{eqn:holman_colevol}), where
\begin{equation}\label{Ko2_nonuniform}
    M(N(z))=\int \limits _0^{N(z)}(\lambda+ x(N'))dN'
\end{equation} 
is the ``effective'' ionization-weighted collisional
column density.

The fractional atmospheric ionization $x$ as a function of column
density $N$ (cm$^{-2}$) changes from $1$ to near $0$ over a small
spatial range in the solar atmosphere.  Therefore, to lowest order,
$x(N)$ can be approximated by a step function $x(N)=1$ for $N <
N_{*}$, and $x(N)=0$ for $N \geq N_{*}$.  This gives $M(N) = (\lambda
+ 1)N$ for $N < N_{*}$ and $M(N) = N_{*} + \lambda N$ for $N \geq
N_{*}$.  Electrons injected into the target with energies less than
$E_{*} = \sqrt{2K'(\lambda + 1)N_{*}} = \sqrt{2KN_{*}}$ experience
energy losses and emit X-rays in the fully ionized plasma with
$x=1$, as in the standard thick-target model.  Electrons injected
with energies higher than $E_{*}$ lose part of their energy and
partially emit X-rays in the un-ionized ($x=0$), or, more generally,
partially ionized plasma.

We can deduce the properties of the X-ray spectrum by substituting
Equation~\ref{Ko2_en_loss} into Equation~\ref{eqn:holman_nu} (with $dN = n{\rm
v}dt$) and comparing $I_{thick}(\epsilon)$ from Equation~\ref{eq:holman_ithick1}
with $I_{thick}(\epsilon)$ from Equation~\ref{eqn_holman_ithick2}.  We
see that for the nonuniformly ionized case the denominator in the
inner integral now contains $\lambda+x(N)$ and $K$ is replaced with
$K'$.  In the step-function model for $x(N)$, photon energies greater
than or equal to $\epsilon_{*} = E_{*}$ are emitted by electrons
in the un-ionized plasma with $E \geq E_{*}$.  Since $\lambda+x(N)$
has the constant value $\lambda$, the thick-target power-law spectrum
is obtained (for injected power-law spectrum), but the numerical
coefficient contains $K'\lambda = 2\pi e^4\Lambda_{eH}$ instead of
$K$.  At photon energies far enough below $\epsilon_{*}$ that the
contribution from electrons with $E \geq E_{*}$ is negligible,
$\lambda+x(N) = \lambda + 1$ and the numerical coefficient contains
$(\lambda + 1)K' = K$.  The usual thick-target spectral shape and
numerical coefficient are recovered.  The ratio of the amplitude
of the high-energy power-law spectrum to the low-energy power-law
spectrum is $(\lambda + 1)/\lambda \simeq 2.8$.  The photon energy
$\epsilon_{*}(\rm{keV}) \simeq 2.3 \times 10^{-9}
\sqrt{N_{*}(\rm{cm}^{-2})}$, between where the photon spectrum
flattens below the high-energy power law and above the low-energy
power law, determines the value of the column density where the
plasma ionization fraction drops from 1 to 0.

\subsection{Application to flare X-ray spectra}
\label{Ko2_RHESSI_nonuniform}

\begin{figure} \begin{center} \psfig{file=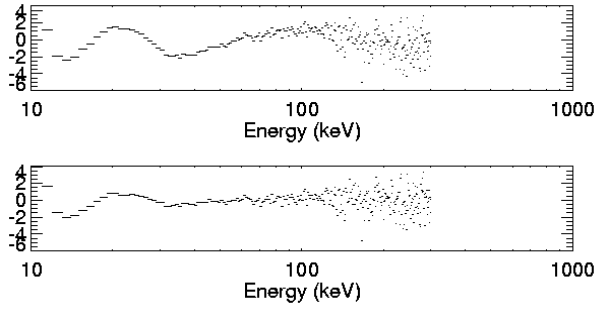,width=8.cm}
\caption{Photon spectrum residuals,
normalized by the statistical error for the spectral fit, for the
time interval 00:30:00 -- 00:30:20~UT, 2002-July-23,
\index{flare (individual)!SOL2002-07-23T00:35 (X4.8)!nonuniform ionization} 
for ({\it upper
panel}) an isothermal Maxwellian 
plus a power-law and ({\it lower
panel}) an isothermal Maxwellian plus the nonuniform ionization
spectrum with $\delta = 4.24$ and $E_*=53$~keV
\citep[from][]{2003ApJ...595L.123K}.} \label{Ko2_ionization_residuals}
\end{center} \end{figure}

The step-function nonuniform ionization model was used by
\citet{2002SoPh..210..419K,2003ApJ...595L.123K} to fit photon spectra
from five flares.  They assume a single power-law distribution of
injected electrons with power-law index $\delta$ and approximate
the bremsstrahlung cross-section with the Kramers cross-section\index{bremsstrahlung!!Kramers approximation}.  First, they
fit the spectra to the sum of a thermal Maxwellian at a single
temperature $T$ plus a single power law of index $\gamma$. For SOL2002-07-23T00:35 (X4.8)\index{flare (individual)!SOL2002-07-23T00:35 (X4.8)!nonuniform ionization} 
\citep[][]{2003ApJ...595L.123K} they limit themselves
to deviations from a power law in the non-thermal component of the
spectrum above $\sim $40~keV. The top panel of Figure
\ref{Ko2_ionization_residuals} shows an example of such deviations,
which represent significant deviations from the power-law fit. These
deviations are much reduced by replacing the power law with the
spectrum from the nonuniform ionization model, with the minimum rms
residuals obtained for values of $\delta = 4.24$ and $E_{*}=53$ keV
(Figure \ref{Ko2_ionization_residuals}, bottom panel).  The
corresponding minimum (reduced) $\chi^2$ value obtained for the
best fit to the X-ray spectrum (10--130~keV) dropped from 1.4 for
the power-law fit to 0.8 for the nonuniform ionization fit. There
are still significant residuals present in the range from 10 to 30
keV; these might be due to photospheric albedo or the assumption
of a single-temperature thermal component.

By assuming that the main spectral feature observed in a hard X-ray
spectrum is due to the increased bremsstrahlung efficiency of the
un-ionized chromosphere, allowance for nonuniform target ionization
offers an elegant direct explanation for the shape of the observed
hard X-ray spectrum and provides a measure of the location of the
transition region.\index{transition region}
Table~\ref{Ko2_table_ionization} shows the best fit parameters
derived for the four flare spectra analyzed by \citet{2002SoPh..210..419K}.
The last column shows the ratio of the minimum $\chi^2$ value
obtained from the nonuniform ionization fit to the minimum $\chi^2$
value obtained from a uniform ionization (single power-law) fit to
the non-isothermal part of the spectrum.  The nonuniform ionization
model fits clearly provide substantially better fits than single
power-law fits.

\begin{table*}[hbt] \caption[]{Best fit nonuniformly ionized target
model parameters for a single power-law ${\cal F}_0(E_0)$, the
equivalent $N_*$ (energy range 20-100 keV), and the ratio of $\chi
^2_{nonuni}/\chi ^2_{uni}$ \citep[from][]{2002SoPh..210..419K}
\index{flare (individual)!SOL2002-02-20T11:07 (C7.5)!nonuniform ionization}
\index{flare (individual)!SOL2002-03-17T19:31 (M4.0)!nonuniform ionization}
\index{flare (individual)!SOL2002-05-31T00:16 (M2.4)!nonuniform ionization}
\index{flare (individual)!SOL2002-06-01T03:58 (M1.5)!nonuniform ionization}}.
\label{Ko2_table_ionization} \centering \begin{tabular}{rcccccc}
\hline \\ Date & \multicolumn{1}{c}{Time, UT} & \multicolumn{1}{c}{$kT({\rm
keV})$} &\multicolumn{1}{c}{$\delta$} & \multicolumn{1}{c}{$E_*$
(keV)} &\multicolumn{1}{c}{$N_*$}(cm$^2$) &\multicolumn{1}{c}{$\chi
^2_{nonuni}/\chi ^2_{uni}$} \\ \hline

20 Feb 2002 & 11:06& 1.47& 5.29 & 37.4 & 2.7 $\times 10^{20}$
&0.032\\ 17 Mar 2002  & 19:27&1.27& 4.99 & 24.4 & 1.1 $\times
10^{20}$ &0.047\\ 31 May 2002 & 00:06&2.02& 4.15 & 56.2 & 6.1 $\times
10^{20}$ &0.041\\ 1 Jun 2002 & 03:53&1.45& 4.46 & 21.0 & 8.4 $\times
10^{19}$ &0.055\\

   \hline
\end{tabular} \end{table*}

\begin{figure}    
\begin{center}
\psfig{file=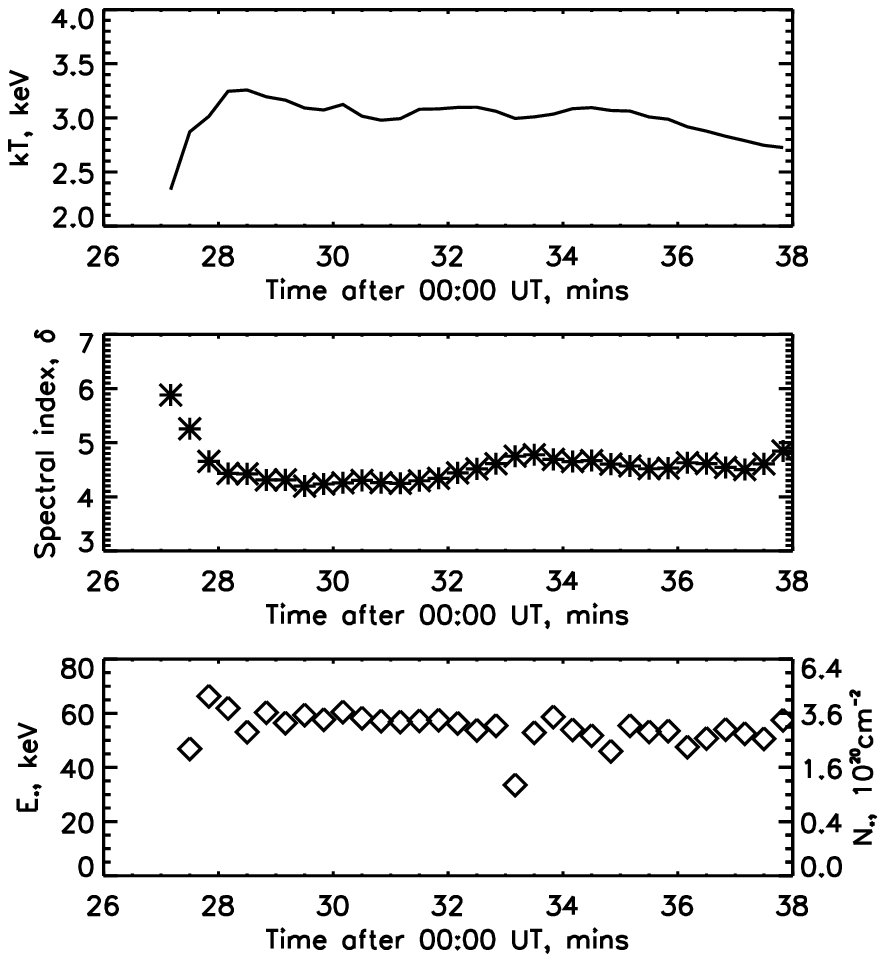,width=10.cm} 
\caption{Variation of $kT$, $\delta$, $E_*$, and
$N_*$ throughout SOL2002-07-23T00:35 (X4.8)
\index{flare (individual)!SOL2002-07-23T00:35 (X4.8)!nonuniform ionization} \citep[][]{2003ApJ...595L.123K}.
The variation of other parameters, such as emission measure, can
be found in \citet{2003ApJ...595L..97H} and \cite{2010ApJ...725L.161C}.
} \label{Ko2_nonuni_temporal}
\end{center} \end{figure}

Values of the fit parameters $kT$~(keV), $\delta$ and $E_*$ as a
function of time for SOL2002-07-23T00:35 (X4.8)
\index{flare (individual)!SOL2002-07-23T00:35 (X4.8)!nonuniform ionization}, 
together with
the corresponding value of $N_*$(cm$^{-2}$) $\simeq 1.9 \times
10^{17}E_*$(keV)$^2$ were obtained by \citet{2003ApJ...595L.123K}.
The results (Figure \ref{Ko2_nonuni_temporal}) demonstrate that the
thermal plasma temperature rises quickly to a value $\simeq 3$~keV
and decreases fairly slowly thereafter. The injected electron flux
spectral index $\delta$ follows a general ``soft-hard-soft'' trend
\index{hard X-rays!soft-hard-soft}
\index{soft-hard-soft}
and qualitatively agrees with the time history of the simple best-fit
power-law index $\gamma$ \citep[][]{2003ApJ...595L..97H}.  $E_*$
rises quickly during the first minute or so from $\sim $40~keV to
$\sim $70~keV near the flare peak and thereafter declines rather
slowly. The corresponding values of $N_*$ are $\sim$2-5~$\times
10^{20}$~cm$^{-2}$.

The essential results of these studies are that (1) for a single
power-law electron injection spectrum, the expression for bremsstrahlung
emission from a nonuniformly-ionized target provides a significantly
better fit to observed spectra than the expression for a uniform
target; and (2) the value of $E_*$ (and correspondingly~$N_*$)
varies with time.

An upper limit on the degree of spectral flattening\index{hard X-rays!spectral flattening}
$\Delta\gamma$
that can result from nonuniform ionization was derived by
\citet{2009ApJ...705.1584S}.  They applied this upper limit to
spectra from a sample of 20 flares observed by {\it RHESSI} in the
period 2002 through 2004.  They found that 15 of the 20 flare spectra
required a downward spectral break at low energies and for each of
these 15 spectra derived the difference $\Delta\gamma$ of the
best-fit power-law spectral indices above and below the break.  A
Monte Carlo method was used to determine the 95\% confidence interval
for each of the derived values of $\Delta\gamma$.  Taking the value
of $\Delta\gamma$ to be incompatible with nonuniform ionization if
the 95\% confidence interval fell above the derived upper limit,
\citet{2009ApJ...705.1584S} found that six of the flare spectra
could not be explained by nonuniform ionization alone.  Thus, for
these six flares some other cause such as a low-energy cutoff or
return-current-associated energy losses
(Section~\ref{sec:zharkova_return_current}) must be at least partially
responsible for the spectral flattening.\index{low-energy cutoff!and return current}\index{return current} 

\section{Return current losses} 
\label{sec:zharkova_return_current}
\index{return current}


The thick-target model assumes that a beam of electrons is injected
at the top of a loop and ``precipitates'' downwards in the solar
atmosphere. Unless accompanied by an equal flux of positively charged
particles, these electrons constitute a current and must create a\index{precipitation}
significant self-induced electric field\index{electric fields!self-field} that
in turn drives a co-spatial return current\index{return current}\index{thick-target model!need for return current}
for compensation
\citep{1976SoPh...48..197H,1977ApJ...218..306K,1980ApJ...235.1055E,1988SoPh..116..119D}.
The return current consists of ambient electrons, plus any primary
electrons that have scattered back into the upward direction. By
this means we have a full electric circuit of precipitating and
returning electrons that keeps the whole system neutral and the
electron beam stable against being pinched off by the self-generated
magnetic field required by Amp{\` e}re's law\index{Amp{\` e}re's Law} for an unneutralized beam
\index{magnetic field!and return current}
current.\index{electron beams!and induced fields}\index{beams!induced fields}
However, the self-induced electric field results in a
potential drop along the path of the electron beam that decelerates
and, therefore, removes energy from the beam electrons.

\subsection{The return current electric field}
\index{electric fields!return current}\index{return current} 

The initial formation of the beam/return-current system has been
studied by \citet{1990A&A...234..496V} and references therein.  We
assume here that the system has time to reach a quasi-steady state.
Van den Oord finds this time scale to be on the order of the thermal
electron-ion collision time.  This time scale is typically less
than or much less than one second, depending on the temperature and
density of the ambient plasma.  In numerical simulations by
\citet{2009A&A...504.1057S}, times to reach a steady state after
injection ranged from 0.07~s to 0.2~s, depending on the initial
beam parameters.

The self-induced electric field\index{electric fields!self-induced} strength at a given location $z$
along the beam and the flare loop, $\mathcal{E}(z)$, is determined
by the current density associated with the electron beam, $j(z)$,
and the local conductivity of the loop plasma, $\sigma(z)$, through
Ohm's law: $\mathcal{E}(z)=j(z)/\sigma(z)$\index{Ohm's Law}. 
Relating the current
density to the density distribution function
\index{electrons!distribution function!density} of the precipitating electrons,
$f(z,E,\theta)$, where $E$ is the electron energy and $\theta$ is
the electron pitch angle, gives
\begin{equation} \mathcal{E}(z)  = \frac{2%
\sqrt{2}\pi }{\sigma
(z)}\frac e{\sqrt{m_e}} \int \limits_{0}^{1} \int \limits_{0}^{\infty
} f(z,E,\theta )\sqrt{E}\mu dE d\mu.  \end{equation}
Here $\mu$ is the cosine of the pitch angle and $e$ and $m_e$ are
the electron charge and mass, respectively.\index{electric fields!self-induced}
The self-induced
electric field strength $\mathcal{E}(z)$ depends on the local
distribution of the beam electrons, which in turn depends on the
electric field already experienced by the beam as well as any Coulomb
energy losses and pitch-angle scattering that may have significantly
altered the beam.  It also depends on the local plasma temperature
(and, to a lesser extent, density) through $\sigma(z)$, which can,
in turn, be altered by the interaction of the beam with the loop
plasma (i.e., local heating and ``chromospheric evaporation'')\index{chromospheric evaporation}.
Therefore, determination of the self-induced electric field and its
impact on the precipitating electrons generally requires self-consistent
modeling of the coupled beam/plasma system.

Such models have been computed by
\citet{2005A&A...432.1033Z,2006ApJ...651..553Z}.  They numerically
integrate the time-dependent Fokker-Planck\index{Fokker-Planck
equation} equation to obtain the self-induced electric field strength
and electron distribution function along a model flare loop.\index{electric fields!self-induced}  
The injected electron beam was assumed to have a single power-law energy
distribution in the energy range from $E_{low}$ = 8~keV to $E_{upp}$
= 384~keV and a normal (Gaussian) distribution in pitch-angle cosine
$\mu$ with half-width dispersion $\Delta \mu =0.2$ about $\mu = 1$.

The model computations show that the strength of the self-induced
electric field is nearly constant at upper coronal levels and rapidly
decreases with depth (column density) in the lower corona and
transition region. The rapidity of the decrease depends on the beam
flux spectral index.  It is steeper for softer beams ($\delta$=5-7)
than for harder ones ($\delta$=3).  The strength of the electric
field is higher for a higher injected beam energy flux
density\index{electrons!energy flux density} (erg cm$^{-2}$ s$^{-1}$),
and the distance from the injection point over which the electric
field strength is highest (and nearly constant) decreases with
increasing beam flux density.

\subsection{Impact on hard X-ray spectra}

Deceleration of the precipitating beam by the electric field most
significantly affects the lower energy electrons ($<$100~keV), since
the fraction of the original particle energy lost to the electric
field is greater for lower energy electrons.\index{electric fields!and beam deceleration} 
This leads to flattening
of the electron distribution function towards the lower energies
and, therefore, flattening of the photon spectrum.\index{hard X-rays!spectral flattening}

\begin{figure} \centering
\scalebox{0.5}{\includegraphics{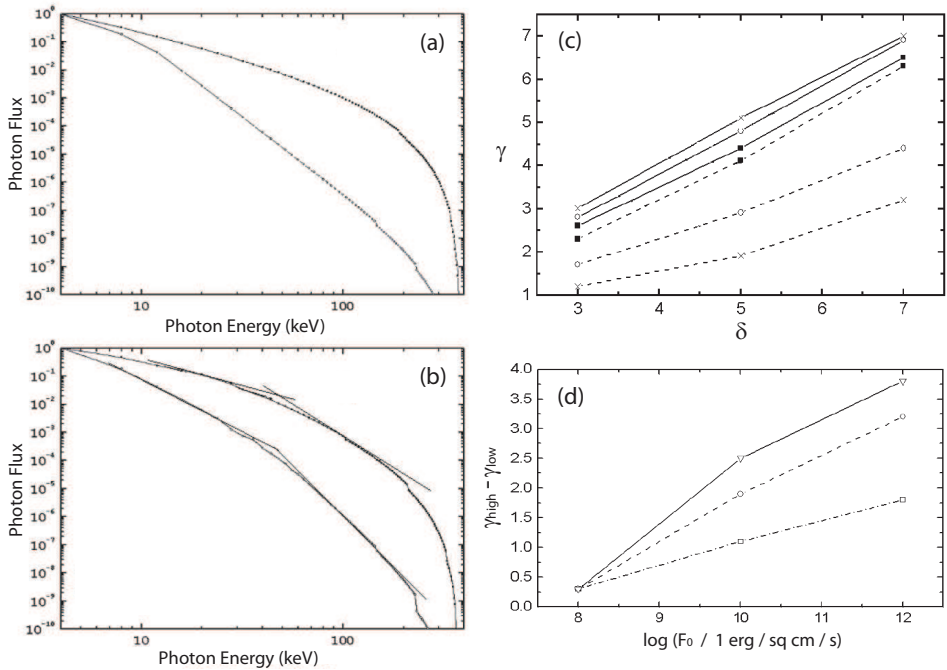}} \vspace{-0.6in}
\caption{(a) Photon spectra computed from full kinetic solutions
including return current losses and collisional losses and scattering.
The top spectrum is for an injected single-power-law electron flux
distribution between 8~keV and 384~keV with an index of $\delta=3$,
and the bottom spectrum is for $\delta=7$.  The injected electron
energy flux density is 10$^8$ erg cm$^{-2}$ s$^{-1}$.  (b) Same as
(a), but for an injected energy flux density of 10$^{12}$ erg
cm$^{-2}$ s$^{-1}$.  The tangent lines at 20 and 100~keV demonstrate
the determination of the low-energy and high-energy power-law
spectral indices $\gamma_{low}$ and $\gamma_{high}$.  (c)  The
photon spectral indices $\gamma_{low}$ (dashed lines) and $\gamma_{high}$
(solid lines) vs. $\delta$ for an injected energy flux density of
10$^8$ (squares), 10$^{10}$ (circles), and 10$^{12}$ erg cm$^{-2}$
s$^{-1}$ (crosses).  (d)  $\gamma_{high} - \gamma_{low}$ vs. the
log of the injected electron energy flux density for $\delta$ equal
to 3 (bottom curve, squares), 5 (middle curve, circles), and 7 (top
curve, triangles) \cite[from][]{2006ApJ...651..553Z}.}
\label{fig:Zharkova_rcspectra} 
\end{figure}
\index{hard X-rays!spectral flattening!illustration}

Photon spectra computed from kinetic solutions that include return
current energy losses and collisional energy losses and scattering
are shown in Figure~\ref{fig:Zharkova_rcspectra} (a) and (b).  Low-
and high-energy spectral indices and their dependence on the power-law
index of the injected electron distribution and on the injected
beam energy flux density are shown in Figure~\ref{fig:Zharkova_rcspectra}
(c) and (d).  The difference between the high-energy and low-energy
spectral indices is seen to increase with both the beam energy flux
density and the injected electron power-law index $\delta$.  The
low-energy index is found to be less than 2 for $\delta$ as high
as 5 when the energy flux density is as high as 10$^{12}$ erg
cm$^{-2}$ s$^{-1}$.

\subsection{Observational evidence for the presence of the return
current} \label{sec:zharkova_rc_evidence}

We have seen that return current energy losses can introduce curvature
into a spectrum, possibly explaining the ``break'' often seen in
observed flare X-ray spectra\index{hard X-rays!spectral break}.  
A difficulty in directly testing
this explanation is that the thick-target model provides the power
(energy flux) in the electron beam (erg s$^{-1}$), but not the
energy flux density\index{electrons!energy flux density} (erg cm$^{-2}$
s$^{-1}$).\index{thick-target model!filamentation}
X-ray images provide information about the area of the
target, but this is typically an upper limit on the area.  Even if
the source area does appear to be well determined, the electron
beam can be filamented so that it does not fill the entire area
(the filling factor\index{hard X-rays!filling factor} is less than 1).
\index{filling factor!hard X-rays}  
Also, if only an upper limit
\index{low-energy cutoff}
on the low-energy cutoff\index{electrons!distribution function!low-energy
cutoff} to the electron distribution is known, as described in
Section~\ref{sec:sainthilaire_Ec}, the energy flux density may be
higher.  Therefore, the observations typically only give a lower
limit on the beam energy flux density.

The non-thermal hard X-ray flux is proportional to the electron beam
flux density, but the return-current energy losses are also
proportional to the beam flux density.  As a consequence,
\citet{1980ApJ...235.1055E} concluded that the flux density of the
non-thermal X-ray emission from a flare cannot exceed on the order
of $10^{-15}$ cm$^{-2}$ s$^{-1}$ above 20~keV.
\citet{2007ApJ...666.1268A} have deduced the photon flux density
from non-thermal electrons in a sample of 10 flares ranging from
\index{satellites!GOES@\textit{GOES}}
\textit{GOES}\index{GOES@\textit{GOES}}
class M1.8 to X17.  
They find that the non-thermal photon flux density
does not monotonically increase with the thermal energy flux, but
levels off (saturates) as the thermal energy flux becomes high.
They argue that this saturation\index{hard X-rays!flux saturation} most
likely results from the growing importance of return current energy
losses as the electron beam flux increases to high values in the
larger flares.  They find that the highest non-thermal photon flux
densities agree with an upper limit computed by Emslie.\index{return current} 

A correlation between the X-ray flux and spectral break
energy\index{hard X-rays!spectral evolution} was found by
\citet{2007ApJ...670..862S} in their study of X-ray spectra in early
impulsive flares\index{flare types!early impulsive} (see
Section~\ref{sec:sainthilaire_Ec}).  They point out that the
increasing impact of return current energy losses on higher energy
electrons as the electron beam energy flux density increases could
be an explanation for this correlation.

\citet{2008A&A...487..337B} studied two flares with non-thermal
coronal hard X-ray sources for which the difference between the
measured photon spectral index at the footpoints and the spectral
index of the coronal source was greater than two, the value expected
for coronal thin-target emission and footpoint thick-target emission
from a single power-law electron distribution (see
Section~\ref{subsec:battaglia_deltagamma}).  They argue that
return-current losses\index{return current!energy losses} between the coronal and footpoint source
regions are most likely responsible for the large difference between
the spectral indices.

The return current can also affect the spectral line emission from
flares.\index{return current!and spectral lines} 
Evidence for the presence of the return current at the
chromospheric level from observations of the linear 
polarization\index{impact polarization}\index{return current!and impact polarization}\index{polarization!H$\alpha$}
of the hydrogen H$\alpha$ and H$\beta$ lines\index{Fraunhofer lines!H$\alpha$}\index{Fraunhofer lines!H$\beta$} has been presented by
\citet{2003A&A...407.1103H}.  \citet{2008SoPh..250..329D} have shown
that the presence of a return current in the corona may have a
distinguishable impact on the relative intensities of spectral lines
emitted from the corona.  
\section{Beam-plasma and current instabilities} 
\label{sec:holman_instabilities}


Interaction of the accelerated electrons with plasma turbulence\index{turbulence!and electron distribution function} as
they stream toward the thick-target emission region can modify the
electron distribution.\index{electrons!distribution function!and turbulence}\index{turbulence}\index{plasma turbulence}
In this section we briefly discuss a likely
source of such turbulence: that generated by the electron beam itself.  
If the beam is or becomes unstable to driving the growth
of plasma waves, these waves can interact with the beam and modify
its energy and/or pitch angle distribution\index{electrons!pitch-angle distribution} 
until the instability is removed or the wave growth is stabilized.  
\index{waves!plasma}
The return current\index{return current!instability}
associated with the beam (Section~\ref{sec:zharkova_return_current})
can also become unstable, resulting in greater energy loss from the
beam.  
\index{plasma instabilities!beam-plasma}\index{plasma instabilities!return-current}\index{return current!and beam stability} 
Beam-plasma and return-current instabilities in solar flares
have been reviewed by \citet{1990SoPh..130....3M} and
\citet{2002ASSL..279.....B}.

A sharp, low-energy cutoff to an electron beam or, more generally,
a positive slope in the beam electron energy distribution is
well-known to generate the growth of electrostatic plasma waves
(Langmuir waves)\index{Langmuir waves}.  
The characteristic time
scale for the growth of these waves is on the order of
$[(N_b/n)\omega_{pe}]^{-1}$, where $\omega_{pe}$ is the electron
plasma frequency and $N_b/n$ the ratio of the density of unstable
electrons in the beam to the plasma density.\index{frequency!plasma}  
This is on the order
of microseconds for a typical coronal loop plasma density and $N_b/n
\approx 10^{-3}$.  This plasma instability is often referred to as the
bump-on-tail instability\index{plasma instabilities!bump-on-tail}.
The result is that on a somewhat longer but comparable time scale
electrons from the unstable region of the electron distribution
lose energy to the waves until the sharp cutoff is flattened so
that the distribution no longer drives the rapid growth of the
waves.  Therefore, the electron energy distribution below the
low-energy cutoff is likely to rapidly become flat or nearly flat
(sufficiently flat to stabilize the instability) after the electrons
escape the acceleration region \citep[see Chapters 9 \& 10 of
][]{1973ppp..book.....K}.\index{acceleration region!escape from}

A recent simulation of the bump-on-tail instability\index{plasma instabilities!bump-on-tail!simulation} for solar flare conditions, including Coulomb collisions and wave damping as the
electrons propagate into an increasingly dense plasma, has been
carried out by \citet{2009ApJ...707L..45H}.  The authors compute
the mean electron flux distribution\index{electrons!distribution
function!mean electron flux} in their model flare atmosphere after
injecting a power-law distribution with a sharp low-energy cutoff.
They find a mean electron flux distribution with no dip (see
Section~\ref{sec:SH_cutoff_shape}) and a slightly negative slope
below the original cutoff energy with a spectral index $\delta$
between 0~and~1.

A beam for which the mean electron velocity parallel to the magnetic
field substantially exceeds the mean perpendicular velocity can
drive the growth of waves that resonantly interact with the beam.
When the electron gyrofrequency\index{frequency!Larmor} exceeds the plasma frequency, these
waves are electrostatic and primarily scatter the electrons in pitch
angle.  Generally known as the anomalous Doppler resonance
instability\index{plasma instabilities!anomalous Doppler resonance},
this instability tends to isotropize the beam electrons.
\cite{1982ApJ...257..354H} showed that under solar flare conditions
this instability can grow and rapidly isotropize the beam electrons
in less than a millisecond.  They found that electrons at both the
low- and high-energy ends of the distribution may remain unscattered,
however, because of wave damping.  This could result in up to two
breaks in the emitted X-ray spectrum.  On the other hand,
\citet{1984A&A...139..263V} have argued that non-thermal tails will
form in the ambient plasma and stabilize the anomalous Doppler
resonance instability by suppressing the growth of the plasma waves.

Electrons streaming into a converging magnetic field can develop a
\index{magnetic structures!loss cone}
loss-cone distribution, with a deficit of electrons at small pitch
angles.  Both classical Coulomb collisions and loss-cone
instabilities\index{plasma instabilities!loss-cone} can relax this
distribution by scattering electrons into the loss cone or extracting
energy from the component of the electron velocities perpendicular
to the magnetic field \citep[e.g.,][]{1990A&AS...85.1141A}.  One
loss-cone instability, the electron-cyclotron
\index{plasma instabilities!electron-cyclotron maser} 
\index{plasma instabilities!loss-cone} 
or gyrosynchrotron
maser\index{plasma instabilities!gyrosynchrotron maser}, produces coherent radiation
observable at radio frequencies
\citep{1980IAUS...86..457H,1982ApJ...259..844M}.

The return current associated with the streaming electrons becomes
unstable to the ion-acoustic instability
\index{plasma instabilities!ion-acoustic} when its drift speed exceeds a value
on the order of the ion sound speed.  
\index{resistivity}
The excited ion sound waves
\index{waves!ion sound}
enhance the plasma resistivity, increasing the electric field
strength associated with the return current, the heating of the
plasma by the current, and the energy loss from the electron beam.

It has been argued that rapid plasma heating and particle acceleration
in the corona should result in the expansion of hot plasma down the
legs of flare loops at the ion sound speed, confined behind a
collisionless ion-acoustic conduction front\index{conduction fronts} \citep{1979ApJ...228..592B}\index{conduction fronts}\index{chromospheric evaporation}.
Electrons with speeds greater than
about three times the electron thermal speed would be able to stream
ahead of the conduction front.  This scenario has not been
observationally verified, but the observational signature may be
confused by the chromospheric evaporation\index{chromospheric
evaporation} produced by the high-energy particles streaming ahead
of the conduction front.

\citet{1985A&A...142..219R} argued that if the electron beam is
unstable to beam plasma interactions, the return current will be
carried by high-velocity electrons.\index{return current!charge carrier}  
This reduces the impact of
collisions on the beam/return-current system and helps stabilize
the system.  In a recent simulation, \citet{2008A&A...486..325K}
have found that for current drift velocities exceeding the electron
thermal speed, the return current is carried by both the primary
(drifting thermal) current and an extended tail of high-velocity
electrons.

The evolution of the electron-beam/return-current system when the
return-current drift speed exceeds the electron thermal speed has
also been simulated by \citet{2008A&A...478..889L}.  They find that
double layers\index{double layer} form in the return current, regions
of enhanced electric field that further increase the energy losses
of the electron beam.  This, in turn, increases the highest electron
energy to which these losses significantly flatten the electron
distribution and corresponding hard X-ray spectrum.\index{return current!and double layers}\index{electric fields!double layers} 

The beam/return-current system has been simulated by
\citet{2009ApJ...690..189K}, with a focus on the role of the Weibel
instability\index{plasma instabilities!Weibel}.  The Weibel instability
tends to isotropize the electron distribution.  \citet{2009A&A...506.1437K}
have computed the thin-target X-ray emission from the evolved
electron distributions for a model with a weak magnetic field and
another model with a strong magnetic field (ratio of the electron
gyrofrequency to the plasma frequency $\sim$0 and $\sim1$,
respectively).\index{magnetization}
They demonstrate that in both cases the electron
distribution is more isotropic and the directivity of the X-ray
emission is lower than when the instability of the system is not
taken into account, with the greatest isotropization occurring in
the weak field limit.

Although we expect plasma instabilities to affect the evolution of
the electron beam, observationally identifying them is difficult.
The bump-on-tail instability and return current losses both lead
\index{plasma instabilities!bump-on-tail}\index{return current!energy losses!and low-energy cutoff} 
to a flat low-energy cutoff.  So far we have not established the
ability to observationally distinguish a flat low-energy cutoff
from a sharp low-energy cutoff.  The bump-on-tail instability may
be distinguishable from return-current losses by its short time
scale and, therefore, the short distance from the acceleration
region over which it effectively removes the unstable positive slope
from the electron energy distribution.  The instabilities that
isotropize the electron pitch-angle distribution may be responsible
for evidence from albedo\index{albedo} measurements that flare
electron distributions are isotropic or nearly isotropic
\citep{2006ApJ...653L.149K}.  
\section{Height dependence and size of X-ray sources with energy and time} 
\label{sec:aschwanden_height}

\subsection{Footpoint Sources}\index{hard X-rays!height dependence}

Hard X-ray footpoint sources\index{hard X-rays!footpoint sources} result
from collisional bremsstrahlung radiated by precipitating electrons,
which produce most of the emission in the chromosphere according
to the collisional thick-target model. Depending on the density
structure in the legs of the coronal magnetic loop, mildly energetic
electrons lose their energy in the lower corona or transition region,
while the more energetic electrons penetrate deeper into the
chromosphere (see Equation~\ref{eqn:holman_colloss}).

\begin{figure} \includegraphics[width=\textwidth]{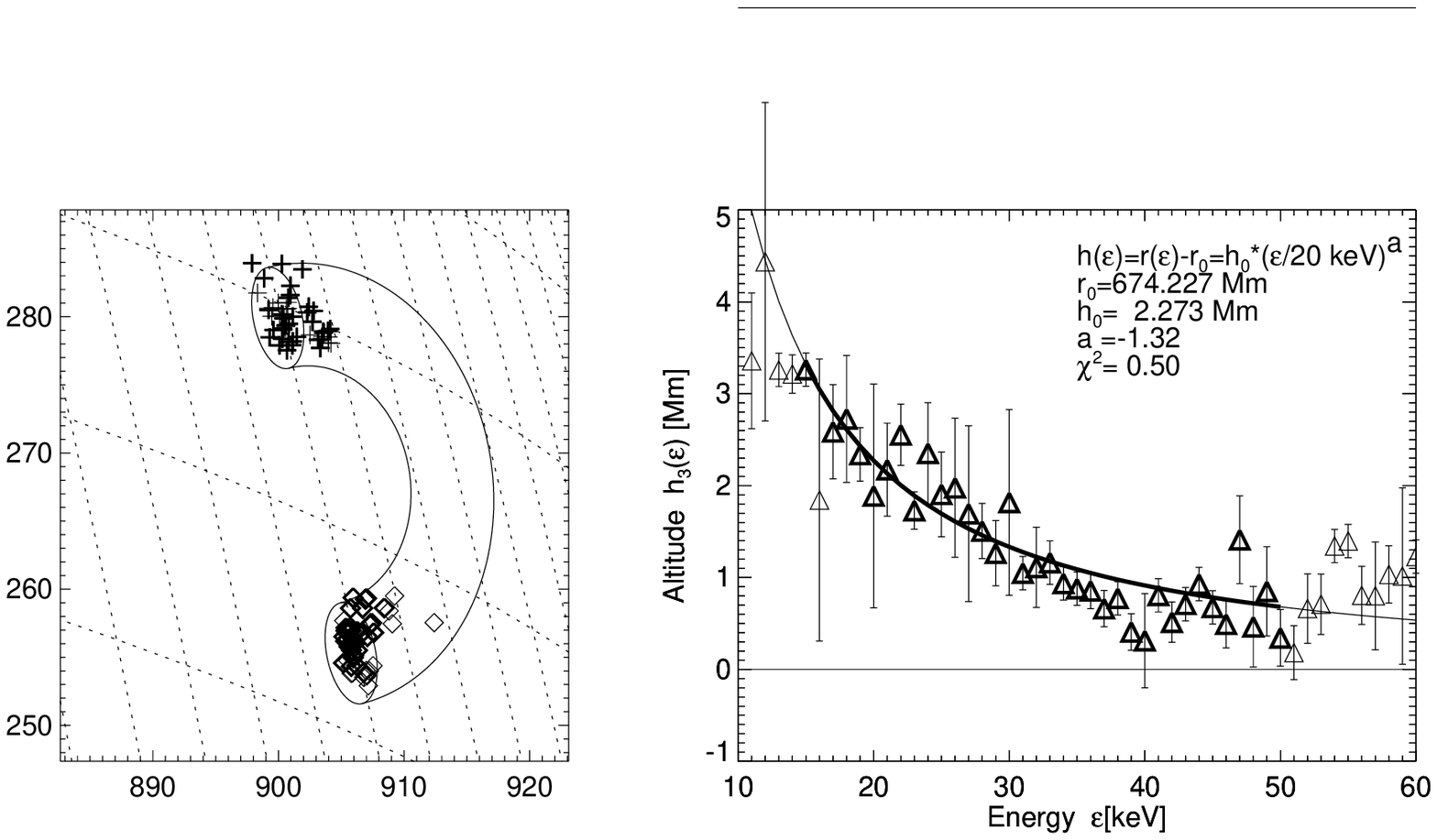}
\caption{The centroids\index{hard X-rays!height dependence} of footpoint
hard X-ray emission are marked for different photon energies between
10 keV and 60 keV for SOL2002-02-20T11:07 (C7.5), which
occurred near the solar west limb and was imaged with {\it RHESSI}
(left panel). The altitude $h(\epsilon)$ as a function of energy
$\epsilon$ shows a systematic height decrease with increasing energy
(right panel) \citep[from][]{2002SoPh..210..383A}.}
\label{fig:aschwanden_footpoint_heights} \end{figure}

The altitude of these hard X-ray footpoint\index{footpoints!altitude of sources} sources could never be
measured accurately before {\it RHESSI}, because of a lack of spatial
and spectral resolution. With {\it RHESSI}, we can measure the
centroid of the footpoint location with an accuracy of order an
arcsecond\index{RHESSI@\textit{RHESSI}!spatial resolution} for every photon energy
in steps as small as 1 keV\index{RHESSI@\textit{RHESSI}!spectral resolution}.  For
a flare near the limb (Figure~\ref{fig:aschwanden_footpoint_heights}),
the centroid location translates directly into an altitude.

\citet{2002SoPh..210..383A} studied such a flare, SOL2002-02-20T11:07 (C7.5)\index{flare (individual)!SOL2002-02-20T11:07 (C7.5)!chromospheric
density structure}.  The heights of the footpoint sources were
fitted with a power-law function of the photon energy.  This yielded
altitudes $h \approx 1000-5000$~km in the energy range
$\epsilon=10-60$~keV, progressively lower with higher energy, as
expected from the thick-target model
(Figure~\ref{fig:aschwanden_footpoint_heights}, right frame).

Since the stopping depth of the precipitating electrons is a function
of column density\index{column density}, the integrated density
along their path in the chromosphere (equation \ref{eqn:holman_colevol}),
the measured height dependence of the hard X-ray centroids can be
inverted to yield a density model of the chromosphere\index{chromospheric
density model} \citep{2002SoPh..210..373B}. Assuming the decrease
in density with height had a power-law dependence and the plasma
is fully ionized, the inversion of the {\it RHESSI} data in the
example shown in Figure~\ref{fig:aschwanden_footpoint_heights} yielded
a chromospheric density model that has a significantly higher
electron density in the $h=2000-5000$ km range than the standard
chromospheric models based on UV spectroscopy and hydrostatic
equilibrium (VAL and FAL models).
\index{atmospheric models!semi-empirical}
\index{atmospheric models!VAL}\index{atmospheric models!FAL} 
The {\it RHESSI}-based chromospheric
density model was therefore found to be more consistent with the
``spicular extended chromosphere,'' similar to the results from
sub-mm\index{atmospheric models!``spicular extended chromosphere''} 
radio observations during solar eclipses carried out at
Caltech \citep{1993ApJ...403..426E}.

Forward fitting {\it RHESSI} X-ray visibilities to an assumed
circular Gaussian source shape, \citet{2008A&A...489L..57K} found
for a limb flare the full width at half maximum (FWHM\index{FWHM})
size and centroid positions of hard X-ray sources as a function of
photon energy with a claimed resolution of $\sim$0$''$.2.\index{visibilities} 
They show
that the height variation of the chromospheric density and of the
magnetic flux density can be found with a vertical resolution of
$\sim$150~km by mapping the $18 - 250$~keV X-ray emission of energetic
electrons propagating in the loop at chromospheric heights of $400
- 1500$~km.  Assuming collisional losses in neutral hydrogen with
an exponential decrease in density with height, their observations
of SOL2004-01-06T06:29 (M5.8)
\index{flare (individual)!SOL2004-01-06T06:29 (M5.8)!chromospheric density \& magnetic structure} suggest that the density of
the neutral gas is in good agreement with hydrostatic models\index{atmospheric models!hydrostatic} with
a scale height of around $140 \pm 30$~km. FWHM sizes of the X-ray
sources decrease with energy, suggesting the expansion (fanning
out) of magnetic flux tubes in the chromosphere with height.\index{magnetic structures!flux tubes}
The magnetic scale height\index{magnetic field!scale height} $B(z)(dB/dz)^{-1}$ is found to be on the order of 300~km and a strong horizontal magnetic field is associated with
noticeable flux tube expansion at a height of $\sim$900~km.  A
subsequent analysis with an assumed elliptical Gaussian source shape
\citep{2010ApJ...717..250K} confirms these results and shows that
the vertical extent of the X-ray source\index{hard X-rays!height dependence} 
decrease with increasing X-ray energy.  
The authors find the vertical source sizes to be
larger than expected from the thick-target model and suggest that
a multi-threaded density structure in the chromosphere is required.
The thick-target model to which the results were compared, however,
did not account for partial occultation of the X-ray sources by the
solar limb.\index{occulted sources}

The flare SOL2002-02-20T11:07 (C7.5)
\index{flare (individual)!SOL2002-02-20T11:07 (C7.5)!chromospheric density structure} has been reanalyzed by
\citet{2009ApJ...706..917P} using both photon maps over a range of
photon energies and mean electron flux maps deduced from {\it RHESSI}
visibilities over a range of electron energies.\index{visibilities!and electron flux maps} 
Using source
centroids\index{hard X-rays!source centroids} computed from the maps and assuming an exponential decrease
in density with height, they found the density scale height to be
an order of magnitude larger than the expected chromospheric scale
height on the quiet Sun, but consistent with the scale height in a
non-static, flaring atmosphere\index{hard X-rays!height dependence}\index{footpoints!height structure}.  
This is also consistent with the
enhanced plasma densities found at $\sim$1000-5000~km altitudes
by \citet{2002SoPh..210..383A}.

If the results for the 400-1500~km height range
\citep{2008A&A...489L..57K} and for the $\sim$1000-5000~km
height range \citep{2002SoPh..210..383A,2009ApJ...706..917P} are
typical of flare loops, they imply that the upper chromosphere and
transition region respond with a non-hydrostatic, expanded atmosphere
while the low chromosphere does not respond to the flare energy
release.\index{transition region!non-hydrostatic}
These results could, of course, depend on the magnitude
of the flare.  More studies of this kind are clearly desirable,
especially in coordination with observations of spectral lines from
the chromosphere and transition region.  


\subsection{Loop Sources and their Evolution} \label{sec:liuw_height}
\index{loops!hard X-rays}\index{hard X-rays!loops}

\begin{figure} \begin{center}
 \includegraphics[width=0.7\textwidth]{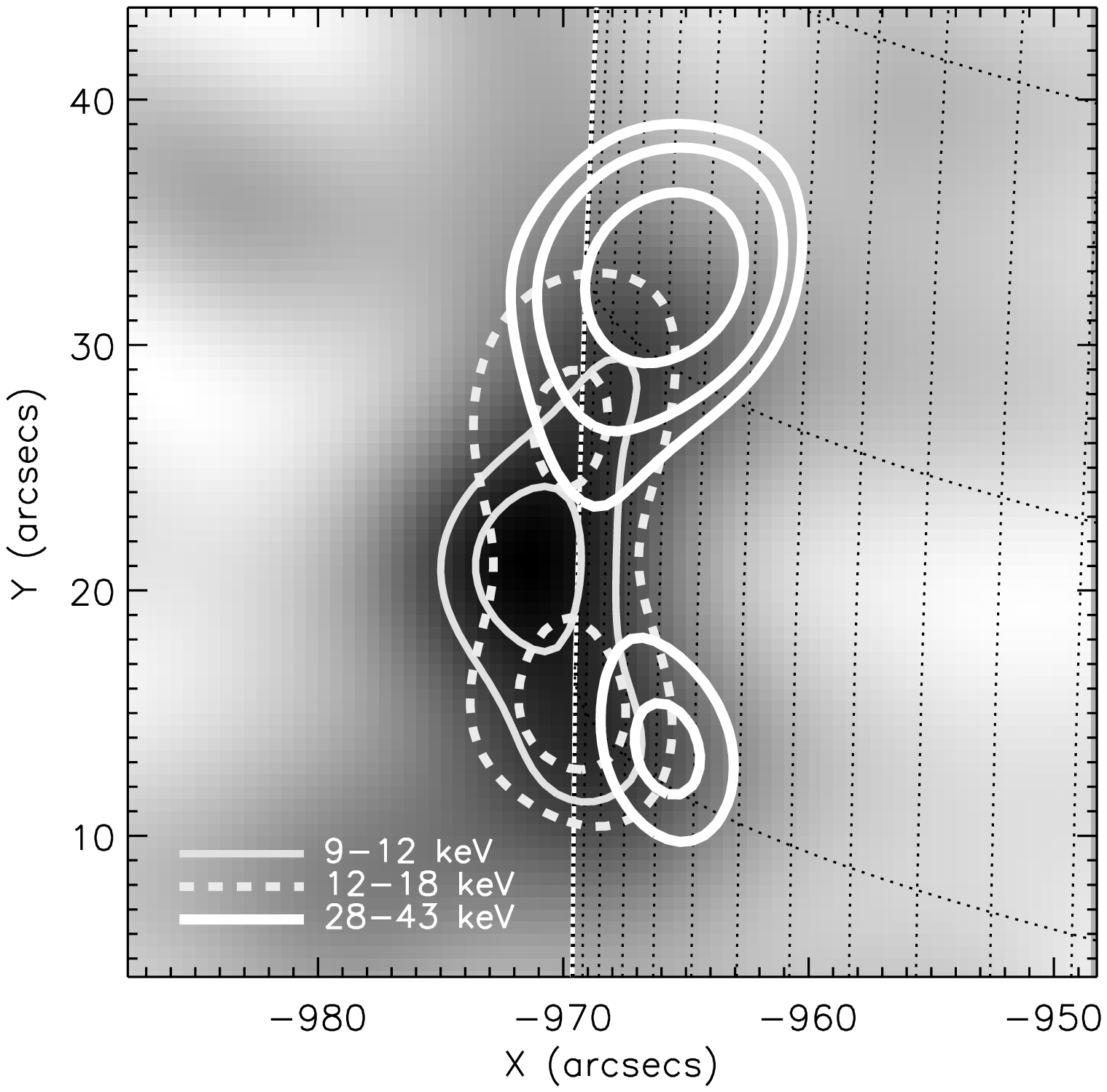} \end{center}
 \caption[]{CLEAN images at 04:58:22-04:58:26 UT during the impulsive
 phase of SOL2003-11-13T05:01 (M1.6).
\index{flare (individual)!SOL2003-11-13T05:01 (M1.6)!illustration}
\index{flare (individual)!SOL2003-11-13T05:01 (M1.6)!source locations vs.\ photon energy}
 The background shows the image at 9-12 keV. The contour levels are
 75\% and 90\% of the peak flux at 9-12 keV ({\it looptop}), 70\%
 and 90\% at 12--18 keV ({\it legs}), and 50\%, 60\%, \& 80\% at
 28--43 keV ({\it footpoints})
 \citep[from][]{2006ApJ...649.1124L}.
 \index{soft X-rays!height dependence}}
 \label{fig:liuw_fig0} 
 \end{figure}

As discussed above, footpoint sources are produced by bremsstrahlung
emission in the thick-target chromosphere\index{footpoints!height structure}.
The compactness of such
sources results from the rapid increase of the density from the
tenuous corona to the much denser chromosphere.  This also gives
rise to the compact height distribution of emission centroids at
different energies as shown in Figure
\ref{fig:aschwanden_footpoint_heights}.  However, if the density
distribution has a somewhat gradual variation, one would expect a
more diffuse height distribution.
Specifically, at some intermediate energies, we expect that HXR
emission would appear in the legs of the loop\index{footpoints!and loop legs}, rather than the
commonly observed looptop sources at low energies and footpoint
sources at high energies.  This has been observed by {\it RHESSI}
in SOL2003-11-13T05:01 (M1.6) \citep{2006ApJ...649.1124L}\index{flare (individual)!SOL2003-11-13T05:01 (M1.6)!X-ray source
motion} (Figure \ref{fig:liuw_fig0}) and in SOL2002-11-28T04:37 (C1.0)\index{flare (individual)!SOL2002-11-28T04:37 (C1.0)!X-ray source motion} 
\citep[][]{2006ApJ...645L.157S}.  %

 \index{footpoints!and loop legs!illustration}

\begin{figure} \begin{center}
 \includegraphics[width=0.32\textwidth]{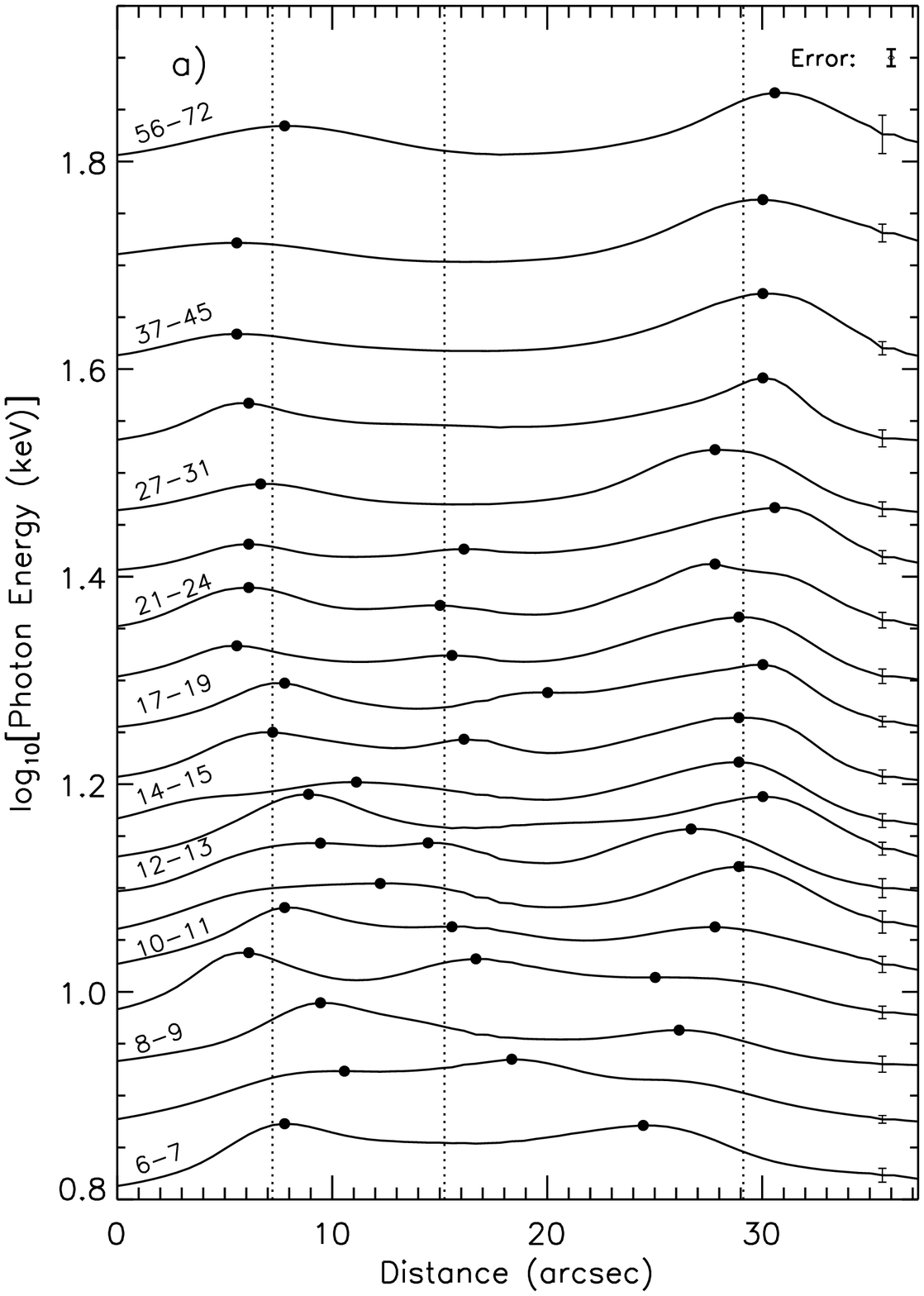}  
\includegraphics[width=0.32\textwidth]{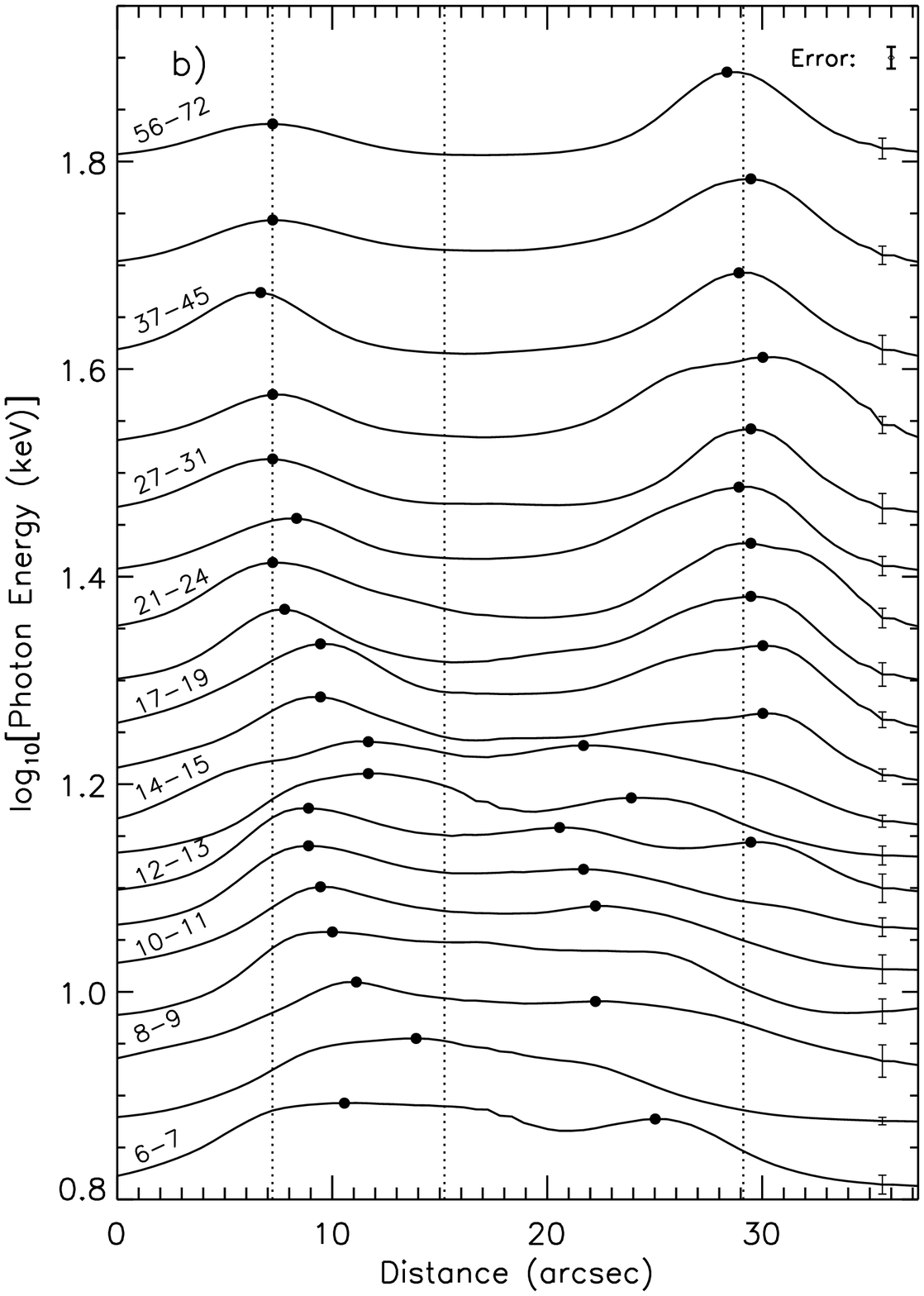}
 \includegraphics[width=0.32\textwidth]{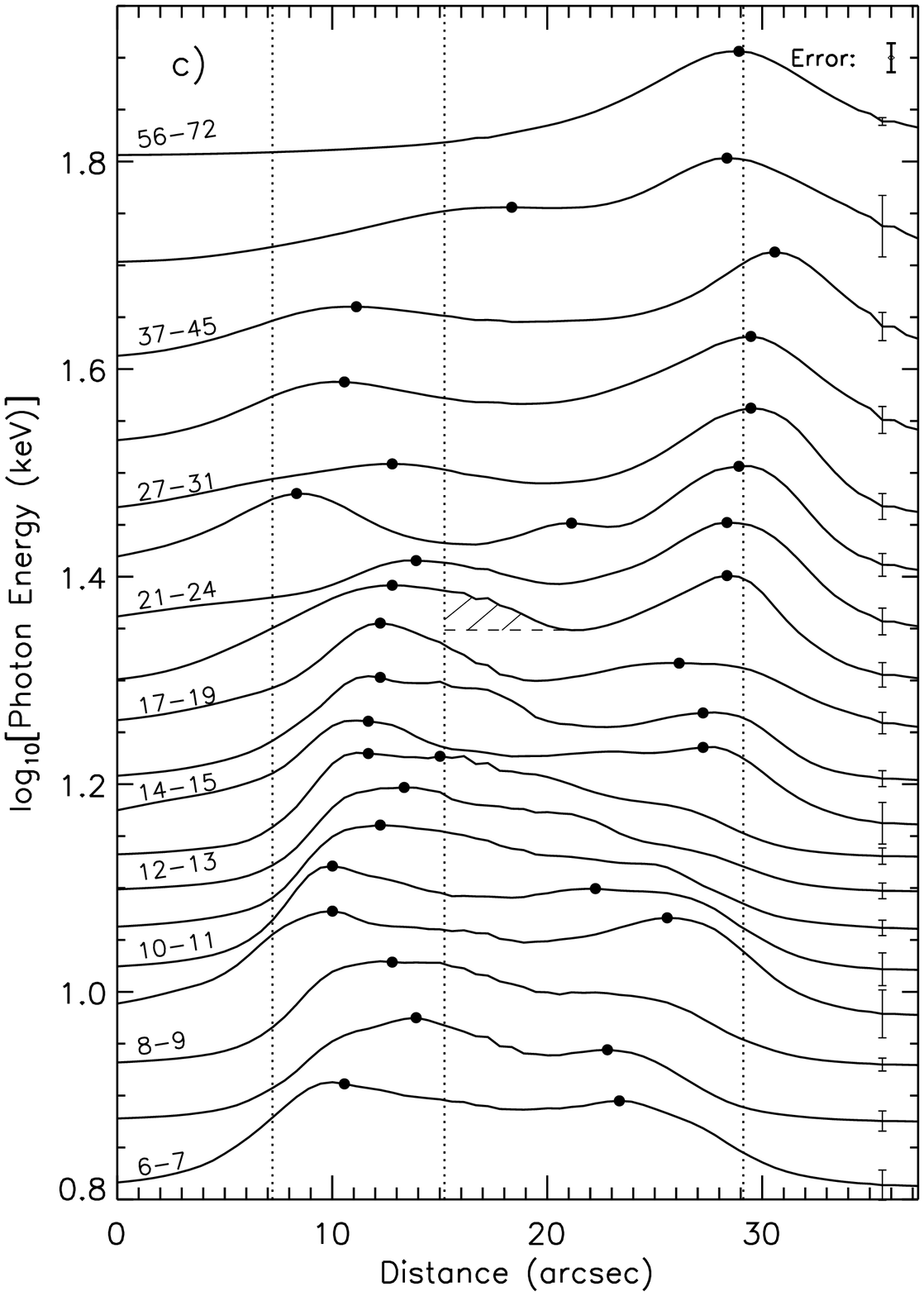} \end{center}
 \caption[]{
({\it a}) Brightness profiles\index{soft X-rays!height dependence} in
different energy bands measured along a semi-circular path fit to
the flaring loop for the time interval 04:58:00--04:58:24~UT of SOL2003-11-13T05:01.
\index{flare (individual)!SOL2003-11-13T05:01 (M1.6)!X-ray
brightness profiles}
The vertical axis indicates the average photon energy (logarithmic
scale) of the energy band for the profile.  Representative energy
bands (in units of keV) are labeled above the corresponding profiles.
The filled circles mark the local maxima, and the vertical dotted
lines are the average positions of the centroids of the looptop and
footpoint sources.  ({\it b, c}) Same as ({\it a}), but for
04:58:24--04:58:48 and 04:58:48--04:59:12 UT, respectively.  The
error bars show the uncertainty of the corresponding profile
(from \citealt{2006ApJ...649.1124L}).
 } \label{fig:liuw_fig1} \end{figure}

To reveal more details of the energy-dependent structure\index{loops!hard X-rays!time-dependent structure}\index{hard X-rays!height dispersion} of
SOL2003-11-13T05:01 (M1.6), Figures \ref{fig:liuw_fig1}{\it
a-c} show the X-ray emission profile along
the flare loop at different energies for three time intervals in
sequence.  The high energy emission is dominated by the footpoints,
but there is a decrease of the separation of the footpoints with
decreasing energy and with time. At later times the profile becomes
a single source, peaking at the looptop. The general trend suggests
an increase of the plasma density in the loop with time
\citep{2006ApJ...649.1124L}, which can be produced by chromospheric
evaporation\index{chromospheric evaporation} and can give rise to
progressively shorter stopping distances for electrons at a given
energy. 
Such a density increase also smooths out to some extent
the sharp density jump at the transition region.\index{transition region} 
This results in the non-thermal bremsstrahlung HXRs at intermediate energies appearing
in the legs of the loop\index{footpoints!height structure}\index{loops!legs}, at higher altitudes than the footpoints,
as shown in Figure \ref{fig:liuw_fig0}.  

\begin{figure} \begin{center}
 \includegraphics[width=0.8\textwidth]{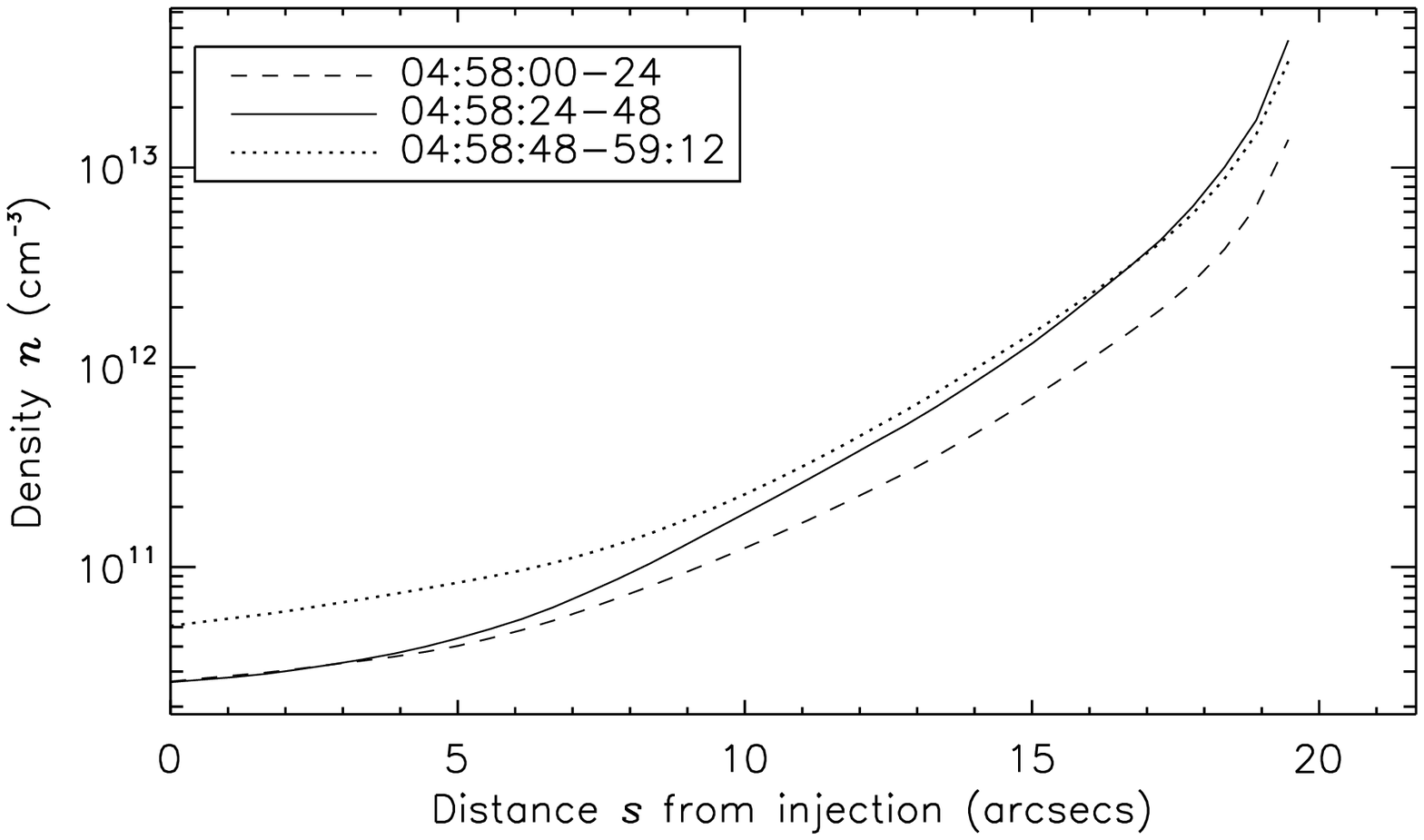} \end{center}
 \caption[]{Averaged density profiles
\index{flare (individual)!SOL2003-11-13T05:01 (M1.6)!density profiles} 
\index{flare (individual)!SOL2003-11-13T05:01 (M1.6)!illustration} 
along one leg of the loop inferred from the
 HXR brightness profiles during the three time intervals in
 Figure~\ref{fig:liuw_fig1}.  The distance is measured along the leg
 extending from the centroid of the thermal looptop source at about
 15~arcsec in Figure~\ref{fig:liuw_fig1} to the end of the fitted
 semi-circle at about 37~arcsec (from \citealt{2006ApJ...649.1124L}).
 } \label{fig:liuw_fig2} 
 \end{figure}

From the emission profiles in the non-thermal regimes of the photon
spectra, \citet{2006ApJ...649.1124L} derived the density distribution
along the loop, using the empirical formula for non-thermal
bremsstrahlung emission profiles given by \citet[][Equation~11]{1983ApJ...269..715L}.  
Leach and Petrosian found that this
formula closely approximates their numerical results for a steady-state,
power-law injected electron distribution with a uniform pitch-angle
distribution, no return-current losses, and a loop with no magnetic
field convergence.  Since this formula is a function of the column
density, one does not need to assume any model form of the density
distribution \citep[cf.][]{2002SoPh..210..383A}.  Figure
\ref{fig:liuw_fig2} shows the density profiles derived from the
emission profiles in the three time intervals shown in Figure
\ref{fig:liuw_fig1}.  Between the first and second intervals, the
density increases dramatically in the lower part of the loop, while
the density near the looptop remains essentially unchanged. The
density enhancement then shifts to the looptop from the second to
the third interval. This indicates a mass flow from the chromosphere
to the looptop, most likely caused by chromospheric evaporation.
For papers studying chromospheric evaporation\index{chromospheric evaporation} using coordinated
{\it RHESSI} HXR and EUV Doppler-shift observations, see
\citet{2006ApJ...638L.117M,2006ApJ...642L.169M} and
\citet{2007ApJ...659L..73B}.  
 \index{hard X-rays!height dependence!illustration}



\begin{figure} \centering
\scalebox{0.5}{\includegraphics{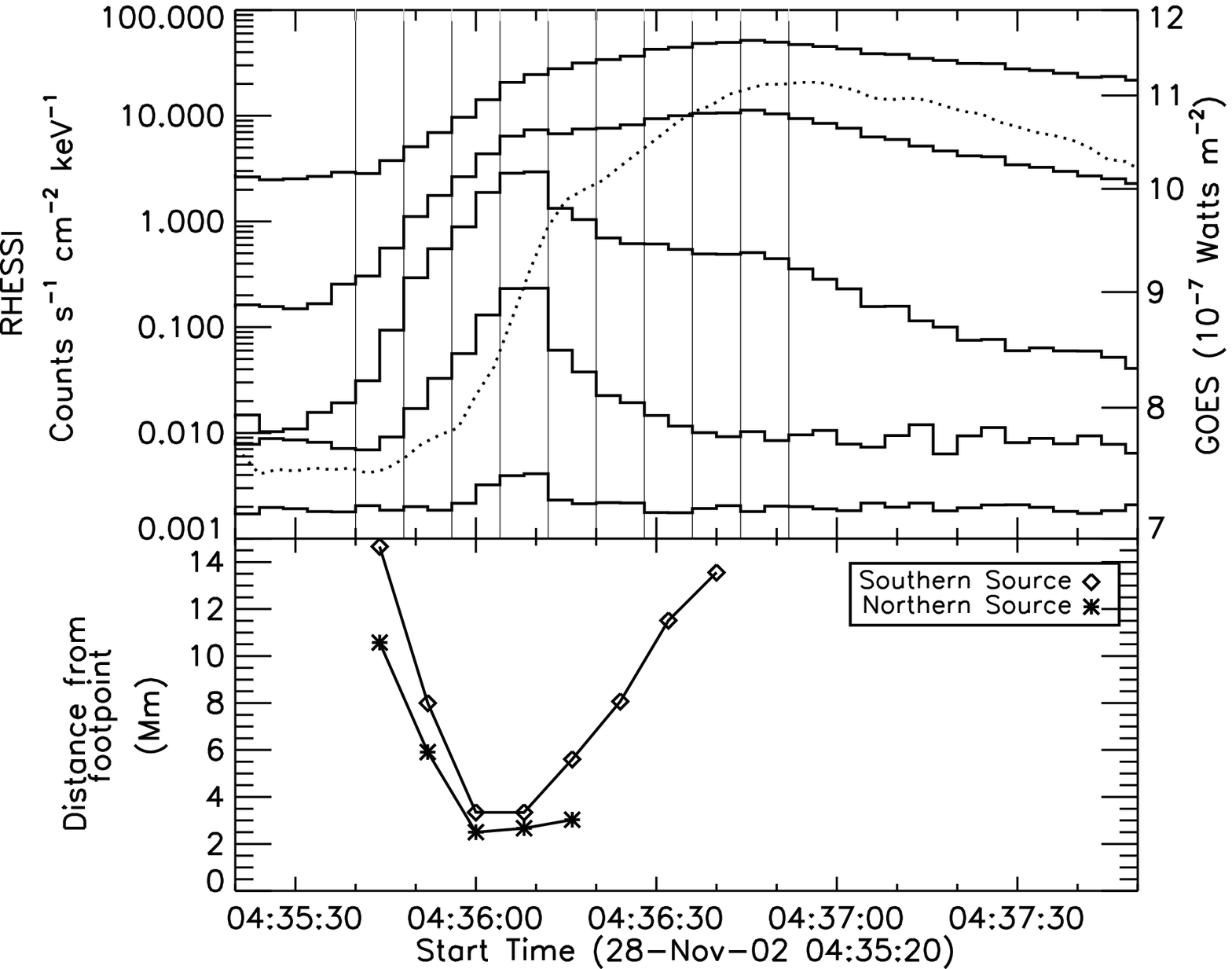}} \vspace{30pt}
\caption{{\it RHESSI} ({\em solid lines}) and {\it GOES} 1-8~\AA~({\em
dotted line}) light curves are shown in the {\em top panel}.  The
{\it RHESSI} energy bands (from top to bottom) are 3--6, 6--12,
12--25, and 50--100~keV, with scaling factors of 5, 1, 4, 3, and
0.5, respectively.  The {\it RHESSI} and {\it GOES} integration
times are 4 and 3~s, respectively.  The {\em bottom panel}
\index{flare (individual)!SOL2002-11-28T04:37 (C1.0)!X-ray source motion} 
shows the distance
between the 3-6~keV moving source centroids\index{soft X-rays!height
dependence} and their corresponding footpoint centroids located in
the 25-50~keV image of the flare at the time of peak emission.
The distances are plane-of-sky values with no correction for motions
away from or toward the observer \citep[from][]{2006ApJ...645L.157S}.}
\label{fig:holman_Nov28motion} \end{figure}

The flare SOL2002-11-28T04:37 (C1.0)\index{flare (individual)!SOL2002-11-28T04:37 (C1.0)!X-ray source motion} 
was an early impulsive flare\index{flare types!early impulsive}, 
meaning that there was minimal pre-heating of plasma
to X-ray-emitting temperatures prior to the appearance of impulsive
hard X-ray emission (see Section~\ref{sec:sainthilaire_Ec}).
{\it RHESSI} observations of this flare showed coronal X-ray sources
that first moved downward and then upward along the legs of the
flare loop \citep{2006ApJ...645L.157S}.  The bottom panel of
Figure~\ref{fig:holman_Nov28motion} shows the motion of the sources
\index{satellites!GOES@\textit{GOES}}
observed in the 3-6~keV band\index{soft X-rays!parallel motions}.  
{\it RHESSI} and {\it GOES}
\index{GOES@\textit{GOES}}
light curves are shown in the top panel.  
The sources originated at the top of
the flare loop and then moved downward along both legs of the loop
until the time of peak emission at energies above 12~keV.  Afterward
the source in the northern leg of the loop was no longer observable,
but the source in the southern leg moved back to the top of the
loop.  Its centroid location at the looptop was slightly but
significantly lower than the centroid position at the beginning of
the flare.  
Higher-energy sources showed a similar evolution, but
they had lower centroid positions than their lower energy counterparts,
again in agreement with the predictions of the thick-target model.\index{thick-target model!height dependence}

\index{flare (individual)!SOL2002-11-28T04:37 (C1.0)!illustration}
\index{soft X-rays!parallel motions!illustration}

The early downward source motion along the legs of the loop is a
previously unobserved phenomenon\index{soft X-rays!downward motions}\index{downflows!soft X-rays}.  
At this time we do not know if
the occurrence is rare, or if it is simply rarely observed because
of masking by the radiation from the thermal plasma.
\citet{2006ApJ...645L.157S} argue that the motion results from the
hardening of the
X-ray spectrum, and possibly an increase in the low-energy cutoff,
as the flare hard X-ray emission rises to its peak intensity.  A
flatter spectrum results in a higher mean energy of the electrons
contributing to the radiation at a given X-ray energy.  In a loop
with a plasma density that increases significantly from the top to
the footpoints, these higher energy electrons will propagate to a
lower altitude in the loop as the spectrum hardens.  The softening
of the spectrum after peak emission would also contribute to the
upward motion of the source after the peak.  However, at that time
chromospheric evaporation\index{chromospheric evaporation} would
likely be increasing the density in the loop, as discussed above
for SOL2003-11-13T05:01 (M1.6)\index{flare (individual)!SOL2003-11-13T05:01 (M1.6)!increasing density} , and thermal emission would be more
important.  All of these can contribute to an increase in the height
of the centroid of the X-ray source.  The downward motion may only
occur in initially cool flare loops, i.e., early impulsive
flares\index{flare types!early impulsive}, because these loops are most
likely to contain the density gradients that are required.

In an attempt to differentiate between thermal and non-thermal X-ray
emission, \citet{2008ApJ...673..576X} modeled the size dependence
with photon energy of coronal X-ray sources observed by {\it RHESSI}
in ten M-class limb flares.  They determined the one-sigma Gaussian
width of the sources along the length of the flare loops by obtaining
forward fits to the source visibilities.  The integration times
ranged from one to ten minutes and the source sizes were determined
in up to eight energy bins ranging in energy from as low as 7~keV
to as high as 30~keV.  They found the source sizes\index{hard X-rays!source sizes} to increase
slowly with photon energy, on average as $\epsilon^{1/2}$.  The
results were compared with several models for the variation of the
source size with energy.  The source size was expected to vary as
$\epsilon^{-1/2}$ for a thermal model with a constant loop density
and a temperature that decreased with a Gaussian profile along the
legs of the loop from a maximum temperature at the top of the loop.
For the injection of a power-law electron flux distribution into a
high-density loop so that the loop is a collisional thick target,
the source size was expected to increase as $\epsilon^{2}$.  Neither
of these models are consistent with the observed $\epsilon^{1/2}$
dependence.  A hybrid thermal/non-thermal model and a non-thermal
model with an extended acceleration region\index{acceleration region!extended} 
at the top of the loop
were found to be consistent with the deduced scaling, however.  The
extended acceleration region was deduced to have a half-length in
the range 10$^{\prime\prime}$ -- 18$^{\prime\prime}$ and density
in the range $(1 - 5) \times 10^{11}$~cm$^{-3}$.  We note that the
extended acceleration region model implies a column density in the
range $0.73 - 6.5 \times 10^{20}$ cm$^{-2}$ along the half length
and, from Equation~\ref{eqn:holman_colevol}, all electrons with
energies less than somewhere in the range of 23~keV -- 68~keV that
traverse this half length will lose all of their energy to collisions.
The acceleration process would therefore need to be efficient enough
to overcome these losses.  On the other hand, the 7 -- 30~keV energy
range is the range in which fits to spatially integrated X-ray
spectra typically show a combination of both thermal and non-thermal
bremsstrahlung emission.

Studies of flare hard X-ray source positions and sizes as a function
of photon energy and time hold great promise for determining the
height structure of flare plasma and its evolution, as well as
providing information about the magnetic structure of the flare
loop.  Such studies are currently in their early stages, in that
they usually assume an over simplified power-law or exponential
height distribution for the plasma and do not take into account the
variation of the plasma ionization state with height.  They also
assume the simple, one-dimensional collisional thick-target model,
without consideration of the pitch-angle distribution of the beam
electrons or the possibility of additional energy losses to the
beam (such as return-current\index{return current!energy losses} losses).  Given
the potential for obtaining a better understanding of flare evolution,
we look forward to the application of more sophisticated models to
the flare hard X-ray data.  

\section{Hard X-ray timing} 
\label{sec:aschwanden_timing}


The analysis of energy-dependent time delays\index{hard X-rays!time
delays} allows us to test theoretical models of physical time scales
and their scaling laws with energy. In the wavelength domain of
hard X-rays, there are at least three physical processes known in
the observation of solar flares that lead to measurable time delays
as a function of energy \citep[for a review, see][]{2004psci.book.....A}:
(1) time-of-flight dispersion of free-streaming electrons, (2)
magnetic trapping\index{trapping!magnetic} with the collisional precipitation of electrons,
and (3) cooling of the thermal plasma.\index{time-of-flight analysis}

\subsection{Time-of-Flight Delays} \label{sec:aschwanden_timing_tof}
\index{electrons!time-of-flight (TOF) analysis}
\index{hard X-rays!energy dispersion}

The first type, the {\sl time-of-flight (TOF)} delay, has a scaling
of $\Delta t(\epsilon) \propto \epsilon^{-1/2}$ and is caused by
velocity differences of electrons that propagate from the coronal
acceleration site to the chromospheric energy-loss region.  The
time differences are of order $\Delta t \approx 10-100$ ms for
non-thermal electrons at energies $E \approx 20-100$ keV
\citep[e.g.,][]{1995ApJ...447..923A,1996ApJ...470.1198A}. The
measurement of such tiny time delays requires high photon statistics
and high time resolution.  
Such data were provided by \textit{CGRO}/BATSE,
\index{satellites!CGRO@\textit{CGRO}}
\index{Compton@\textit{Compton Gamma Ray Observatory
(CGRO)}} which had 8 detectors, each with an effective collecting
area of $\sim$2000~cm$^2$ and oriented at different angles to the
Sun so that detector saturation at high count rates was not a
problem.  (For comparison, the total effective collecting
area\index{RHESSI@\textit{RHESSI}!effective collecting area} of {\it RHESSI}'s
detectors is less than 100~cm$^2$.)

These studies of TOF delays have provided important evidence that
electrons are accelerated\index{acceleration region} in the corona,
above the top of the hot flare loops observed in soft X-rays.  The
fine structure in the light curves of most, but not all, of the
studied flare bursts showed energy-dependent time delays consistent
with the free streaming of electrons to the footpoints of the flare
loops from an origin somewhat more distant than the half-length of
the loops
\citep{1995ApJ...447..923A,1995ApJ...455..699A,1996ApJ...470.1198A}.

\subsection{Trapping Delays} \label{sec:aschwanden_timing_trap}

The second type, the {\sl trapping delay},
\index{magnetic trapping}
\index{trapping!magnetic} 
\index{trapping!time scale} 
\index{magnetic structures!trapping}
is caused by magnetic mirroring of coronal electrons which precipitate toward
the chromosphere after a collisional time scale $\Delta t(\epsilon)
\propto \epsilon^{3/2}$.  This is observable for time differences
of $\Delta t \approx 1-10$ s for non-thermal electrons at $E \approx
20-100$ keV \citep[e.g.,][]{1982A&A...108..306V,1997ApJ...487..936A}.
For trapping delays the higher energy X-rays lag the lower energy
X-rays, as opposed to time-of-flight delays where the higher energy
X-rays precede the lower energy X-rays.

\citet{1997ApJ...487..936A} filtered variations on time scales
\index{satellites!CGRO@\textit{CGRO}}
$\sim$1~s or less out of {\it CGRO}\index{Compton@\textit{Compton Gamma Ray Observatory
(CGRO)}} BATSE flare HXR light curves. 
They found time delays in the
remaining gradually varying component to be consistent with magnetic
trapping and collisional precipitation of the particles.  Trap
plasma densities $\sim 10^{11}$ cm$^{-3}$ were deduced.  No evidence
was found for a discontinuity in the delay time as a function of
energy and, therefore, for second-step (two-stage)
acceleration\index{acceleration!two-stage} of electrons at energies
$\leq 200$~keV.

\subsection{Thermal Delays} \label{sec:aschwanden_timing_thdelays}
\index{soft X-rays!cooling!and time delays}

The third type, the {\sl thermal delay}, can be caused by the
temperature dependence of cooling processes, such as by thermal
conduction\index{thermal conduction}\index{cooling!conductive}, $\tau_c(T) \propto T^{-5/2}$
\citep[e.g.,][]{1978ApJ...220.1137A,1994SoPh..153..307C}, or by
radiative cooling\index{radiative cooling}\index{cooling!radiative}, $\tau_r(T) \propto
T^{5/3}$ \citep[e.g.,][]{1990ApJ...357..243F,1995ApJ...439.1034C}.
The observed physical parameters suggest that thermal conduction
dominates in flare loops at high temperatures as observed in soft
X-ray wavelengths, while radiative cooling dominates in the later
phase in postflare loops as observed in EUV wavelengths
\citep{1978ApJ...220.1137A,1994SoPh..153..307C,2001SoPh..204...91A}. When
the temperature drops in the decay phase of flares, the heating
rate can justifiably be neglected and the conductive or radiative
cooling rate dominate the temperature evolution. Before {\it RHESSI},
the cooling curve $T(t)$ in flare plasmas had been studied in only
a few flares
\citep[e.g.,][]{1993ApJ...416L..91M,1994SoPh..153..307C,2001SoPh..204...91A}.

The high spectral resolution\index{RHESSI@\textit{RHESSI}!spectral resolution} of
{\it RHESSI} data is particularly suitable for any type of thermal
modeling, because we can probe the thermal plasma from $\sim$3~keV up to $\sim$30~keV with a FWHM resolution of $\sim$1~keV thanks to the cooled germanium detectors
\citep[][]{2002SoPh..210....3L,2002SoPh..210...33S}. This allows
us to measure flare temperatures with more confidence. A statistical
study of flare temperatures measured in the range of $T\approx 7-20$
MK indeed demonstrates some agreement between the values obtained
\index{satellites!GOES@\textit{GOES}}
from spectral fitting of {\it RHESSI} data with those obtained from
{\it GOES}\index{GOES@\textit{GOES}} flux ratios \citep[][]{2005A&A...439..737B}, although
{\it RHESSI} has a bias for the high-temperature tail of the
differential emission measure (DEM)\index{differential emission measure}\index{emission measure!differential} distribution \citep{2008ApJ...672..659A,2007SoPh..241..279V}.
Of course, we expect an agreement between the deduced
emission-measure-weighted temperatures only when both instruments
are sensitive to a temperature range that covers the flare DEM peak.

A close relationship between the non-thermal and thermal time profiles
was found early on, in the sense that the thermal emission often
closely resembles the integral of the non-thermal emission, a
relationship that is now known as the {\sl Neupert effect}\index{Neupert
effect}
\citep[][]{1968ApJ...153L..59N,1991BAAS...23R1064H,1993SoPh..146..177D}.
This relationship is, however, strictly only expected for the
asymptotic limit of very long cooling times, while a physically
more accurate model would quantify this effect by a convolution of
the non-thermal heating with a finite cooling time.  The deconvolution
of the e-folding cooling time in such a model has never been attempted
statistically and as a function of energy or temperature. Theoretical
discussions of the Neupert effect\index{Neupert effect!theory}, including multiple energy release
events, chromospheric evaporation, and cooling, can be found in
\citet{2004ApJ...611L..49W}, \citet{2010ApJ...709...58L}, and
\citet{2010ApJ...712..429R}.

The cooling time at a given energy can be estimated from the decay
time of a flare time profile. For instance, the decay times measured
with {\it GOES}\index{GOES@\textit{GOES}} in soft X-rays were found to have a median of $\tau_{decay}
\approx 6$ min \citep{2002A&A...382.1070V,2002A&A...392..699V}.
The observed cooling times have typically been found to be much
longer than predicted from classical conduction\index{thermal
conduction}, but shorter than the radiative cooling time\index{cooling!radiative}\index{cooling!conductive}
\citep[e.g.,][]{1993ApJ...416L..91M,2006ApJ...638.1140J,2007ApJ...659..750R}.
This discrepancy could result from either continuous heating or
suppression of conduction during the decay phase, or a combination
of both.

The Neupert effect\index{Neupert effect} was tested by correlating
the soft X-ray peak flux with the (time-integrated) hard X-ray flux.
A high correlation and time coincidence between the soft X-ray peak
and hard X-ray end time was generally found, but a significant
fraction of events also had a different timing \citep{2002SoPh..208..297V}.
A delay of 12~s was found in the soft X-ray flux time derivative
with respect to the hard X-ray flux in SOL2003-11-13T05:01 (M1.6)\index{flare (individual)!SOL2003-11-13T05:01 (M1.6)!Neupert Effect}
\citep[][also see Section~\ref{sec:liuw_height}]{2006ApJ...649.1124L}.
Time delays such as this could be related to the hydrodynamic flow
time during chromospheric evaporation\index{chromospheric evaporation}.
Tests of the ``theoretical Neupert effect,''\index{Neupert effect!theoretical} i.e., comparisons of
the beam power supply of hard X-ray-emitting electrons and the
thermal energy of evaporated plasma observed in soft X-rays, found
it to strongly depend on the low-energy cutoff to the non-thermal
\index{low-energy cutoff}
electron distribution\index{electrons!distribution function!low-energy
cutoff} \citep{2005ApJ...621..482V}.  This provides another approach
to deducing the energy at which the low-energy cutoff in the electron
distribution occurs in individual flares.  The Neupert effect\index{Neupert
effect} has also been studied in several flares by
\citet{2008SoPh..248...99N,2009ScChG..52.1686N}, who finds a high
correlation between the hard X-ray flux and the time derivative of
the thermal energy deduced from X-ray spectral fits
\citep{2008SoPh..248...99N} and an anti-correlation between the
hard X-ray spectral index and the time rate of change of the UV
\index{satellites!TRACE@\textit{TRACE}}
\index{hard X-rays!correlation with EUV}
flare area observed by {\em TRACE}\index{TRACE@\textit{TRACE}}
\citep{2009ScChG..52.1686N}.

\begin{figure} \includegraphics[width=\textwidth]{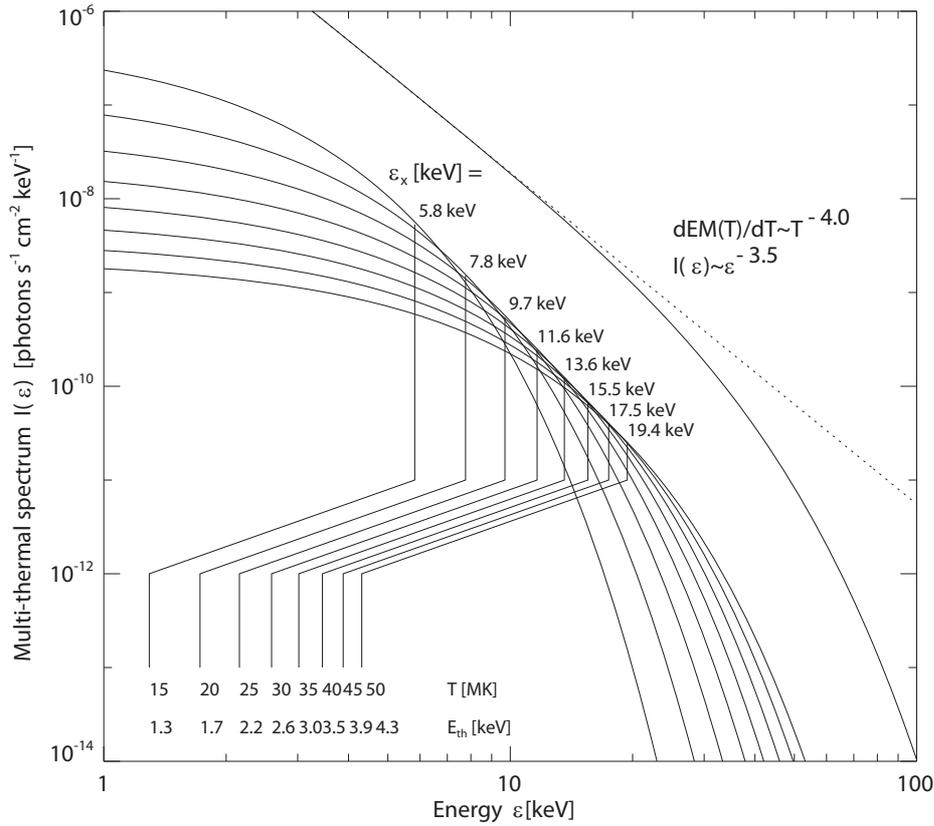}
\caption{Example of a multi-thermal spectrum with contributions
from plasmas with temperatures of $T=15, 20, ..., 50$ MK and a DEM
distribution of $dEM(T)/dT \propto T^{-4}$. The individual thermal
spectra and their sum are shown with thin linestyle, where the sum
represents the observed spectrum. Note that the photons in the
energy range $\epsilon=5.8-19.4$ keV are dominated by temperatures
of T=15--50 MK, which have a corresponding thermal energy that is
about a factor of $(4+1/2)=4.5$ lower than the corresponding photon
energy ($\epsilon_{th}=1.3-4.3$ keV). The summed photon spectrum
without the high-temperature cutoff approaches the power-law function
$I(\epsilon) \propto \epsilon^{-3.5}$ ({\em dotted line})
\citep[from][]{2007ApJ...661.1242A}.} \label{fig:aschwanden_multi_f1}
\end{figure}

\begin{figure} \includegraphics[width=\textwidth]{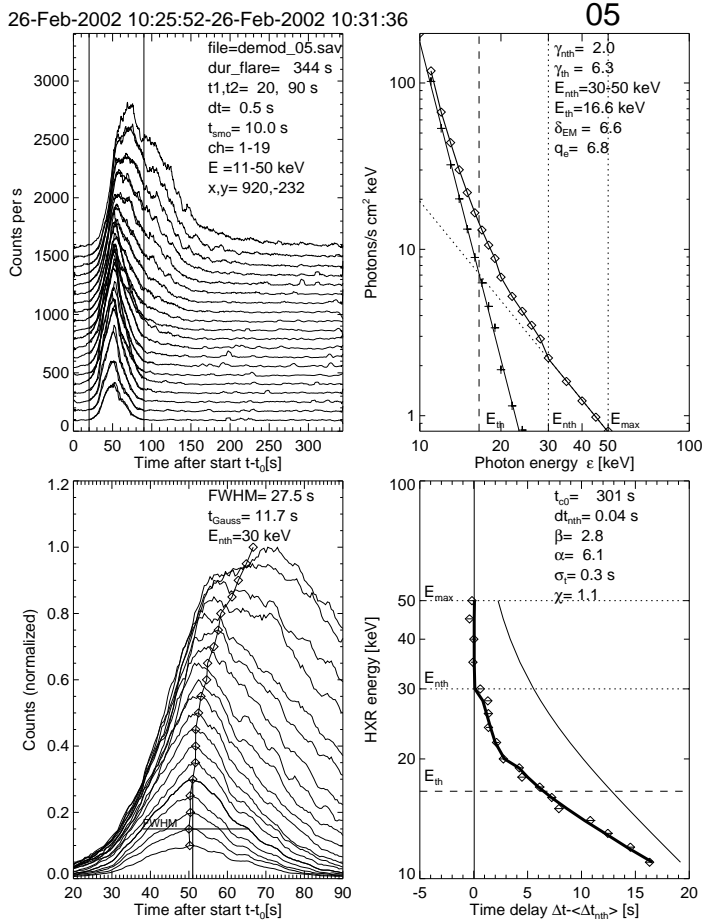}
\caption{X-ray light curves are shown for SOL2002-02-26T10:27 (C9.6)\index{flare (individual)!SOL2002-02-26T10:27 (C9.6)!multi-thermal
time delays}, for energies of 10~keV to 30~keV in intervals of 1~keV, 
observed with {\it RHESSI} (left panels). The spectrum is
decomposed into thermal and non-thermal components (top right panel)
and the delay of the peaks at different energies is fitted with a
thermal conduction cooling time model that has a scaling of
$\tau_{cond}(T) \approx T^{-\beta}$ (right bottom panel). The best
fit shows a power index of $\beta=2.8$, which is close to the
theoretically expected value of $\beta=5/2$
(Equation~\ref{eq:aschwanden_conduction}). The full delay of the thermal
component is indicated with a thin curve (bottom left panel), while
the weighted (thermal+non-thermal) fit is indicated with a thick
curve \citep[from][]{2007ApJ...661.1242A}.} \label{fig:aschwanden_Feb26_f2}
\end{figure}
\index{soft X-rays!multi-thermal modeling!illustration}

\subsection{Multi-Thermal Delay Modeling with {\it RHESSI}}
\label{sec:aschwanden_timing_mthdelays}
\index{soft X-rays!multi-thermal modeling}

Since major solar flares generally produce a large number of
individual postflare loops, giving the familiar appearance of loop
arcades lined up along the flare ribbons, it is unavoidable that
each loop is heated up and cools down at different times, so that
a spatially integrated spectrum always contains a multi-thermal
differential emission measure distribution
\citep[cf.][]{2006ApJ...637..522W}\index{magnetic structures!arcades}\index{ribbons}\index{flare types!two-ribbon}.
The resulting multi-thermal
bremsstrahlung spectrum (for photon energies $\epsilon$) observed
in soft X-rays (neglecting the Gaunt factor of order unity),
\begin{equation}
    I(\epsilon ) = I_0 \int {\exp(-\epsilon/k_B T) \over T^{1/2}}
	{dEM(T) \over dT} dT \ ,
\end{equation} is then a function of a multi-thermal {\sl differential
emission measure} distribution $dEM(T) = n^2(T) dV$\index{emission measure!differential!model}. 
An example of
a multi-thermal spectrum from a differential emission measure
proportional to $T^{-4}$ up to a maximum temperature of 50~MK is
shown in Figure~\ref{fig:aschwanden_multi_f1}.

As discussed above, the initial cooling of the hot flare plasma
(say at $T \gapprox 10$ MK) is generally dominated by conductive
cooling (rather than by radiative cooling, which can dominate later
after the plasma cools to EUV-emitting temperatures of $T\lapprox
2$ MK).  
\index{cooling!conductive}
The thermal conduction\index{thermal conduction} time has
the following temperature dependence: 
\begin{equation}
    \tau_{cond}(T) = {\epsilon_{th} \over dE/dt_{cond}}
	= {3 n_e k_B T \over {d\over ds} \kappa T^{5/2} {dT\over
	ds}} \approx {21 \over 2} {n_e L^2 k_B \over \kappa} T^{-5/2}
	= \tau_{c0} \left({T \over T_0}\right)^{-5/2} \ ;
\label{eq:aschwanden_conduction} 
\end{equation} 
see \cite{2007ApJ...661.1242A} for parameter definitions.
Since the thermal
bremsstrahlung at decreasing photon energies is dominated by radiation
from lower temperature flare plasma, the conductive cooling\index{cooling!conductive} time
is expected to become longer at lower temperatures ($\tau_{cond}
\propto T^{-5/2}$).  Thus, the soft X-ray peak is always delayed
with respect to the harder X-ray peaks, reflecting the conductive
cooling of the flare loops.

\citet{2007ApJ...661.1242A} has measured and modeled this conductive
cooling delay\index{cooling!conductive!delay} $\tau_{cond}(\epsilon)$ for a comprehensive set of
short-duration ($\leq 10$~min) flares observed by {\it RHESSI}. One
example is shown in Figure~\ref{fig:aschwanden_Feb26_f2}. He finds
that the cooling delay $\Delta t$ expressed as a function of the
photon energy $\epsilon$ and photon spectral index $\gamma$ can be
approximated by \begin{equation}
	\Delta t(\epsilon, \gamma) \approx \tau_g {7\over 4}
	\left[\log{\left(1 + {\tau_{c0} \over \tau_g} \left({\epsilon
	\over (\gamma-1) \epsilon_0}\right)^{-\beta}\right)}\right]^{3/4}
	\ ,
\end{equation} (where $\tau_g$ is the Gaussian width of the time
profile peak) and yields a new diagnostic of the process of conductive
cooling in multi-thermal flare plasmas\index{soft X-rays!multi-thermal}. 
In a statistical study of
65 flares \citep[][]{2007ApJ...661.1242A}, 44 (68\%) were well fit
by the multi-thermal model with a best fit value for the exponent
of $\beta=2.7\pm 1.2$, which is consistent with the theoretically
expected value of $\beta=2.5$ according to
Equation~\ref{eq:aschwanden_conduction}.  The conductive cooling time
at $T_0 = 11.6$~MK ($\epsilon_0 = 1$~keV) was found to range from
2 to 750~s, with a mean value of $\tau_{c0} = 40$~s.  

We note that these timing data, as well as thick-target fits to the
non-thermal part of spectra that reveal the evolution of the energy
content in accelerated electrons, provide additional constraints
on models such as the multithread flare model\index{flare models!multithread} of
\citet{2006ApJ...637..522W}.


\section{Hard X-ray spectral evolution in flares} 
\label{sec:grigis_spectral_evolution}


\subsection{Observations of spectral evolution} \label{Gr_shs_observations}

\index{hard X-rays!spectral evolution} The non-thermal hard X-ray emission
from solar flares, best observed in the 20 to 100 keV range, is
highly variable. Often several emission spikes with durations ranging
from seconds to minutes are observed. In larger events, sometimes
a more slowly variable, long duration emission can be observed in
the later phase of the flare. Hence, most flares start out with an
\emph{impulsive} phase\index{impulsive phase}, while some
events, mostly large ones, show the presence of a late \emph{gradual}
phase\index{gradual phase} in the hard X-ray time profile.

While these two different behaviors can already be spotted by looking
at lightcurves, they also are distinct in their spectral evolution.
The impulsive spikes tend to be harder at the peak time, and softer
both in the rise and decay phase. The spectrum starts soft, gets
harder as the flux rises and softens again after the maximum of the
emission. This pattern of the spectral evolution is thus called
\emph{soft-hard-soft} (SHS)\index{hard X-rays!soft-hard-soft}. 
On the other hand, in the gradual phase,
the flux often slowly decreases, while the spectrum stays hard or
gets even harder. This different kind of spectral evolution is
called \emph{soft-hard-harder} (SHH)\index{hard X-rays!soft-hard-harder}.

Historically, both the SHS\index{hard X-rays!soft-hard-soft}
\citep{1969ApJ...155L.117P,1970ApJ...162.1003K} and the SHH
behavior\index{hard X-rays!soft-hard-harder}
\citep{1971ApJ...165..655F} were observed in the early era of hard
X-ray observations of the Sun.  Subsequent investigation confirmed
both the SHS \citep{1977ApJ...211..270B,1985MNRAS.212..245B,
1987ApJ...312..462L,1998Ap&SS.260..515G,2002SoPh..210..307F,
2002ESASP.506..261H} and the SHH
\citep{1986ApJ...305..920C,1995ApJ...453..973K,2008ApJ...673.1169S,2008ApJ...683.1180G}
behavior.  The SHH behavior has been found to be correlated with
proton events\index{proton events} in interplanetary space
\citep{1995ApJ...453..973K,2008ApJ...673.1169S,2009ApJ...707.1588G}.

Evidence for hard-soft-hard (HSH) spectral evolution\index{hard X-rays!hard-soft-hard} at energies above $\sim$50~keV has been
reported for multiple spikes in SOL2004-11-03T03:35 (M1.6)\index{flare (individual)!SOL2004-11-03T03:35 (M1.6)!hard-soft-hard}
\citep{2009ApJ...694L.162S}.  
\index{hard X-rays!hard-soft-hard}
SHS behavior was observed at lower
energies.  
This HSH behavior might be explained by albedo\index{albedo},
which typically peaks around 30--40~keV \citep[see][]{Chapter7},
but the authors corrected for albedo from isotropically emitted
photons.  A likely explanation is that the spikes overlie a harder,
gradually varying component, possibly emission from trapped\index{magnetic
trapping} electrons (Section~\ref{sec:aschwanden_timing_trap}).

While all these observations established the qualitative properties
of the spectral evolution, a statistical analysis of the quantitative
relation between the flux and spectral index had not been performed
in the pre-{\it RHESSI} era\index{eras!pre-\textit{RHESSI}}.  
Here, we summarize {\it RHESSI} results
investigating quantitatively the spectral evolution of the non\-thermal
component of the hard X-ray emission, as well as the theoretical
implications. More details can be found in
\citet{2004A&A...426.1093G,2005A&A...434.1173G,2006A&A...458..641G}.

%
\begin{figure}[t!]
\centering\includegraphics[width=14cm]{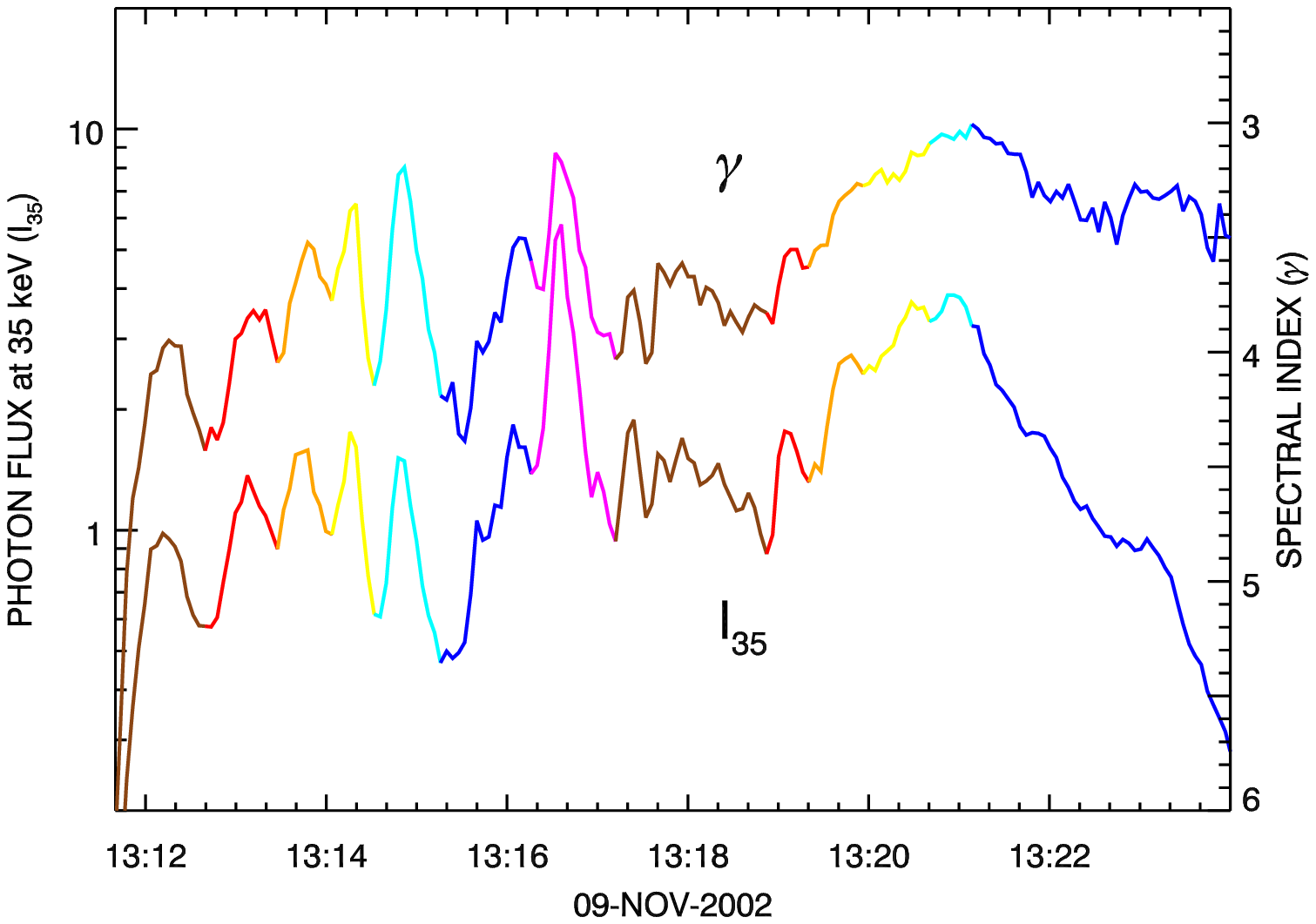} \caption{Time
evolution of the spectral index $\gamma$ ({\it upper curve, linear
  scale on right}) and the flux normalization $I_{35}$ ({\it lower
  curve, logarithmic scale on left}) of the non-thermal component
  in SOL2002-11-09T13:23 (M4.9).
\index{flare (individual)!SOL2002-11-09T13:23 (M4.9)!spectral evolution}
\index{flare (individual)!SOL2002-11-09T13:23 (M4.9)!illustration}
Different emission spikes are shown in
  different colors \citep[after][]{2004A&A...426.1093G}.}
\label{Gr_gflux_timedep} \end{figure} 
%

To quantify the spectral evolution, a simple parameterization for
the shape of the non\-thermal spectrum is needed. Luckily, in solar
flares the spectrum is well described by a power law in energy,
which often steepens above 50 keV\index{hard X-rays!spectral parametrization}.  
Such a softening of the spectrum
can be modeled by a broken power-law model. However, it is difficult
to observe such a downward bending at times of weak flux, because
the high-energy region of the spectrum is lost in the background.
As a compromise, \citet{2004A&A...426.1093G} fitted the data to a
single power-law function at all times. Although the single power
law does not always provide a good fit to the spectra, it provides
a characteristic spectral slope and ensures a uniform treatment of
the spectra at different times.

The two free parameters of the power-law model are the spectral
index\index{hard X-rays!spectral index} $\gamma$ and the power-law
normalization $I_{\epsilon_0}$ at the reference energy $\epsilon_0$.
The reference energy $\epsilon_0$ is arbitrary, but fixed, usually
near the logarithmic mean of the covered energy range.  In the {\it
RHESSI} spectral analysis software, OSPEX, $\epsilon_0 = 50$~keV
by default.  The time dependent spectrum is given by  
\begin{equation}
\label{Gr_spectrum_timedep}
  I(\epsilon,t)=I_{\epsilon_0}(t)\left(\frac{\epsilon}{\epsilon_0}\right)^{-\gamma(t)}\,.
\end{equation} 

A representative sample of 24 solar flares of {\it GOES}
\index{satellites!GOES@\textit{GOES}}
\index{GOES@\textit{GOES}} 
magnitudes between M1 and X1 was selected by \citet{2004A&A...426.1093G}.  
The spectral
model (Equation~\ref{Gr_spectrum_timedep}), with the addition of an
isothermal emission component at low energies, was fitted with a
cadence of one {\it RHESSI} spin period (about 4 s). This delivered
a sequence of measurements of the quantities $I_{\epsilon_0}(t)$
and $\gamma(t)$ for each of the 24 events, covering a total time
of about 62 minutes of non-thermal hard X-ray emission.  For these
events, $\epsilon_0=35$~keV was chosen, a meaningful energy which
lies about in the middle of the range where the non-thermal emission
is best observed in these M-class flares.

An example of the measured time evolution of the spectral index
$\gamma$ and the flux normalization $I_{35}$ for the longer-lasting
event of the set is shown in Figure~\ref{Gr_gflux_timedep}.  
A correlation in time between the two curves can be readily
seen.  Single emission spikes are plotted in different colors, so
that the soft-hard-soft\index{hard X-rays!soft-hard-soft} evolution can be observed during each spike
(with the exception of the late, more gradual phase, where the
emission stays hard as the flux decays).

\begin{figure}[t!]
\centering\includegraphics[width=\hsize]{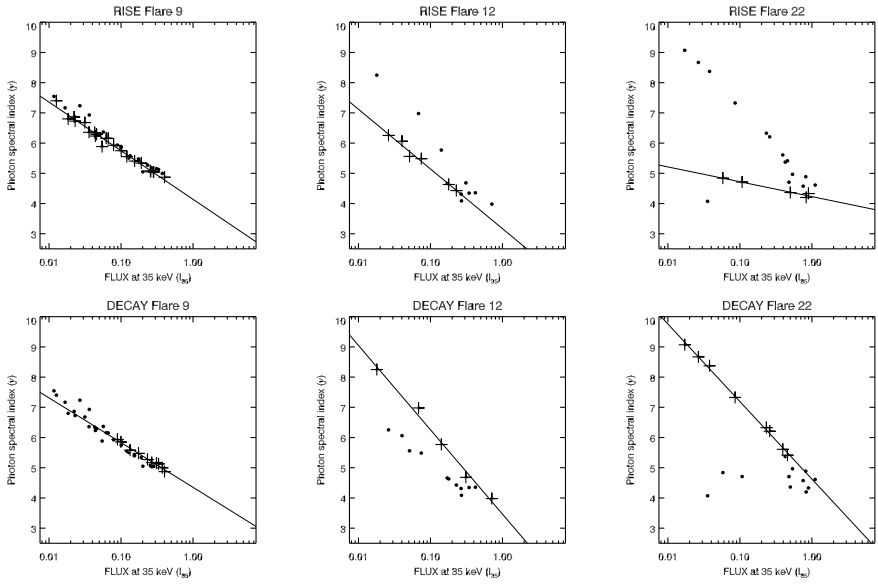} \caption{Spectral
index $\gamma$ vs. flux normalization $I_{35}$ for three events,
showing the linear dependence of single rise and decay phases of
emission spikes on a log-linear scale.  Dots mark results from
individual spikes, while pluses mark the longer rise or decay phase
\citep[from][]{2004A&A...426.1093G}.} \label{Gr_singlephases}
\end{figure}

As there is an anti-correlation in time between $\log I_{35}(t)$
and $\gamma(t)$, a plot of one parameter as a function of the other,
eliminating the time dependence, shows the relationship between
them. Figure \ref{Gr_singlephases} shows plots of $\gamma$ vs.
$I_{35}$ for 3 events where there are only one or two emission
peaks. The points in the longer uninterrupted rise or decay phase
during each event are marked by plus symbols. A linear relationship
between $\log I_{35}$ and $\gamma$ can be seen during each phase,
although it can be different during rise and decay.

%
%

\begin{figure}[t!]
\centering\includegraphics[width=10cm]{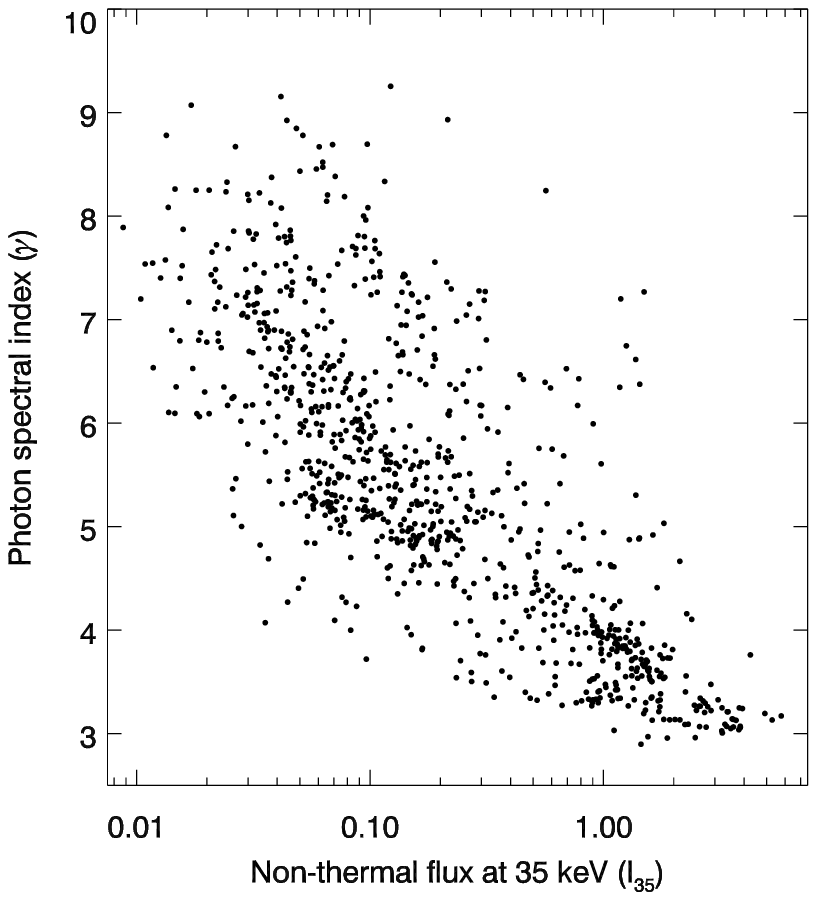} \caption{Plot
of the spectral index $\gamma$ versus the fitted non-thermal flux
at 35~keV (given in photons $\mathrm{s}^{-1}$ $\mathrm{cm}^{-2}$
  $\mathrm{keV}^{-1}$).  All 911 data points from the 24 events are
  shown \citep[from][]{2004A&A...426.1093G}.%
} \label{Gr_shs_overall} \end{figure}

On the other hand, a plot of all the 911 fitted model parameters
for all the events show a large scatter, as shown in Figure~\ref{Gr_shs_overall}. 
The large scatter can be understood as
originating from the superposition of data from a large numbers of
different emission spikes, each featuring linear trends with different
parameters.  This plot does demonstrate, however, the tendency for
flatter spectra to be associated with more intense flares.

%

\textit{RHESSI} observations of the gradual phase\index{gradual phase}
of large solar flares \citep{2008ApJ...683.1180G} and its relation
with proton events\index{proton events!and soft-hard-harder}
\citep{2008ApJ...673.1169S,2009ApJ...707.1588G} have shown that the
hardening behavior\index{hard X-rays!soft-hard-harder}
is complex and cannot be characterized by a continuously increasing
hardness during the event. Therefore the soft-hard-harder (SHH)
denomination does not accurately reflect the observed spectral
evolution. Rather, phases of hardening (or even approximatively
constant hardness) are often seen in larger events as the flux
decays \citep{1995ApJ...453..973K}. The start of the hardening phase
can happen near the main peak of the flare, or later. The end of
hardening can even be followed by new impulsive SHS peaks.  The
most recent statistical study of the correlation of SHH behavior
with proton events\index{proton events} \citep{2009ApJ...707.1588G}
found that in a sample of 37 flares that were magnetically
well-connected to Earth, 18 showed SHH behavior and 12 of these
produced solar energetic particle (SEP) events.  None of the remaining
19 flares that did not show SHH behavior produced SEP events.

\subsection{Interpretation of spectral evolution} \label{Gr_shs_theory}
\index{hard X-rays!spectral evolution!interpretation}

Can we explain the soft-hard-soft spectral behavior theoretically?
The problem here is that many effects contribute to the properties
of the high-energy electron distribution whose bremsstrahlung hard
X-rays are observed by {\it RHESSI} and similar instruments. We can
identify three main, closely related classes of physical processes
that affect the distribution of the electrons and the spectrum of
the X-ray photons they generate: (1) the \emph{acceleration} of
part of the thermal ambient plasma\index{electrons!acceleration from thermal plasma}, (2) the \emph{escape}\index{electrons!escape} from the
acceleration region\index{acceleration region}, and (3) the
\emph{transport} to the emitting region\index{electrons!transport}\index{transport!electrons}.  
The photon spectrum also
depends on the properties of the bremsstrahlung emission mechanism.

%
\begin{figure}[t!] \centering\includegraphics[width=10cm]{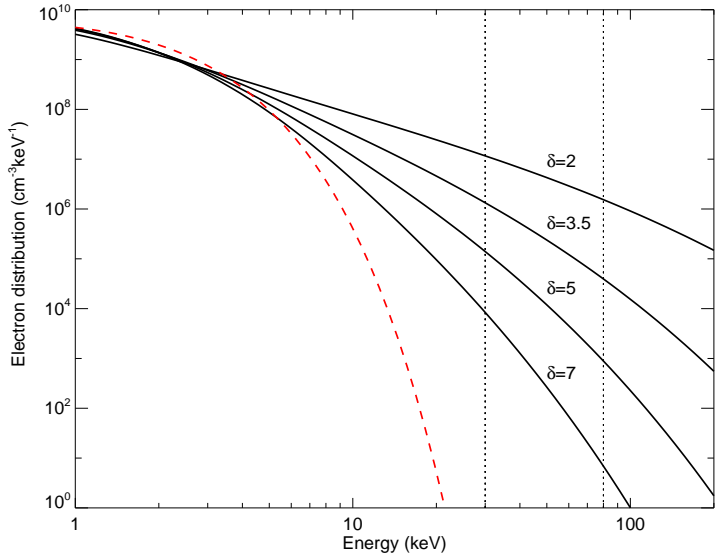}
\caption{
  Accelerated electron density distributions with different values
  of the power-law index resulting from changes in
  $I_\tau=I_\mathrm{ACC}\cdot\tau$.  The dashed curve represents
  the ambient Maxwellian distribution. The two dotted lines indicate
  the energy range used for the computation of the power-law index
  $\delta$ shown above each spectrum. Harder spectra have a larger
  $I_\tau$ value \citep[from][]{2006A&A...458..641G}.
} \label{Gr_acceq} \end{figure} 
%

\cite{1996ApJ...461..445M} proposed a stochastic
acceleration\index{acceleration!stochastic} mechanism where electrons
are energized by small-amplitude turbulent fast-mode waves, called
the transit-time damping model.\index{acceleration!transit-time
damping}\index{turbulence!particle acceleration}\index{transit-time damping}
They showed that their model could successfully account
for the observed number and energy of electrons accelerated above
20~keV in subsecond spikes or energy release fragments in impulsive
solar flares. 
However, they made no attempt to explain the observed
hard X-ray spectra (which are softer than predicted by the transit-time
damping model) and did not consider spectral evolution.  
Furthermore, this approach does not account for particle escape.
\citet{2006A&A...458..641G} extended the model with the addition
of a term describing the escape of the particles from the acceleration
region, as in the model of \citet{1999ApJ...527..945P}. To ensure
conservation of particles, they also add a source term of cold
particles coming into the accelerator (such as can be provided by
a return current\index{return current}).

The stochastic nature of this acceleration model implies that the
electrons undergo a diffusion process in energy space.  Mathematically,
the acceleration is described by the following convective-diffusive
equation\index{convective-diffusive equation}:  
\begin{equation}
  \label{Gr_eq:maindiff}
\diff{f}{t} =
\frac{1}{2}\difff{}{E}\Big[\left(D_\mathrm{COLL}+D_\mathrm{T}\right)f\Big]
	     -\diff{}{E}\Big[ \left(A_\mathrm{COLL}+A_\mathrm{T}\right)f
	     \Big] -S(E)\cdot f + Q(E)\,,
\end{equation}  
where $f(E)$ is the electron density distribution
function\index{electrons!distribution function!density}, $D_\mathrm{T}$
and $A_\mathrm{T}$ are, respectively, the diffusion and convection
coefficients due to the interactions of the electrons with the
accelerating turbulent waves, $D_\mathrm{COLL}$ and $A_\mathrm{COLL}$
are, respectively, the diffusion and convection coefficients due
to collisions with the ambient plasma, $S(E)$ is the sink (escape)
term, and $Q(E)$ is the source (return current) term.
\index{magnetic trapping} 
The escape term is proportional to $ v(E)/\tau$, where
$v(E)$ is the electron speed, and $\tau$ is the escape time. The
escape time\index{electrons!escape time} can be energy-dependent, but for simplicity it is
initially kept constant. The longer the escape time, the better the
particles are trapped in the accelerator. The source term is in the
form of a Maxwellian distribution of electrons with the same
temperature as the ambient plasma.

The coefficients $D_\mathrm{T}$ and $A_\mathrm{T}$ are proportional
to the dimensionless acceleration parameter \begin{equation}
I_\mathrm{ACC}=\displaystyle\frac{U_\mathrm{T}}{U_\mathrm{B}}\cdot
\frac{c\langle k\rangle}{\Omega_\mathrm{H}}, \end{equation} where
$U_\mathrm{T}$ and $U_\mathrm{B}$ are, respectively, the energy
densities of the turbulent waves and of the ambient magnetic field,
$\langle k\rangle$ is the average wave vector, and $\Omega_\mathrm{H}$
is the proton gyrofrequency.\index{frequency!Larmor!proton}

Equation~\ref{Gr_eq:maindiff} can be solved numerically until an
equilibrium state ($\partial{f}/\partial{t}=0$) is reached.
The equilibrium electron spectra from the model are controlled by
two parameters: the acceleration parameter $I_\mathrm{ACC}$ described
above and the escape time $\tau$. Above 10-20 keV, the collision
and source terms in Equation~(\ref{Gr_eq:maindiff}) can be neglected,
since they apply to the ambient Maxwellian, and thus the equilibrium
spectra depend to a first approximation only on the product
$I_\tau=I_\mathrm{ACC}\cdot\tau$.

Figure \ref{Gr_acceq} shows the equilibrium electron spectra for
different values of $I_\tau=I_\mathrm{ACC}\cdot\tau$. As $I_\tau$
increases, the spectrum gets harder and harder. To explain the
soft-hard-soft effect, either the acceleration or the trapping
efficiency (or both) must increase until the peak time, and then
decrease again.
\index{soft-hard-soft!and trapping efficiency}\index{trapping!in turbulence}
We note that this model does not include magnetic
trapping\index{magnetic trapping} (other than in the magnetic
turbulence itself), which can alter the computed electron spectra
and their time evolution \citep[e.g., ][]{1999ApJ...522.1108M}.

To see whether this produces the linear relation between the spectral
index and the log of the flux normalization, \citet{2006A&A...458..641G}
computed the hard X-ray emission from these model electron spectra.
Since these are equilibrium spectra, thin-target emission was
computed.  They then plotted the spectral index vs. the flux
normalization of the resulting photon spectra.  Since the spectra
are not power-law, but bend down, they fit a power-law model to the
model photon spectrum in a similar range as the one used for the
observations.

\begin{figure}[t!]
\centering\includegraphics[width=12cm]{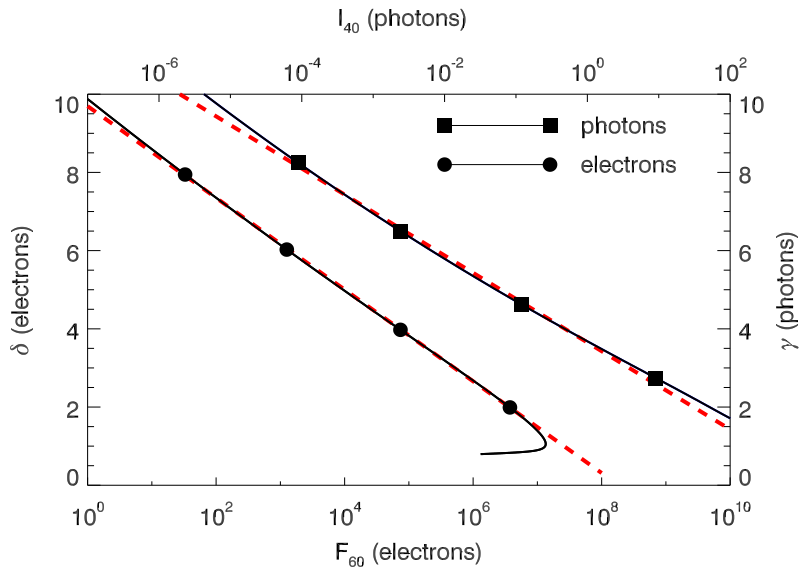} \caption{Model
results for the spectral index and flux normalization for
  electrons and photons. The dashed line is the best straight-line
  fit to the model results (in the range of spectral indices from
  2 to 8 for the electrons, and 3 to 9 for the photons), corresponding
  to a pivot-point behavior \citep[from][]{2006A&A...458..641G}.}
\label{Gr_manysp} \end{figure}

Figure \ref{Gr_manysp} shows the computed values for the spectral
indices and flux normalizations for both the electron and the photon
spectrum from the model.  The results show that there is indeed a
linear relation between the spectral index and the log of the flux
normalization.

%

An alternative mechanism that could be responsible for soft-hard-soft
spectral evolution is return current\index{return current!energy losses} losses
as the electrons propagate to and within the thick-target footpoints
of the flare loop \citep{2006ApJ...651..553Z}.  The highest electron
energy to which return current losses are significant is proportional
to the return current electric field strength, which is in turn
proportional to the electron beam flux density (see
Section~\ref{sec:zharkova_return_current}).\index{electric fields!and soft-hard-soft pattern} 
Therefore, as the
electron flux density increases and then decreases, the low-energy
part of the X-ray spectrum flattens to higher and then lower energies
as the return current electric field strength increases and then
decreases.  The net effect is SHS spectral evolution below the
maximum energy for which return current losses are significant
during the flare.  The observation of SHS behavior in coronal X-ray
sources, however, indicates that this spectral evolution is a
property of the acceleration mechanism rather than a consequence
of energy losses during electron propagation
(\citeauthor{2006A&A...456..751B} \citeyear{2006A&A...456..751B},
see Section~\ref{Ba_shs_corn}).

Are there two stages of electron
acceleration\index{acceleration!two-stage}, one responsible for the
impulsive phase and one for the gradual phase?  \textit{RHESSI} spectroscopy
and imaging of a set of 5 flares with hardening phases
\citep{2008ApJ...683.1180G} showed that there is no discontinuity
in the motion of footpoints at the onset of hardening and no clear
separation between the impulsive and the gradual phase: the former
seems to smoothly merge into the latter. This supports the view
that the same acceleration mechanism changes gradually in the later
phase of the flare, rather than a two stage acceleration theory.
The hardening phase may in fact be caused by an increase in the
efficiency of trapping\index{magnetic trapping} of the electrons
above 100~keV.

The underlying cause of the SHS spectral evolution has been addressed
in terms of the stochastic acceleration\index{acceleration!stochastic}
model by \citet{2009ApJ...692L..45B} and \citet{2009ApJ...701L..34L}.
\index{turbulence!MHD}
Bykov and Fleishman consider acceleration in strong, long-wavelength
MHD turbulence, taking into account the effect of the accelerated
particles on the turbulence.  They argue that the electron spectrum
flattens during the linear acceleration phase, while the spectrum
steepens during the nonlinear phase when damping of the turbulence
because of the particle acceleration is important, giving SHS
spectral evolution.\index{soft-hard-soft!and turbulence}
They argue that SHH evolution will be observed
when the injection of particles into the acceleration region is
strong.\index{acceleration region!and SHH pattern}
Liu~\&~Fletcher also argue that the SHS evolution results
from dependence of the electron distribution power-law index on the
level of turbulence as it increases and subsequently decreases.
They attribute changes in the SHS correlation during a flare to
changes in the background plasma, likely due to chromospheric
evaporation.

We note that simple direct-current (DC) electric field
acceleration\index{acceleration!DC electric field} of electrons out
of the thermal plasma can produce the SHS spectral evolution.\index{electric fields!DC} 
The
flux of accelerated electrons and the maximum energy to which
electrons are accelerated and, therefore, the high-energy
cutoff\index{electrons!distribution function!high-energy cutoff} to
the electron distribution, increase and decrease together as the
electric field strength increases and decreases
\citep{1985ApJ...293..584H}.  The X-ray spectrum is steeper at
energies within one to two orders of magnitude below the high-energy
cutoff \citep{2003ApJ...586..606H}.  In large flares, however, where
the X-ray spectrum is observed to continue to MeV energies or higher,
there is no evidence for a high-energy cutoff in the appropriate
energy range.  Therefore, at least for large flares with spectra
extending to high energies, a simple DC electric field acceleration
model does not appear to be appropriate.

\section{The connection between footpoint and coronal hard X-ray sources} 
\label{sec:battaglia_footpoint_coronal}
\index{footpoints!and coronal sources}\index{hard X-rays!coronal sources}


Hard X-ray (HXR) sources at both footpoints of a coronal loop
structure have been observed since \citet{1981ApJ...246L.155H}. As
reviewed in Sections \ref{sec:holman_introduction} \&
\ref{sec:holman_thick}, they are understood to be thick-target
bremsstrahlung emission produced by precipitating electrons,
accelerated somewhere in or above the loop.  
\index{coronal sources}\index{hard X-rays!above-the-looptop source}
A third HXR source situated above the looptop 
\index{hard X-rays!coronal sources}
\citep[see][for a review]{2008A&ARv..16..155K} was first
noted by \citet{1994Natur.371..495M} in {\it Yohkoh}\index{Yohkoh@\textit{Yohkoh}}
observations. 
\index{satellites!Yohkoh@\textit{Yohkoh}}
The nature of this coronal HXR source has remained
uncertain, but in simple solar flare models with reconnection and
particle acceleration in the corona, we expect some relation between
coronal HXR sources and footpoints. 
\index{reconnection!and particle acceleration}
{\it RHESSI} has enabled us to
study events featuring coronal HXR sources and footpoints simultaneously.
By studying the behavior of the sources in time and the relations
between them, we can address questions such as:  Are both coronal
and footpoint emissions caused by the same electron population? How
is such an electron beam modified in the loop (collisions, return
currents, trapping, etc.)\index{electrons!transport}?  
Is SHS behavior (Section~\ref{Gr_shs_observations}) a transport effect produced by collisions
or return currents, or is it a feature imposed by the acceleration
mechanism?\index{return current!and soft-hard-soft} 

\subsection{{\it RHESSI} imaging spectroscopy} 
\index{RHESSI@\textit{RHESSI}!imaging spectroscopy}\index{imaging spectroscopy}

{\it RHESSI} has
provided the possibility of obtaining simultaneous, high-resolution
imaged spectra\index{RHESSI@\textit{RHESSI}!imaging spectroscopy} at different
locations on the Sun. One can therefore study each source separately
in events with several contemporaneous HXR sources. The high spectral
resolution has allowed a reliable differentiation between thermal
and non-thermal emission to be made in many flares.  Furthermore,
{\it RHESSI}'s imaging spectroscopy has allowed differences in
individual flare source spectra and their evolution to be studied
in considerable detail.

Imaged spectra and the relative timing of sources in three flares,
including the limb flare SOL2002-02-20T11:07\index{flare (individual)!SOL2002-02-20T11:07 (C7.5)!coronal source}, were studied by
\citet{2002SoPh..210..229K}.   
\citet{2002SoPh..210..245S} analyzed
and modeled the two footpoint sources and a high, above-the-looptop
hard X-ray source observed in this flare.  \citet{2003ApJ...595L.107E}
analyzed SOL2002-07-23T00:35 (X4.8)
\index{flare (individual)!SOL2002-07-23T00:35 (X4.8)!coronal hard X-ray source} flare with four 
HXR sources observed
by {\it RHESSI}. They found a coronal source with a strong thermal
component, but the non-thermal component could not be studied due
to severe pulse pile-up.   \citet{2006A&A...456..751B} studied five
M-class events. Due to the smaller pile-up amount in those events,
studying the non-thermal coronal emission was possible. The results
of these studies are summarized below.

\begin{figure}[h] {\includegraphics[width=12cm]{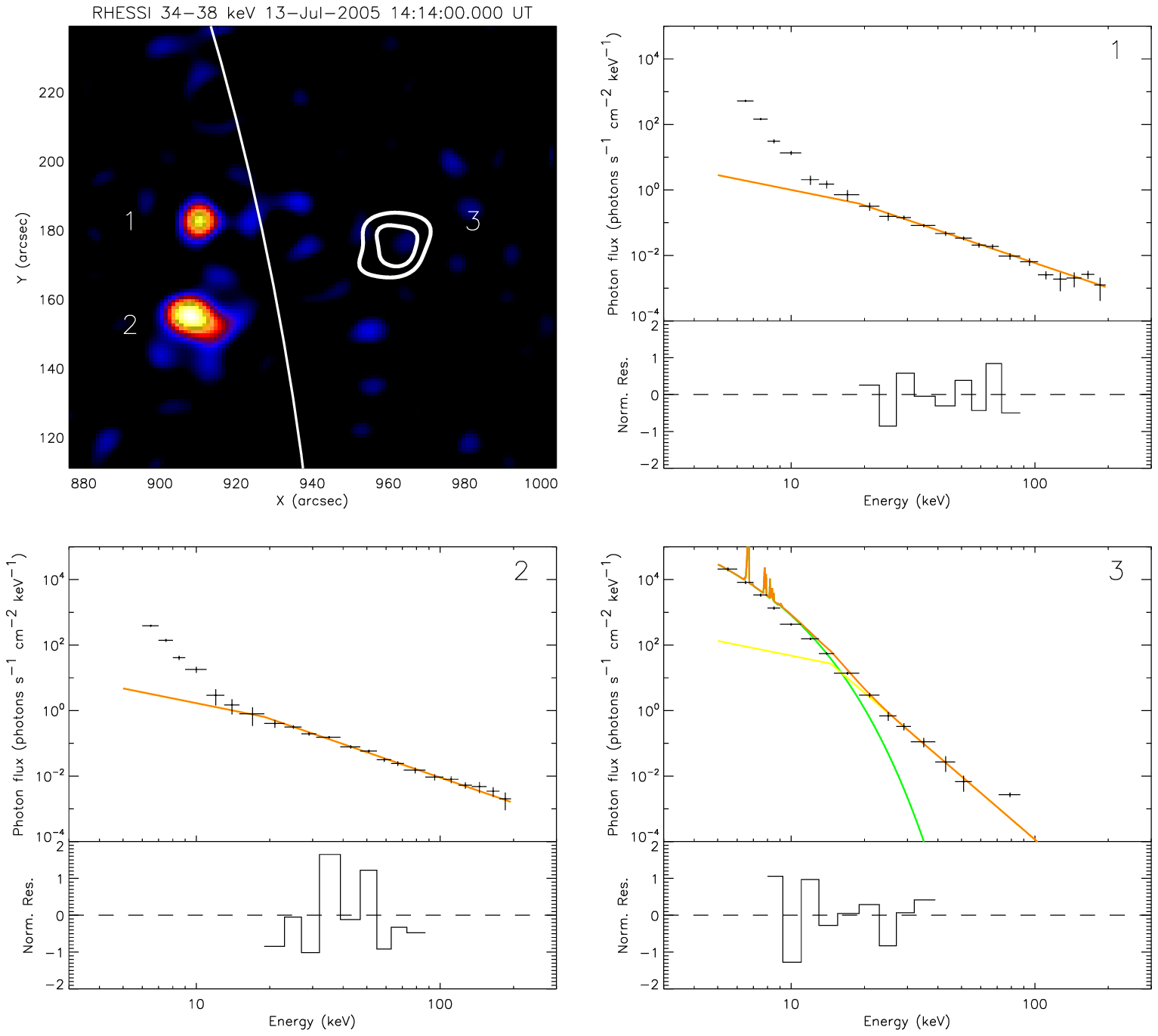}}
\caption {\textit{Top left} Composite CLEAN image of a {\it RHESSI}
event\index{flare (individual)!SOL2005-07-13T14:49 (M5.0)!coronal hard X-ray
source} with three hard X-ray sources. The footpoints (labeled 1
\& 2) are visible on the solar disc in an image made at 34--38 keV.
The position of the coronal source (labeled 3) high above the limb
is indicated by the 50 and 80\% white contours taken from a 10--12
keV image.  \textit{Plots 1-3} show spectra and normalized residuals
over the fitted energy range for the north footpoint (1), south
footpoint (2), and coronal source (3)
\citep[after][]{2006A&A...456..751B}.} \label{Ba_fig1} 
\end{figure}

\subsection{Relation between coronal and footpoint sources}
\label{subsec:battaglia_deltagamma} 
\index{hard X-rays!relation between footpoint and coronal sources}

The quantitative relations
between the footpoints and the coronal source and between the two
footpoints can give information about the physical mechanisms at
work in a solar flare. Simple models envision a beam of accelerated
electrons encountering a low-density region in the corona, leading
to thin-target bremsstrahlung\index{hard X-rays!thin-target}. When the
same electron beam reaches the chromosphere, the particles are fully
stopped in the dense material, producing thick-target
emission\index{hard X-rays!thick-target}. Assuming an electron power-law
distribution for the electron energy $E$ of the form 
\begin{equation}
\label{Ba_electroneq} {\cal F}(E)=AE^{-\delta} 
\end{equation}
producing thin-target bremsstrahlung\index{bremsstrahlung!thin-target} in the coronal source, the
observed photon spectrum has spectral index\index{hard X-rays!spectral
index} $\gamma_{thin}=\delta+1$ (Equation~\ref{eqn:holman_ithinpl}).
Reaching the chromosphere, the accelerated electrons will be fully
stopped, producing thick-target bremsstrahlung\index{bremsstrahlung!thick-target} with a photon spectral
index $\mathrm{\gamma_{thick}=\delta-1}$
(Equation~\ref{eqn:holman_ithickpl}). In such a simple scenario one
would therefore expect a difference in the photon spectral index
$\gamma_{thin}-\gamma_{thick}=2$ between the coronal source and the
footpoints. Further, the two footpoints should be of equal hardness
and intensity if one assumes a symmetric loop and symmetric injection
of particles into the legs of the loop.

\subsubsection{Observed difference between coronal and footpoint
spectral indices} 

A sample of flares observed with {\it
Yohkoh}\index{Yohkoh@\textit{Yohkoh}} to have coronal HXR 
sources\index{hard X-rays!coronal sources} was studied by
\index{satellites!Yohkoh@\textit{Yohkoh}}
\citet{2002ApJ...569..459P}.  They found that the spectral index
of the coronal sources was, on the average, steeper by~1 than the
spectral indices of the footpoint sources.  \citet{2002SoPh..210..245S}
also found a spectral index difference of~1 for SOL2002-02-20T11:07 (C7.5)
\index{flare (individual)!SOL2002-02-20T11:07 (C7.5)!coronal hard X-ray
source} observed with {\it RHESSI}.

\citet{2006A&A...456..751B} found that the coronal source was softer
than both footpoints for all of their five events in nearly all
analyzed time bins. Figure~\ref{Ba_fig1} (top left) shows an image
of SOL2005-07-13T14:49 (M5.0)
\index{flare (individual)!SOL2005-07-13T14:49 (M5.0)!coronal hard X-ray source} 
in the 34-38 keV energy band. The
two footpoints are visible, as well as the 50 and 80\% contours of
the coronal source taken from a 10--12 keV image.  Spectra and
spectral fits are shown for the two footpoints and the coronal
source.  The steepness of the coronal source spectrum (number 3 in
the figure) relative to the spectra from the footpoints is apparent.
However, the quantitative difference between the values of the
spectral index obtained for the coronal source and the footpoints
often differed significantly from~2.  For the five flares analyzed,
the smallest mean difference in the spectral indices, averaged over
time, was 0.59$\pm$0.24. The maximum mean difference, averaged over
time, was 3.68$\pm$0.14. These clearly contradict the theoretical
expectation summarized above.  Simple thin-thick target scenarios
do not seem to work in most cases and additional effects need to
be considered.  

Evidence for two populations of coronal source non-thermal spectra
was found by \citet{2009ApJ...691..299S}.  They compare coronal and
footpoint spectral indices at 28 hard X-ray peaks from 13 single-loop
flares observed by {\it RHESSI}.  The spectral index in the coronal
sources was determined from an isothermal plus power-law fit below
30~keV, while the footpoint spectral indices were determined from
a power-law fit at 30-60~keV photon energies.  They argue that
the coronal spectra can be divided into two groups.
One, for which the coronal spectral index is greater than~5, is well correlated
with the footpoint spectral index, and the difference in the indices
ranges from 2-4.
For the other, where the spectral indices are
anticorrelated, the coronal spectral index is less than~5, and the
difference in the indices ranges from 0-2.  
For the group of
anticorrelated spectral indices, the coronal spectral index is
correlated with the photon flux, while the footpoint spectral index
is anticorrelated with the photon flux for both groups.  These are
intriguing results if confirmed by future studies.

\subsubsection{Differences between footpoints} 

No significant
difference was found in the spectral indices for the two
footpoints\index{hard X-rays!footpoint sources} in SOL2002-02-20T11:49 (C7.5)
\index{flare (individual)!SOL2002-02-20T11:49 (C7.5)!footpoint differences}
by \citet{2002SoPh..210..229K} and \citet{2002SoPh..210..245S}.
\citet{2007ApJ...665..846P} inverted count visibility spectra for
this flare to obtain mean electron flux distributions
\index{electrons!distribution function!mean electron flux} for the footpoints.  They
found the mean electron flux distribution function at the northern
footpoint to be somewhat steeper ($\Delta\delta \approx 0.8$) than
that derived for the southern footpoint.  They also found the
distribution function for the region between the footpoints (not
the coronal source studied by Sui et al.) to be steeper than the
footpoint distribution functions ($\Delta\delta \approx 1.6$ relative
to the southern footpoint) and to substantially steepen at energies
above $\sim$60~keV.

\citet{2002SoPh..210..229K} found that, when a connection between
footpoints could be determined, the footpoints brightened simultaneously
(to within the $\sim$1~s time resolution of the observations) and
had similar spectra\index{footpoints!simultanteity}\index{footpoints!spectral similarity}.

Differences of 0.3 -- 0.4 between the spectral indices of two
footpoints in SOL2002-07-23T00:35 (X4.8)\index{flare (individual)!SOL2002-07-23T00:35 (X4.8)!footpoint differences} were reported by \citet{2003ApJ...595L.107E}.

For the flares analyzed by \citet{2006A&A...456..751B}, a significant
difference was found in only one out of five events. For all other
events, the mean difference in $\mathrm{\gamma_{fp}}$ was zero
within the statistical uncertainty.

Different spectra at the two footpoints imply an asymmetric loop.
Such an asymmetry can result, for example, from different column
densities\index{column density} or different beam fluxes and
corresponding return current\index{return current} energy losses
in the legs of the loop.  It could also result from asymmetric
magnetic trapping\index{magnetic trapping} within the loop
\citep[e.g.,][]{2002SoPh..210..323A}.  In a study of 53 flares
showing two HXR footpoints, \citet{2008SoPh..250...53S} found that
footpoint asymmetry was greatest at the time of peak HXR flux and
the difference in the footpoint spectral indices $\Delta\gamma$
rarely exceeded 0.6.  In most cases they found the footpoint asymmetry
to be inconsistent with different column densities in the two legs
of the loops\index{footpoints!asymmetry}.

In SOL2003-10-29T20:49 (X10.0)
\index{flare (individual)!SOL2003-10-29T20:49 (X10.0)!footpoint differences} \citet{2009ApJ...693..847L} found that
the brighter HXR footpoint was marginally, but consistently harder
than the dimmer footpoint by $\Delta \gamma= 0.15\pm 0.13$. They
concluded that neither asymmetric magnetic mirroring nor asymmetric
column density {\it alone} can explain the full time evolution of
the footpoint HXR fluxes and spectral indices.  However, a
self-consistent explanation might be obtained by considering these
two effects together and/or in combination with one or more additional
transport effects, such as nonuniform target ionization, relativistic
beaming, and return current losses.

\subsection{Spectral evolution in coronal sources} \label{Ba_shs_corn}
\index{hard X-rays!coronal sources!spectral evolution}

\begin{figure}[h] \begin{center}
{\includegraphics[height=12cm]{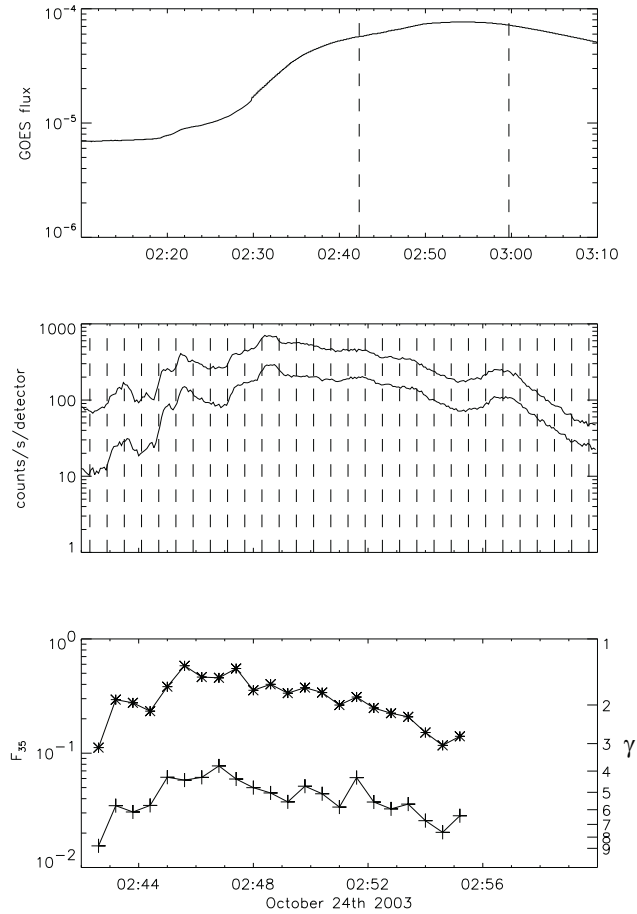}} \caption
{\textit{Top:} \textit{GOES} 1--8~\AA\ light curve of SOL2003-10-24T02:54 (M7.6).
\index{flare (individual)!SOL2003-10-24T02:54 (M7.6)!spectral evolution}
\textit{Middle:} {\it RHESSI} 25--50 and 50--100~keV light curves
near the peak of the
{\it GOES} flare.
\textit{Bottom:} time evolution
of fitted coronal source flux at 35 keV ($F_{35}$, * symbols, left
log scale) and spectral index ($\gamma$, $+$~symbols, right log scale)
displaying SHS evolution \citep[after][]{2006A&A...456..751B}.}
\label{fig:Ba_figshs} \end{center} \end{figure}

Previous observations of SHS\index{hard X-rays!soft-hard-soft} spectral evolution (see
Section~\ref{Gr_shs_observations}) were made with full-Sun spectra
which, except for over-the-limb events\index{hard X-rays!occulted}, are typically dominated by
footpoint emission.  \citet{2006A&A...456..751B}, in their imaging
spectroscopy study, found that the coronal source itself shows SHS
evolution. This is illustrated in Figure~\ref{fig:Ba_figshs}. This
finding implies that SHS is not caused by transport effects within
the flare loop, but is rather a property of the acceleration mechanism
itself. Indeed, \citet{2006A&A...458..641G} showed that SHS can be
reproduced for electron spectra in a transit-time-damping,
stochastic-acceleration model (Section~\ref{Gr_shs_theory}).

\subsection{Interpretation of the connection between footpoints and
the coronal source}\label{sec:battaglia_theory}

In the above account, emphasis was given to the difference in the
spectral index between the coronal source
\index{hard X-rays!coronal sources} and footpoints. Assuming a thin target in the corona and
a thick target at the footpoints, one would expect a difference of
two. However, whether the coronal source acts as thin- or thick-target depends on the energy of the accelerated electrons and the
column density\index{column density} in the corona.
\citet{2004ApJ...603L.117V}, for example, found coronal sources\index{hard X-rays!coronal sources!thick-target}
with column densities high enough to act as thick targets for
electrons with energies up to 60~keV.

As early as 1976, \citet{1976MNRAS.176...15M} showed that magnetic
trapping\index{magnetic trapping} with collisional scattering of
electrons out of the trap can lead to a thick-target coronal source.
The coronal source transitions through a thin-thick period, with
the time scale for this transition depending on the electron energy
and the plasma density in the trap.  The trapping essentially
increases the effective column density in the corona.
\citet{1999ApJ...522.1108M} analyzed six flares with coronal sources
\index{satellites!Yohkoh@\textit{Yohkoh}}
\index{hard X-rays!coronal sources!Yohkoh@\textit{Yohkoh}}
observed by {\it Yohkoh}\index{Yohkoh@\textit{Yohkoh}} and found that three of the
six flares showed properties consistent with trapping.

A simple 1-D model 
that described
the coronal emission as intermediate thin-thick, depending on
electron energy, was developed by \citet{1995SoPh..158..283W}. In
this model a high-density region ($\gtrsim$10$^{12}$ cm$^{-3}$) is
hypothesized to be present at or above the top of the flare loop.
The model makes predictions for the shape of the coronal and footpoint
spectra and the relations between them.  \citet{1995A&A...303L...9F}
obtained Monte Carlo solutions to the Fokker-Planck
equation\index{Fokker-Planck equation} to show that, with the
inclusion of high electron pitch angles and collisional scattering,
a compact coronal X-ray source is produced at the top of a loop
with a constant coronal density $\sim$$3 \times 10^{10}$ cm$^{-3}$.
\citet{1996AAS...188.7005H} showed that, even in the simple 1-D
model, a compact coronal source is produced when electrons are
injected into a loop with a constant coronal density $\sim$$2 \times
10^{11}$ cm$^{-3}$ (see \textit{hesperia.gsfc.nasa.gov/sftheory/loop.htm}).
A compact coronal HXR source can also be produced if there is a
compact magnetic trap\index{magnetic trapping} at or above the top
of the loop.  \citet{1998ApJ...505..418F} showed that, with such a
trap, a significant coronal X-ray source\index{hard X-rays!coronal sources} 
can be produced at plasma
densities as low as $\sim$$4 \times 10^{9}$ cm$^{-3}$.
\citet{1999ApJ...527..945P} showed that the coronal HXR source can
be a consequence of acceleration and trapping by turbulence or
plasma waves.\index{turbulence!and trapping}
In their stochastic acceleration
model\index{acceleration!stochastic}, the difference between the
coronal and footpoint spectra is explained by the energy-dependent
time scale for electrons to escape\index{electrons!escape!energy-dependent} 
the acceleration region.\index{acceleration region!escape from}

\begin{figure}[h] 
{\includegraphics[width=12cm]{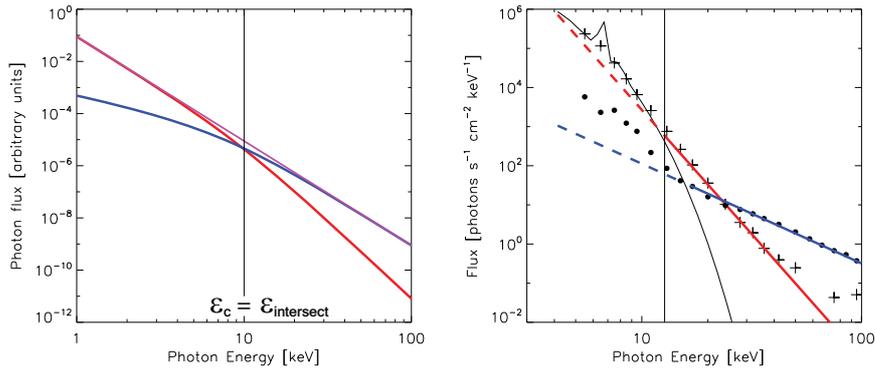}}
\caption{ \textit{Left:} spectra for coronal source (red) and
footpoints (blue) according to the model of \citet{1995SoPh..158..283W}.
The spatially integrated spectrum is shown in violet.  \textit{Right:}
observed {\it RHESSI} spectra for the event SOL2003-10-24T02:54 (M7.6).
\index{flare (individual)!SOL2003-10-24T02:54 (M7.6)!coronal source vs.\
footpoint spectra}  
Isothermal and power-law fits to the coronal
(crosses) and footpoint (dots) spectra are shown.  The vertical
line indicates the predicted critical energy for the transition
between thin and thick target \citep[after][]{2007A&A...466..713B}.}
\label{Ba_figwm} 
\end{figure}

The left panel of Figure~\ref{Ba_figwm} illustrates the model of
\citet{1995SoPh..158..283W}.  The spatially integrated spectrum
(violet) is the power-law spectrum (thick-target,
$\gamma_{thick}=\delta-1$) expected for a single-power-law electron
distribution with no low- or high-energy cutoffs and no thermal
component.  For $\epsilon \ll \epsilon_c = \sqrt{2KN}$ (see
Equation~\ref{eqn:holman_colevol}), the spectrum is dominated by
thick-target radiation from the coronal source (red).  There is a
low-energy cutoff in the electron distribution at the footpoints
at $E \approx \sqrt{2KN}$ because of the energy losses in the coronal
source.  The spectrum is dominated by thick-target radiation from
the footpoints (blue) where $\epsilon \gg \epsilon_c$.  It is in
this regime that the radiation from the coronal source is thin-target
and the spectral index of the coronal source is steeper by 2 than
that of the footpoints.  These spectra are characteristic of all
the models reviewed above.

\citet{2002SoPh..210..245S} compared the {\it RHESSI} observations
of SOL2002-02-20T11:07 (C7.5)
\index{flare (individual)!SOL2002-02-20T11:07 (C7.5)!coronal source vs.\ footpoint spectra} to a model with a
constant-coronal-density loop and no magnetic trapping.  They used
a finite difference method
\citep[e.g.,][]{1990ApJ...359..524M,2002mwoc.conf..405H} to obtain
steady-state solutions to the Fokker-Planck equation\index{Fokker-Planck
equation} with collisional scattering and energy losses.  Model
images were convolved with the {\it RHESSI} response to produce
simulated {\it RHESSI} observations for direct comparison with the
SOL2002-02-20T11:07 flare images and imaged spectra.  They found that, after
obtaining a power-law model spectrum with an index of $\gamma = 3$
that agreed with the observed footpoint spectra, the effective
spectral index of the coronal source from the model ($\gamma = 4.7$)
was significantly steeper than that obtained for the flare ($\gamma
= 4$).

\citet{2007A&A...466..713B} compared the model of
\citet{1995SoPh..158..283W} to the results of their study of five
flares observed by {\it RHESSI}. The right panel of Figure~\ref{Ba_figwm}
shows observed spectra and spectral fits for one particular event.
The observed spectra were dominated by thermal coronal emission at
low energies.  Therefore, not all of the model predictions could
be tested. However, the observed relations between the spectra did
not agree with the predictions of the model.  For the flare in
Figure~\ref{Ba_figwm}, for example, the difference between the
coronal source and footpoint spectral indices at the higher photon
energies is $3.8\pm0.1$, not 2.  Also, an estimate of the column
density in the coronal source gives $\sqrt{2KN} \approx\ $10-15~keV,
while the intersection of the coronal and footpoint spectra is found
to be at $\epsilon \approx 23$~keV.  \citet{2008A&A...487..337B}
have found that this large difference in the spectral indices is
consistent with spectral hardening caused by return current\index{return
current!energy losses} losses (see Section~\ref{sec:zharkova_return_current}).


\section{Identification of electron acceleration sites from radio observations} 
\label{sec:aurass_implications}


\noindent While energetic electrons excite hard X-ray emission
during their precipitation into the dense layers of the solar
atmosphere, they can also excite decimeter and meter wave radio
emission\index{radio emission} during propagation and trapping in
magnetic field structures in the dilute solar corona. The radio
emission pattern in dynamic spectrograms can give information about
the electron acceleration process, the locations of injection of
electrons in the corona, and the properties of the coronal magnetoplasma
structures.


\begin{figure}[b,h!] \center
\includegraphics[height=4.5cm]{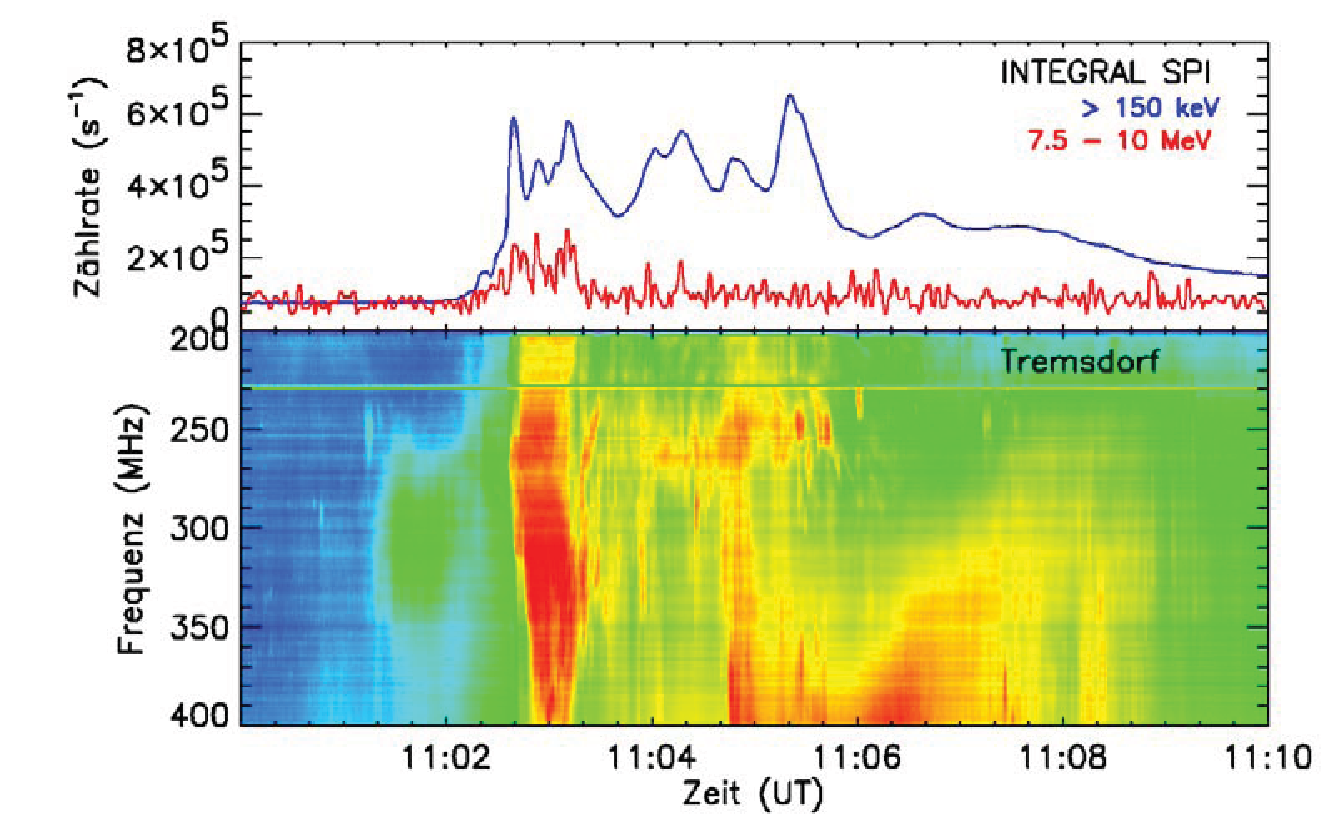}%
\raisebox{0mm}{\includegraphics[height=4.5cm]{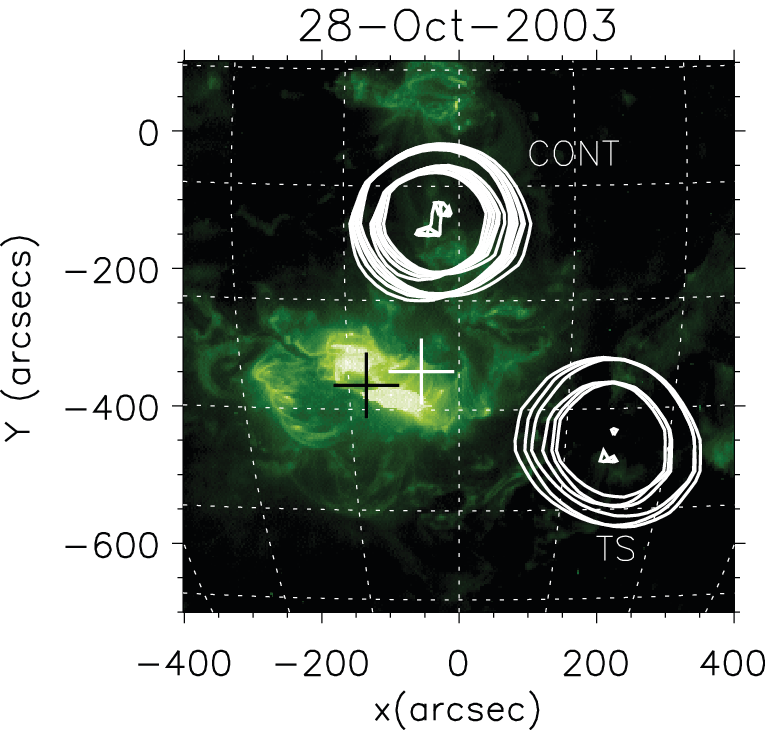}}
\caption{SOL2003-10-28T11:10 (X17.2).\index{flare (individual)!SOL2003-10-28T11:10 (X17.2)!radio emission}
{\it Left, bottom:} (see
\citeauthor{2007CEAB...31..135W}~\citeyear{2007CEAB...31..135W})
200-400~MHz radio spectrum (Astrophysical Institute
Potsdam) showing the signature of the outflow termination shock
(TS, starting at 11:02:47~UT). {\it Left, top}: {\it
INTEGRAL}\index{INTEGRAL@\textit{INTEGRAL}} count rates at 150~keV and 7.5--10~MeV.
\index{satellites!INTEGRAL@\textit{INTEGRAL}}
{\it Right:} (after
\citeauthor{2007A&A...471L..37A}~\citeyear{2007A&A...471L..37A}):
radio source positions (Nan\c{c}ay Radio Heliograph, 327~MHz)
overlaid on a {\it SOHO}-EIT image (11:47 UT 195~\AA{}). The bright
areas are EUV flare ribbons in AR10486.  {\it RHESSI} HXR centroids
are shown as ``+''.  The integration time intervals are: for the
TS source SW of AR10486 11:02:45--11:03:15~UT, for the continuum
source CONT~N of AR10486 11:13--11:17~UT, respectively (see also
Figure~\ref{fig:aurass_fig2}). The radio contours are at 50, 70, and
99.5\% of the peak flux value.} \label{fig:aurass_fig1} 
\end{figure}
\index{flare (individual)!SOL2003-10-28T11:10 (X17.2)!illustration}
\index{termination shock!illustration}


Here we take as an example SOL2003-10-28T11:10 (X17.2)
(shown in Figure~\ref{fig:aurass_fig1}).  Different acceleration
sites\index{acceleration region} can be discriminated during the
impulsive and the gradual flare phases. Radio spectral data from
the Astrophysical Institute Potsdam \citep[AIP;][]{1992ESASP.348..129M},
imaging data from the Nan\c{c}ay Radio Heliograph
\citep[NRH,][]{1997LNP...483..192K}, and hard X-ray ({\it RHESSI},
{\it INTEGRAL}) data were combined in the analysis of this event.
The conclusion was reached that a nondrifting, high-frequency type
II\index{radio emission!type II burst} radio burst signature in the radio
spectrum coincided with a powerful electron acceleration stage.
Simultaneously with the nondrifting type II signature, highly
relativistic ($\ge$10~MeV) electrons\index{electrons!relativistic} were observed in the impulsive
phase of the flare (Figure~\ref{fig:aurass_fig1}, upper left).  The
radio spectrum suggests that this can be due to
\index{acceleration!shock}
\index{reconnection!termination shock}
acceleration\index{shocks!particle acceleration} at the reconnection outflow
termination shock\index{shocks!termination} \citep*{2004ApJ...615..526A},
as predicted for a classical two-ribbon flare
(\citeauthor{1986ApJ...305..553F} \citeyear{1986ApJ...305..553F},
\citeauthor*{1998ApJ...495L..67T} \citeyear{1998ApJ...495L..67T},
\citeauthor{2002A&A...384..273A} \citeyear{2002A&A...384..273A}).
The radio source site is observed about 210~Mm to the SW of the
flaring active region (TS in  Figure~\ref{fig:aurass_fig1}, right).
In this direction, {\it TRACE}\index{TRACE@\textit{TRACE}} and 
{\it SOHO}/LASCO\footnote{Large Angle and Spectrometric Coronagraph}
C2\index{SOHO@\textit{SOHO}} images reveal dynamically evolving magnetoplasma
\index{satellites!TRACE@\textit{TRACE}}
\index{satellites!SOHO@\textit{SOHO}}
\index{satellites!SOHO@\textit{SOHO}!LASCO}
structures in an erupting arcade \citep{2006A&A...457..681A}. For
realistic parameters derived from these observations (the geometry,
density, temperature, and low magnetic field values of $\sim$5~Gauss),
\citet{2006A&A...454..969M} have found that a fully relativistic
treatment of acceleration at the fast-mode outflow shock can explain
the observed fluxes of energetic particles \citep[see][]{Chapter8}.


\begin{figure}[t,h] \center
\includegraphics[width=.6\hsize]{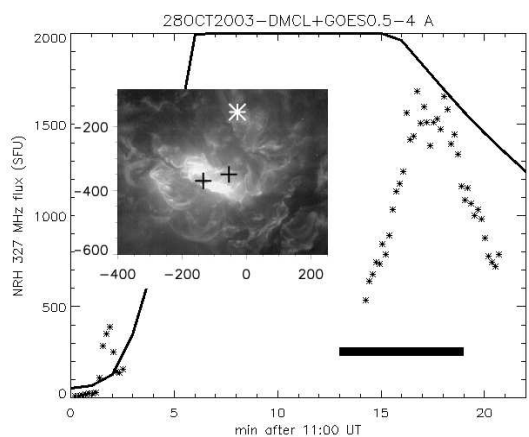} \caption{Timing
of the source CONT in Figure~\ref{fig:aurass_fig1}: the NRH 327~MHz
flux curve (asterisks) versus the 
{\it GOES} 0.5-4~\AA\ flux curve (solid
line, partly off-scale). Inset: {\it SOHO}/EIT image showing the
radio source centroid (white asterisk) and {\it RHESSI} HXR centroids
as in Figure~\ref{fig:aurass_fig1}. Thick bar: the start time of
GeV-energy proton injection in space (after
\citeauthor{2007A&A...471L..37A}~\citeyear{2007A&A...471L..37A}).
}
\label{fig:aurass_fig2} 
\index{flare (individual)!SOL2003-10-28T11:10 (X17.2)!illustration} 
\end{figure}

In the main flare phase of the same event, an additional radio
source (CONT in Figure~\ref{fig:aurass_fig1}) was found,  lasting for
$\sim$10~min, indicating the presence of another acceleration
site.  No X-ray, EUV, or H$\alpha$ emission was observed at the
location of this radio source.  Figure~\ref{fig:aurass_fig2} gives
the timing and the source position with respect to the flaring
active region. CONT is a m-dm-continuum source with fiber burst
fine structure. Fiber bursts\index{radio emission!fiber burst} are
\index{waves!whistlers}
excited by whistler waves\index{waves!whistlers} propagating along
field lines of the coronal magnetic field. As marked by a bold bar
in the Figure, the time of the CONT emission is also the start time
of GeV proton injection in space. \cite{2006A&A...457..681A} have
shown that this source site is not far from an open field (particle
escape) region in the potential coronal magnetic field. The source
briefly flashes up already in the early impulsive phase. Based on
a new method of fiber burst analysis
(\citeauthor{2005A&A...435.1137A}~\citeyear{2005A&A...435.1137A};
\citeauthor{2007SoPh..245..327R}~\citeyear{2007SoPh..245..327R}),
\citeauthor{2007A&A...471L..37A}~(\citeyear{2007A&A...471L..37A})
argue that this source most likely indicates
acceleration\index{acceleration region} at a contact between
separatrix surfaces\index{magnetic structures!separatrix} of different magnetic flux systems.\index{acceleration region!separatrix structure}

Radio observations of flares and their implications are further
addressed in \citet{Chapter5}.

\section{Discussion and Conclusions} 
\label{sec:holman_discussion}

\subsection{Implications of X-ray observations for the collisional
thick-target model}\label{sec:holman_discussion_tt}\index{thick-target model!implications}

As discussed in Section~\ref{sec:holman_thick}, the core assumption
of the collisional thick-target\index{hard X-rays!thick-target} model
is that the spatially integrated hard X-ray emission from non-thermal
electrons is bremsstrahlung (free-free radiation) from electrons
that lose all their suprathermal energy through collisional losses
in the ambient plasma as they simultaneously radiate the hard X-rays.\index{suprathermal populations}
``Simultaneously'' means within the observational integration time.
This implies that all electrons that contribute significantly to
the observed radiation reach a plasma dense enough or, more precisely,
traverse a high enough column density\index{column density} for all
of their suprathermal energy above the observed photon energies to
be collisionally lost to the ambient plasma within the integration
time.  For typical $\gtrsim$1~s integration times, these conditions
are met when the electrons stream downward from the corona into the
increasingly dense plasma of the solar transition region and
chromosphere.

Since the thick-target model is often implicit in our interpretation
of the hard X-ray emission from flares, it is important to keep the
underlying assumptions in mind and test the model while at the same
time applying it to flare observations.  We have discussed above
several physical processes that, if significant, change the conclusions
of the simple collisional thick-target model regarding the electron
distribution produced in the acceleration region.\index{acceleration region!distinguished from energy-loss region}
These processes
occur in either the thick-target region itself, or during the
propagation of the electrons from the acceleration region to the
thick-target region.  Only with the high spectral resolution and imaging
of {\it RHESSI} has it become possible to observationally address
these processes.  Even with the {\it RHESSI} observations, however,
it is difficult to conclusively determine the importance of each
process.

A physical process that distorts the emitted X-ray spectrum is
albedo\index{albedo}\index{hard X-rays!albedo} \citep[Section~\ref{sec:sainthillaire_caveats}
and][]{Chapter7}.  
Fortunately, the albedo contribution to the X-ray
spectrum can be easily corrected on the assumption that the X-ray
photons are isotropically emitted.  This correction is available
in the {\it RHESSI} spectral analysis software.  If the photons are
significantly beamed downward, however, the distortion of the
spectrum can be substantially greater than that from isotropically
emitted photons.  An anisotropic photon distribution results from
emitting electrons with an anisotropic pitch-angle distribution.
The degree of anisotropy of the electron pitch-angle distribution
also quantitatively affects conclusions from the thick-target model
concerning the acceleration process.  Therefore, it is important
to better determine the pitch-angle distribution of the emitting
electrons and the contribution of albedo to the hard X-ray spectrum
\citep[see][]{Chapter7}.

The simple collisional thick-target model assumes that the target
plasma is fully ionized.  We have seen, however, that a nonuniformly
ionized\index{non-uniform ionization} target region can produce an
upward kink, or ``chicane,''\index{hard X-rays!chicane}\index{chicane}
in an otherwise power-law X-ray spectrum
(Section~\ref{sec:kontar_nonuniform}).  This spectral shift can
provide a valuable diagnostic of the ionization state
\index{ionization state} of the target plasma and its evolution.  It is likely,
however, that the power-law spectrum below the chicane is hidden
by thermal radiation.  The chicane is then observed only as a
downward break in the spectrum at energies above those dominated
by the thermal emission. The upper limit on the magnitude of the
break provides a method for ruling out nonuniform ionization as the
sole cause of large spectral breaks.  To further distinguish this
break from spectral breaks with other causes, it is important to
better determine the degree of ionization as a function of column
density at the thick-target footpoints.

Return-current\index{return current} energy losses can also produce
a downward break in the X-ray spectrum
(Section~\ref{sec:zharkova_return_current}).  The break energy
depends on both the thermal structure of the plasma in the flare
loop and on the non-thermal electron flux density distribution.
These spectral modifications and their evolution throughout flares
provide an important test for the presence of initially un-neutralized
electron beams and the return currents they must drive to neutralize
them.\index{beams!un-neutralized} 
Although {\it RHESSI} observations provide substantial
information about the structure and evolution of flare spectra,
only a lower limit on the electron flux density can usually be
determined.  Observations and analysis sufficiently accurate and
comprehensive to verify the presence of return current energy losses
as the cause of a spectral break are yet to be obtained.  On the
other hand, significant evidence exists
(Section~\ref{sec:zharkova_rc_evidence}) indicating that return
current losses do have an impact on flare hard X-ray emission.

A thorough comparison of measured flare spectra with theoretical
spectra computed from models incorporating collisional and return
current energy losses\index{return current!energy losses} (including their effect on the angular
distribution of the non-thermal electrons), as well as nonuniform
target ionization and albedo, is still needed.  
Spectral fitting alone, however, is not likely to distinguish the importance of these
different mechanisms.  
Comparison of the time evolution of the spectra,
\index{hard X-rays!spectral evolution}
as well as of the spatial
structure of the X-ray emission, with expectations would certainly
enhance the possibility of success for such an endeavor.

The analysis of the evolution of X-ray source positions and sizes
with photon energy and time provides another important test of the
collisional thick-target model
\index{hard X-rays!height dependence}
(Section~\ref{sec:aschwanden_height}).  
For these flares that show
non-thermal source evolution in the corona and upper transition
region, the source position and size are sensitive to the energy
losses experienced by the non-thermal electrons.  They are, in fact,
sensitive to the very assumption that the sources are produced by
electrons as they stream downward from an acceleration region higher
in the corona.\index{acceleration region!height}
Further studies of the evolution of these coronal
X-ray sources should substantially clarify the applicability of the
collisional thick-target model.

For completeness, we note that under some circumstances other
radiation mechanisms may significantly contribute to the X-ray
emission from non-thermal electrons.  The possibility that recombination
\index{free-bound emission} (free-bound) radiation
\index{recombination radiation (non-thermal)}
from the non-thermal electrons is sometimes important is discussed
in \citet{2008A&A...481..507B,2009ApJ...697L...6B} \citep[also
see][]{Chapter7}.  However, the contribution of non-thermal free-bound
radiation has recently been found to be less significant than
originally estimated \citep{2010A&A...515C...1B}.
\citet{2010A&A...510A..29M} have concluded that inverse Compton
radiation\index{inverse Compton radiation}\index{hard X-rays!inverse Compton radiation} may significantly
contribute to the X-ray/$\gamma$-ray emission from low-density
coronal sources.

Another testable aspect of the collisional thick-target model is
the heating of the flare plasma by the non-thermal electrons.  If
the flare plasma is primarily heated by these electrons and the
thick-target region is primarily in the chromosphere and lower
transition region, heating originating in the footpoints and expanding
into the rest of the flare loop through ``chromospheric
evaporation''\index{chromospheric evaporation} should be observed.
On the other hand, if the loop is dense enough for the thick-target
region to extend into the corona or if return-current heating is
important, localized coronal heating and different ion abundances
should be observed.\index{abundances!evaporation}

It has generally been difficult to establish a clear connection
between the location and evolution of X-ray sources produced by
non-thermal electrons and by thermal plasma at different temperatures.
This is largely because of a lack of high-cadence EUV images covering
a broad range of coronal and transition region temperatures prior to the
launch of  \textit{SDO}.  
\index{satellites!SDO@\textit{SDO}}
Future studies of the coevolution of non-thermal X-ray sources and thermal
sources in flares will be important in determining the extent to
which heating mechanisms other than collisional heating by non-thermal
electrons is significant.

Predicting the expected evolution of the heated plasma is hampered
by insufficient knowledge of the dominant heat transport mechanisms.
We have seen evidence that many flares cool by classical thermal
conduction\index{thermal conduction} or radiation once the heating
has subsided (Section~\ref{sec:aschwanden_timing_mthdelays}), but
this is not likely to be the dominant transport mechanism during
rapid heating.  Nevertheless, the spatial evolution of flare X-ray
sources has so far been found to be consistent with chromospheric
evaporation\index{chromospheric evaporation} (Section
\ref{sec:liuw_height}).  Also, the Neupert effect,
\index{Neupert effect} 
observed in most flares, and Doppler-shift measurements
qualitatively support the thick-target model
(Section~\ref{sec:aschwanden_timing_thdelays}), but these do not
rule out the possibility of other heating mechanisms temporally
correlated with the electron beam collisional heating.  As discussed
in Section~\ref{sec:holman_cutoffs}, substantial progress has been
made in deducing the energy flux (total power)
\index{electrons!distribution function!total power} carried by non-thermal
electrons, but we usually can deduce only a lower limit to this
energy flux.  
Continuing studies of flares similar to SOL2002-04-15T03:55 (M1.2)
\index{flare (individual)!SOL2002-04-15T03:55 (M1.2)!electron energy flux}
and the initially cooler, early-impulsive flares\index{flare types!early
impulsive} (Section~\ref{sec:sainthilaire_Ec}) may provide a better
handle on this energy flux for comparison with thermal evolution.
The thermal properties, energetics, and evolution of flares are
discussed further in \citet{Chapter2}\index{flare types!early impulsive}.

\subsection{Implications of X-ray observations for electron
acceleration mechanisms and flare models}
\label{sec:holman_discussion_models}
\index{flare models!and X-ray observations}

In Section~\ref{sec:battaglia_footpoint_coronal} we addressed the
X-ray spectra of hard X-ray sources sometimes observed above the
top of the hot loops or arcades of loops observed in
flares.
\index{soft X-rays!coronal sources}  
We reviewed results indicating that the spectra are qualitatively, 
but not quantitatively consistent
with expectations for electrons passing through a thin-target or
quasi-thick-target region on their way to the thick-target footpoints
of the flare loops.  The apparent failure of these relatively simple
models is probably a manifestation of the more complex above-the-looptop
X-ray source structure revealed by {\it RHESSI} observations.

Before {\it RHESSI}, time-of-flight delays\index{hard X-rays!time delays}
in hard X-ray timing indicated that electrons were accelerated in
a region\index{acceleration region} somewhat above the looptops of
the hot flare loops in most flares
(Section~\ref{sec:aschwanden_timing_tof}).\index{time-of-flight analysis}
Also, cusps were observed
at the top of flare loops by {\it Yohkoh} (e.g.,
Section~\ref{sec:battaglia_footpoint_coronal}), indicating a magnetic
connection to the region above the hot loops.

{\it RHESSI} images have revealed flares with double coronal
sources,
\index{soft X-rays!coronal sources!double} 
one at or just above
the top of the hot loops and the other at a higher altitude above
the lower source.  The centroid of the lower source is higher in
altitude at higher X-ray energies, while the centroid of the upper
source is lower in altitude at higher X-ray energies, indicating
that energy release occurred between these coronal sources
\citep{2003ApJ...596L.251S,2004ApJ...612..546S,2008ApJ...676..704L,2009ApJ...698..632L}.
In one flare, the upper source accelerated outward to the speed of
a subsequent coronal mass ejection\index{coronal mass ejections (CMEs)}\index{satellites!SMM@\textit{SMM}}\index{satellites!SOHO@\textit{SOHO}}\index{satellites!TRACE@\textit{TRACE}}.
The white-light coronagraph on the {\it Solar Maximum
Mission}\index{Solar Maximum Mission@\textit{Solar Maximum Mission}} \citep{2003JGRA..108.1440W},
the Large Angle and Spectrometric Coronagraph (LASCO) and the
Ultraviolet Coronagraph Spectrometer (UVCS) on {\it SOHO}\index{SOHO@\textit{SOHO}}
\citep{2003ApJ...594.1068K, 2005ApJ...622.1251L}, {\it RHESSI}
\citep{2005ApJ...633.1175S}, and {\it TRACE}\index{TRACE@\textit{TRACE}}
\citep{2006ApJ...646..605S} have all provided direct evidence for
the presence of an extended, vertical current sheet above the hot
\index{current sheets}
flare loops and below the coronal mass ejection associated with
eruptive flares.  These and related observations are discussed
further in \citet{Chapter2}.

These recent observations strongly support the ``standard'' model
of eruptive solar flares, in which the hot flare loops build up
below a vertical current sheet where inflowing magnetic fields
reconnect and a magnetic flux rope forms above the current sheet
to become a coronal mass ejection \citep[see][]{Chapter2, Chapter8}.
The rate of electron acceleration has been observed to be correlated
with the rate at which magnetic flux is swept up by the expanding
footpoints of flare loops and with the rate of looptop expansion
\citep{2004ApJ...604..900Q, 2004ApJ...612..546S, 2005AdSpR..35.1669H},
indicating that the electron acceleration rate is correlated with
the rate of magnetic reconnection\index{reconnection!rate correlated with acceleration}.
On the other hand, the observations also indicate that the rate of
electron acceleration in the impulsive phase of flares is greatest
{\em before} a large-scale current sheet or soft X-ray cusp is
observed \citep{2008AdSpR..41..976S}\index{magnetic structures!cusps}\index{flares!model}\index{flare models!standard}.

Initially, when the electron acceleration rate is highest, the
current sheet may be short and associated with slow-mode shock
waves, as in Petschek reconnection\index{shocks!slow mode}\index{reconnection!Petschek}.
Fast reconnection jets\index{reconnection!outflow}\index{jets!reconnection outflow}
\citep[e.g.,][]{2007ApJ...661L.207W} can stream upward and downward
from the current sheet, possibly ending in fast-mode shock waves
where they collide with slower magnetized plasma at the flare loop
tops and the lower boundary of the magnetic flux rope (termination
shocks).\index{shocks!termination}\index{above-the-looptop sources}\index{looptop sources}
The pair of above-the-looptop
X-ray sources may be associated with these fast-mode shock waves.
We have described possible evidence for these shock waves from radio
observations\index{shocks!radio emission} in
Section~\ref{sec:aurass_implications}.

The most difficult task is determining the dominant acceleration
mechanism
\index{acceleration!list of mechanisms} 
or mechanisms responsible for the energetic particles.  
The region above the flare loops contains or
can contain quasi-DC electric fields, plasma turbulence, slow- and
fast-mode shock waves, and collapsing magnetic traps, allowing for
almost any acceleration mechanism imaginable.  
\index{magnetic structures!collapsing traps}\index{plasma turbulence}
The problem is as
much one of ruling out mechanisms as of finding mechanisms that
work \citep[cf.][]{1997JGR...10214631M}.  
Acceleration mechanisms are addressed in \citet{Chapter8}.

In Section~\ref{sec:grigis_spectral_evolution} we addressed the
soft-hard-soft\index{hard X-rays!soft-hard-soft}
evolution of flare X-ray spectra.  This spectral evolution could
occur during the propagation of the electrons from the acceleration
region to the thick-target footpoints.  Return current
\index{return current} losses, with their dependence on the electron beam flux
(Section~\ref{sec:zharkova_return_current}), for example, could be
responsible for this evolution.  However, the observation that
above-the-looptop sources also show this spectral evolution
(Section~\ref{Ba_shs_corn}) indicates that it is a property of the
acceleration process rather than electron beam propagation.  We saw
in Section~\ref{Gr_shs_theory} that the soft-hard-soft behavior can
be reproduced in the acceleration region if the acceleration or
trapping efficiency first increases and then decreases.\index{trapping!magnetic}\index{acceleration region!and SHS pattern}

Flares displaying soft-hard-harder
\index{hard X-rays!soft-hard-harder} 
spectral evolution are of special
interest, because they have been shown to be associated with
high-energy proton events\index{proton events} in space
\citep[][]{1995ApJ...453..973K,2008ApJ...673.1169S,2009ApJ...707.1588G}.
\index{proton events}
\index{solar energetic particles (SEPs)!associated with hard X-rays}
What is the connection between the appearance of energetic protons
in space and X-ray spectral hardening late in flares?  The answer
to this question is important to both space weather\index{space weather} prediction and
understanding particle acceleration in flares.

\subsection{Implications of current results for future flare studies
in hard X-rays}

What characteristics should a next-generation hard X-ray telescope
have to make substantial progress in understanding electron propagation
and acceleration in flares?  The advances made with {\it
RHESSI}\index{RHESSI@\textit{RHESSI}!spectral resolution} 
have depended on its high-resolution count
spectra that could generally be convolved with the detector response
to obtain reliable photon flux spectra.  These have been the first
observations to allow detailed information about the evolution of
accelerated electrons and associated hot flare plasma to be deduced
for many flares.  Equally important has been the ability to produce
hard X-ray images in energy bands determined by the user during the
data analysis process.  This imaging capability has been critical
to determining the origin of the X-ray emission at a given photon
energy and in obtaining spectra for individual imaged source
regions.
\index{RHESSI@\textit{RHESSI}!imaging spectroscopy}\index{imaging spectroscopy} 
These high-resolution
imaging spectroscopy capabilities will remain important for continued
progress.

\textit{RHESSI}'s X-ray imaging capability has allowed a clear spatial
separation to be made for many flares between footpoint sources
with non-thermal spectra at higher energies and looptop sources with
thermal spectra at lower energies. However, in the energy range of
overlap between $\sim$10 keV and 50 keV, where both types of sources
may coexist, it is often difficult to distinguish weaker coronal
sources (both thermal and non-thermal) in the presence of the stronger
footpoint sources. This is because of the limited dynamic
range\index{image dynamic range}\index{RHESSI@\textit{RHESSI}!dynamic range} of $<$100:1 (and significantly
less for weaker events) that is possible in any one image made from
\textit{RHESSI} data. This is a consequence of the particular form of the
Fourier-transform imaging technique that is used. Thus, in most
flares the usually intense footpoints mask the much weaker coronal
hard X-ray sources that can sometimes be seen in over-the-limb
flares when the footpoints are occulted
\citep[e.g.,][]{2008ApJ...673.1181K}.\index{occulted sources}
In fact, these coronal hard
X-ray sources can extend to high energies \citep[up to $\sim$800
keV,][]{2008ApJ...678L..63K} and seem to be non-thermal in origin,
thus making them of great interest in locating and understanding
the particle acceleration process. It is important to study these
non-thermal coronal sources in comparison with the footpoint sources,
something that is currently not possible with \textit{RHESSI}'s limited dynamic
range except in the few cases with exceptionally strong coronal
emission (see Section 10). In addition, again because of the \textit{RHESSI}
dynamic range and sensitivity limitations,  it has not generally
been possible to observe the thin-target bremsstrahlung
emission\index{hard X-rays!thin-target} from the corona that must be
present from the electrons streaming down the legs of magnetic loops
and also from electrons streaming out from the Sun and producing
type-III bursts \citep[see,
however,][]{2008ApJ...681..644K,2009ApJ...696..941S}. For all of
these reasons, a significantly greater dynamic range will be an
important goal for future advanced solar hard X-ray instruments.

Flares at the solar limb for which the hard X-ray footpoints are
\index{occulted sources}
occulted\index{hard X-rays!occulted} by the disk provide an important
way of observing coronal hard X-ray sources\index{hard X-rays!coronal
sources}, but these flares do not allow a comparison to be made
between the coronal emission and the thick-target footpoint emission.
A possible substitute for a high-dynamic-range instrument is hard
X-ray observations from two or more spacecraft.  Under the right
conditions, one spacecraft can observe all the flare emission while
the other observes only the coronal emission, with the footpoint
emission occulted by the solar disk.  Multi-spacecraft observations
would also be important for deducing the directivity of the flare
emission \citep[especially in conjunction with X-ray
\index{polarization}
polarization\index{hard X-rays!polarization} measurements -- see][]{Chapter7}
and 3-D source structure.  This multi-spacecraft approach, however,
limits the number of flares for which the coronal and footpoint
emissions can be compared.

Hard X-ray timing\index{hard X-rays!time delays} studies have provided
valuable information about electron propagation and the location
of the acceleration region (Section~\ref{sec:aschwanden_timing}).
Since the time of flight of energetic electrons from a coronal
acceleration region to the thick-target loop footpoint is typically
$\sim$10--100~milliseconds, the photon count rate must be high
enough to distinguish differences in flux on these time scales.
Time-of-flight studies have not been successful with {\it RHESSI},
because of its relatively low collecting area and, therefore, count
rate.  
\index{satellites!CGRO@\textit{CGRO}}
An instrument with the collecting area and pulse-pileup
\index{pulse pileup} avoidance of {\it CGRO}/BATSE\index{Compton@\textit{Compton Gamma Ray Observatory (CGRO)}}, and the imaging and spectral resolution of {\it RHESSI}, would provide a new generation of studies on the characteristic time scales of propagation for the hard-X-ray-emitting electrons
accelerated in flares.  
Alternatively, smaller instruments sent
\index{satellites!Solar Orbiter@\textit{Solar Orbiter}}
\index{satellites!Solar Probe@\textit{Solar Probe}}
closer to the Sun on, for example, \textit{Solar Orbiter}\index{Solar Orbiter@\textit{Solar
Orbiter}} or \textit{Solar Probe}\index{Solar Probe@\textit{Solar Probe}} could achieve the
required sensitivity.  
Flare studies on these time scales would
provide important insights into the physical processes that impact
the acceleration and propagation of energetic electrons in flares.

\begin{acknowledgements}

We thank the chapter editor, Brian Dennis, and the two reviewers
for comments that led to many improvements to the text.  GDH
acknowledges support from the \textit{RHESSI} Project and NASA's Heliophysics
Guest Investigator Program.  MJA acknowledges support from NASA
contract NAS5-98033 of the \textit{RHESSI} mission through University of
California, Berkeley (subcontract SA2241-26308PG), and NASA contract
NAS5-38099 for the \textit{TRACE} mission.  HA acknowledges support by the
German Space Agency {\it Deutsches Zentrum f\"ur Luft- und Raumfahrt
} (DLR), under grant No. 50 QL 0001.  MB acknowledges support by
the Leverhulme Trust.  PCG acknowledges support from NASA contract
NNM07AB07C.  EPK acknowledges support from a Science and Technology
Facilities Council Advanced Fellowship.  NASA's Astrophysics Data
System Bibliographic Services have been an invaluable tool in the
writing of this article.

\end{acknowledgements}

\clearpage

\bibliography{%
book_chapters,%
rhessi_ads,%
aschwanden_ads,%
aurass_ads,%
battaglia_footpoint_coronal,%
grigis_spectral_evolution,%
holman_ads,%
kontar_ads,%
liuw_leg,%
sainthilaire_ads,%
zharkova_ads%
}

\bibliographystyle{ssrv}

\printindex

\end{document}